\documentclass[a4paper,fleqn,usenatbib]{mnras}

\usepackage{newtxtext,newtxmath}

\usepackage[T1]{fontenc}
\usepackage{ae,aecompl}

\usepackage{graphicx}	
\usepackage{amsmath}	
\usepackage{amssymb}	

\newcommand{\kms}{\,km\,s$^{-1}$}	

\newcommand{\Msun}{\rm M_{\odot}}
\newcommand{\Lsun}{\rm L_{\odot}}

\newcommand{\feh}{\rm [Fe/H]}

\def\tablefoot#1{\par\vspace*{2ex}%
 \parbox{\hsize}{\leftskip0pt\rightskip0pt
 {\noindent\small\textbf{Notes.}~#1\par}}}
\def\tablefootmark#1{$^{#1}$\,\ignorespaces}
\def\tablefoottext#1#2{$^{(#1)}$~#2}

\title[Star formation histories of NSCs]{Stellar populations and star formation histories of the nuclear star clusters in six nearby galaxies
\thanks{Based on observations made with ESO telescopes at the La Silla Paranal Observatory under programme IDs 084.B-0499(C) and 086.B-0651(C)}
}

\author[N. Kacharov et al.]{
Nikolay Kacharov,$^{1}$\thanks{E-mail: kacharov@mpia.de}
Nadine Neumayer,$^{1}$
Anil C. Seth,$^{2}$
Michele Cappellari,$^{3}$
\newauthor{}
Richard McDermid,$^{4,5}$
C. Jakob Walcher,$^{6}$
and Torsten B\"{o}ker$^{7}$
\\
$^{1}$Max Planck Institut f\"{u}r Astronomie, K\"{o}nigstuhl 17, 69117 Heidelberg, Germany \\
$^{2}$Department of Physics and Astronomy, University of Utah, 115 South 1400 East, Salt Lake City, Utah 84112, USA \\
$^{3}$Sub-department of Astrophysics, Department of Physics, University of Oxford, Denys Wilkinson Building, Keble Road, Oxford OX1 3RH, UK \\
$^{4}$Department of Physics and Astronomy, Macquarie University, Sydney, NSW 2109, Australia \\
$^{5}$Australian Gemini Office, Australian Astronomical Observatory, PO Box 915, Sydney, NSW 1670, Australia \\
$^{6}$Leibniz-Institut f\"{u}r Astrophysik Potsdam (AIP), An der Sternwarte 16, 14482, Potsdam, Germany \\
$^{7}$European Space Agency, c/o STScI, 3700 San Martin Drive, Baltimore, MD 21218, USA
}

\date{Accepted XXX; Received YYY; in original form ZZZ}

\pubyear{2018}

\begin{document}
\label{firstpage}
\pagerange{\pageref{firstpage}--\pageref{lastpage}}
\maketitle

\begin{abstract}
The majority of spiral and elliptical galaxies in the Universe host very dense and compact stellar systems at their centres known as nuclear star clusters (NSCs).
In this work we study the stellar populations and star formation histories (SFH) of the NSCs of six nearby galaxies with stellar masses ranging between $2$ and $8\times10^9\,\Msun$ (four late-type spirals and two early-types) with high resolution spectroscopy.
Our observations are taken with the X-Shooter spectrograph at the VLT.
We make use of an empirical simple stellar population (SSP) model grid to fit composite stellar populations to the data and recover the SFHs of the nuclei.
We find that the nuclei of all late-type galaxies experienced a prolonged SFH, while the NSCs of the two early-types are consistent with SSPs.
The NSCs in the late-type galaxies sample appear to have formed a significant fraction of their stellar mass already more than $10$\,Gyr ago, while the NSCs in the two early-type galaxies are surprisingly younger.
Stars younger than $100$\,Myr are present in at least two nuclei: NGC\,247 \& NGC\,7793, with some evidence for young star formation in NGC\,300's NSC.
The NSCs of the spirals NGC\,247 and NGC\,300 are consistent with prolonged {\it in situ} star formation with a gradual metallicity enrichment from $\sim-1.5$\,dex more than $10$\,Gyr ago, reaching super-Solar values few hundred Myr ago.
NGC\,3621 appears to be very metal rich already in the early Universe and NGC\,7793 presents us with a very complex SFH, likely dominated by merging of various massive star clusters coming from different environments.
\end{abstract}




\section{Introduction}

Nuclear star clusters (NSC) are among the densest stellar objects in the Universe.
They reside in the dynamical centres \citep{neumayer+2011} of more than $75\%$ of spiral galaxies \citep{boeker+2002,seth+2006,georgiev+boeker2014}.
A similar nucleation fraction is also found for dwarf ellipticals and early-type galaxies \citep{cote+2006,turner+2012,denBrok+2014}.
Their sizes are comparable to those of globular clusters \citep[${\rm r_{eff}}\sim2-5$\,pc][]{boeker+2004,cote+2004,cote+2006} and are intrinsically very luminous with typical absolute magnitudes ${\rm M_I}\sim-12$\,mag and masses in the range $10^5-10^7\,\Msun$ \citep{ho+filippenko1996,gordon+1999,matthews+gallagher2002,walcher+2005,georgiev+2016}.

Several studies \citep{ferrarese+2006,rossa+2006,wehner+harris2006,turner+2012,georgiev+2016} have shown that NSCs' masses are coupled to the mass of their host galaxy, following similar scaling relations with galaxy mass as super massive black holes (SMBH) but with possibly different slopes \citep{balcells+2007,leigh+2012,graham2012,scott+graham2013}.
It is worth noting that the NSCs scaling relations vary with galaxy morphology \citep{seth+2008a,erwin+gadotti2012,georgiev+2016} and more concentrated galaxies tend to host brighter NSCs \citep{denBrok+2014}.
This has lead to the conclusion that the formation and evolution of both types of central compact objects are somehow linked to the evolution of their host galaxies, but the observed differences suggest this link is not the only process shaping NSC growth.

NSCs also often co-exist with SMBHs in the centres of galaxies \citep[e.g.][]{seth+2008a,graham+spitler2009,seth+2010,neumayer+walcher2012,nguyen+2018}, including in the Milky Way \citep{genzel+2010,schoedel+2014}.
Unlike SMBHs, however, the formation history of NSCs is directly visible through their stellar populations, which allows for much more detailed studies of the mass accretion history, that takes place in the galactic nuclei.
However, spectroscopic studies of the stellar content of NSCs have been very scarce to date.
Often high extinction and contamination from a prominent bulge component make such studies very challenging.
\citet{walcher+2006,rossa+2006,seth+2006,lyubenova+2013,carson+2015,spengler+2017} reveal that they consist of multiple stellar populations, the majority hosting both young ($<100$\,Myr) and old ($>1$\,Gyr) stars.
Related to the stellar populations, the morphologies of NSCs also appear to be complex.
\citet{seth+2006} find that in edge-on spirals, the NSCs are typically elongated and aligned with the disk of the galaxy, and some appear to host a younger disk component with an old spheroidal core.
Strong rotation is also a common kinematic feature of NSCs \citep[e.g.][]{seth+2008b}.
There is also evidence for such complexity in the stellar populations and morphologies of the NSCs of early-type galaxies \citep{monaco+2009,seth+2010,lyubenova+2013,nguyen+2018}.
\citet{georgiev+boeker2014,carson+2015} find that the sizes of the NSCs in spiral galaxies vary with wavelength.
The more common case is a positive correlation, indicating that the younger population is more centrally concentrated than the old one \citep[this is also the case in the Milky Way NSC][]{do+2013,lu+2013,feldmeier+2015}, but there are also exceptions that show circumnuclear star formation.
In early-type galaxies, NSCs have been found to be younger on average than their hosts \citep{lotz+2004,butler+martinez-delgado2005,cote+2006,chilingarian+2007,chilingarian2009,paudel+2011,guerou+2015}.

The morphologies, sizes, and masses of NSCs are similar to those of the most massive globular clusters and ultra-compact dwarf galaxies \citep[UCD, e.g.][]{norris+2014}, the latter being most likely the remnant nuclei of tidally dissolved galaxies \citep{bekki+2001,bekki+2003,boeker2008,goerdt+2008,pfeffer+baumgardt2013,mieske+2013,seth+2014,ahn+2017}.
The nucleus of the Sgr dwarf spheroidal (M\,54), currently in the process of tidal disruption by the Milky Way, is a prime such example.
Several discrete multiple populations are present in its colour magnitude diagram \citep[CMD;][]{siegel+2007}.
The stellar populations of UCDs are typically old \citep{janz+2016}, but in some cases do seem to show evidence of an extended star formation history (SFH) as expected if they are stripped nuclei \citep{norris+2015}.
As UCDs may have been stripped at some time in the past, understanding the link between these objects and NSCs requires understanding the complete evolution of NSCs over a Hubble time.

In this paper we study the stellar populations and SFHs of four nearby bulgeless spiral galaxies (NGC\,247, NGC\,300, NGC\,3621, and NGC\,7793) and two dwarf early-type galaxies (NGC\,5102 and NGC\,5206) through integrated light spectroscopic observations.
\citet{ruiz-lara+2015,ruiz-lara+2018} have demonstrated that this approach leads to comparable results to recovering SFHs from CMDs, and it is especially valuable when studying distant objects, where individual stars cannot be resolved.
We aim to put constraints on the formation mechanisms and chemical evolution of these nuclei disentangling between {\it in situ} formation and star clusters accretion scenarios.
In the former scenario gas funnels into the nucleus and fuels new star formation episodes \citep{mihos+hernquist1994,milosavljevic2004,schinnerer+2008,bekki2015}.
In this scenario we expect mixing of the gas in the centre, which would observationally manifest as a continuous chemical enrichment, where older populations appear more metal poor, and younger populations increasingly more metal rich.
In the star clusters accretion scenario \citep{tremaine+1975,oh+lin2000,lotz+2001,capuzzo-dolcetta+miocchi2008a,capuzzo-dolcetta+miocchi2008b,antonini+2012,gnedin+2014}, on the other hand, we expect more stochastic mixing of the stellar populations, as stars are born in presumably different environments than the central region of the galaxy.
In reality, the two scenarios most likely operate simultaneously \citep{neumayer+2011,denBrok+2014,antonini+2015,cole+2017}.

\section{Sample selection}

\begin{table*}
	\centering
	\caption{Host galaxy properties.}
	\resizebox{\textwidth}{!}{
	\label{tab:galaxies}
	\begin{tabular}{ccccccc} 
		\hline
Galaxy  & Distance\tablefootmark{1} & Galaxy $\rm M_B$\tablefootmark{2} & Galaxy (B$-$V)\tablefootmark{2} & Stellar mass    &   $\rm r_{eff}^{NSC}$& $E(B-V)_{\rm MW}$\tablefootmark{6} \\
        & [Mpc]                     & [mag]                             & [mag]                           & [$10^9\,\Msun$] & [pc]     	           & [mag] \\	     
		\hline   
NGC\,247 &  $3.4\pm0.7$ & $-19.24\pm0.01$ & $0.56$ & $2.8$\tablefootmark{3} & $1.0$\tablefootmark{3} & $0.016$ \\
NGC\,300 &  $2.0\pm0.4$ & $-18.14\pm0.21$ & $0.89$ & $2.2$\tablefootmark{3} & $1.9$\tablefootmark{3} & $0.011$ \\
NGC\,3621 & $6.7\pm1.3$ & $-20.07\pm0.23$ & $0.68$ & $8.1$\tablefootmark{3} & $1.8$\tablefootmark{3} & $0.071$ \\
NGC\,5102 & $3.7\pm0.7$ & $-17.91\pm0.23$ & $0.68$ & $6.8$\tablefootmark{4} & $1.6$\tablefootmark{4} & $0.048$ \\
NGC\,5206 & $3.2\pm0.6$ & $-16.07\pm0.63$ & $0.87$ & $2.6$\tablefootmark{4} & $3.4$\tablefootmark{4} & $0.105$ \\
NGC\,7793 & $3.8\pm0.8$ & $-18.67\pm0.16$ & $0.54$ & $4.4$\tablefootmark{5} & $10.2$\tablefootmark{5}& $0.017$ \\
		\hline
	\end{tabular}}
\tablefoot{
\tablefoottext{1}{Median distance from NED with assumed $20\%$ uncertainty;}
\tablefoottext{2}{HyperLeda \url{http://leda.univ-lyon1.fr/};}
\tablefoottext{3}{\citet{georgiev+2016};}
\tablefoottext{4}{\citet{nguyen+2018};}
\tablefoottext{5}{\citet{bell+2003}}
\tablefoottext{6}{Values from NED: \citet{schlafly+finkbeiner2011} assuming the extinction law of \citet{ccm1989}.}
}
\end{table*}

We target prominent nearby nuclear star clusters in the southern hemisphere with minimal obscuration or bulge contamination.
These galaxies are chosen as part of an adaptive optics targeted sample, which is roughly complete within 4\,Mpc, and the galaxy NGC\,3621 is also included due to prior evidence for an accreting central SMBH in this object \citep{satyapal+2007,barth+2009}.
A brief description of each galaxy in our sample is given below and their properties are summarized in Table \ref{tab:galaxies}.

{\bf NGC\,247} is a bulgeless dwarf spiral galaxy (SAB(s)d) at a distance $\sim3.6$\,Mpc (the quoted distances in this work are the median estimates from NED\footnote{\url{https://ned.ipac.caltech.edu/}}) and a member of the Sculptor group.
\citet{carson+2015} report a colour gradient in its NSC - the effective radius is smallest in UV and largest in IR, which suggests the presence of multiple stellar populations, and that the young stars are centrally concentrated and likely formed {\it in situ}.

{\bf NGC\,300} is a bulgeless field spiral galaxy (SA(s)d) projected in front of the Sculptor group at a distance $\sim2$\,Mpc and an inclination angle of $42^{\circ}$.
Similarly to NGC\,247, the size of the NSC of NGC\,300 also increases monotonically from the UV to the IR \citep{carson+2015}.
\citet{walcher+2005} measure an average velocity dispersion $\sigma = 13.3$\kms for the NSC, using UVES spectroscopy, and estimate a dynamical mass of $\sim10^6\,\Msun$ of the NSC.

{\bf NGC\,3621} is a bulgeless field spiral galaxy (SA(s)d) in the constellation Hydra at a distance $\sim6.7$\,Mpc and an inclination angle of $25^{\circ}$.
NGC\,3621 hosts an active galactic nucleus \citep[AGN][]{satyapal+2007,gliozzi+2009} and it is classified as a Seyfert 2, fueled by a SMBH with a mass of $\lesssim3\times10^6\,\Msun$ \citep{barth+2009}.
\citet{carson+2015} note that the NSC is more extended in the UV than in the IR, at a difference with the other two spirals, which could either mean that young stars are situated in the outskirts of the cluster, or that the UV emission from the central AGN is stronger at larger radii due to dust obscuration in the inner regions.
\citet{carson+2015} also note that the UV emission is strongly elongated along the north-south axis, while the IR flux has mostly spherical morphology.

{\bf NGC\,5102} is a nucleated dwarf lenticular (S0) galaxy in the M\,83 group of galaxies at a distance of $\sim3.7$\,Mpc.
Previous studies based on resolved stars and spectral synthesis have shown that the galaxy is dominated by an intermediate age ($\sim0.1 - 3$\,Gyr) stellar population towards its centre \citep{kraft+2005,davidge2008,davidge2015,mitzkus+2017}.
\citet{mitzkus+2017} find a mass weighted age of $0.8$\,Gyr and their composite population fit favours the existence of a young population of $\sim0.3$\,Gyr at solar metallicity together with an old ($>10$\,Gyr), metal poor ($\feh<-1$) population.
Their kinematic model suggests the presence of counter-rotating disks.
\citet{mitzkus+2017,nguyen+2018} also measure a flat velocity dispersion profile at $\sigma \sim 44$\kms in the circumnuclear region ($0.7\arcsec$ to $0.4\arcsec$ from the centre), that peaks at $60$\kms in the very centre.
They also find a strong rotation $\sim30$\kms in the nucleus.
\citet{nguyen+2018} detect a central SMBH with a mass of $8.8^{+4.2}_{-6.6}\times10^5\,\Msun$ and estimated the dynamical mass of the NSC to be $7.30\pm2.34\times10^7\,\Msun$.

{\bf NGC\,5206} is a nucleated dwarf elliptical (dE/S0) galaxy that belongs to the M\,83 group, and is located at a distance of $\sim3$\,Mpc.
\citet{nguyen+2018} measure similar to NGC\,5102 flat velocity dispersion profile in the central region (from $0.7\arcsec$ to $0.4\arcsec$) at $33$\kms, that peaks at $46$\kms at the very centre.
They measure a slow degree of rotation ($\sim10-15$\kms) and an overall roundness of the NSC.
This galaxy also hosts a central SMBH with a mass of $4.7^{+2.3}_{-3.4}\times10^5\,\Msun$, and the dynamical mass of its NSC is estimated at $1.54\pm0.51\times10^7\,\Msun$ \citep{nguyen+2018}.

{\bf NGC\,7793} is a bulgeless spiral galaxy (SA(s)d) and a member of the Sculptor group located at $\sim3.9$\,Mpc.
Interestingly, the NSC is largest in the UV and its size monotonically decreases towards the IR \citep{carson+2015}, which means that younger stars are predominantly located in the outskirts of the nucleus.
HST images reveal young stars and dust filaments indicative of ongoing star formation in the disk in the immediate vicinity of the NSC.
\citet{walcher+2005} measure an average velocity dispersion of $24.6$\kms of the NSC and make a dynamical mass estimate of $7.8\times10^6\,\Msun$.

\section{Data and models}

\subsection{Observations and data reduction}\label{sec:obs}

The observations were taken with the X-Shooter instrument in service mode during two observing semesters in 2009 - 2011 - programme IDs: 084.B-0499(C) and 086.B-0651(C).
X-Shooter is a long slit, medium resolution, \'{E}chelle spectrograph mounted at the Cassegrain focus of Unit Telescope 2 (Kueyen) of the VLT \citep{vernet+2011}.
The instrument is designed to cover the entire optical and near-infrared wavelength range simultaneously through dichroic splitting of the light in three arms: UVB ($294 - 593$\,nm, spectral resolution $R\sim6200$), VIS ($525 - 1049$\,nm, $R\sim11000$), and NIR ($983 - 2481$\,nm, $R\sim6200$).
The $11$\arcsec~long slit was positioned along the minor axis of the galaxy with widths of $0.8$\arcsec, $0.7$\arcsec, and $0.6$\arcsec~for the UVB, VIS, and NIR arms, respectively.
This paper is focused on the analysis of the UVB and VIS arm data that cover the optical wavelength range.
The spatial pixel scale is $0.16$\arcsec\,px$^{-1}$ and the dispersion scale is $0.2$\,\AA\,px$^{-1}$.
The six NSCs were observed in offset mode adopting a strategy ``object - sky - object'' with typical individual exposure times $\sim10 - 15$\,min.
Along with the scientific targets, we also scheduled calibration observations that include telluric calibration stars (taken every night in stare mode), flux calibration stars (typically taken within 10 days from the science targets in stare mode), and velocity calibration stars (no specific time allocation; observed in nod-on-slit mode).
The latter are used to infer the line spread function (LSF) of the spectrograph, which is presumed to be rather stable over long periods of time.
The observing log is presented in Table \ref{tab:observing-log}.

\begin{table*}
	\centering
	\caption{Observing log.}
	\label{tab:observing-log}
	\begin{tabular}{cccccc} 
		\hline
Object  & Type & Date & Exp. time (UVB, VIS, NIR) & Seeing & Flux calib. date \\
		\hline
	&      &      & P84       &          &                  \\
{\bf NGC\,7793} & {\bf NSC} & {\bf 17 Oct. 2009} & \bf {2$\times$530\,s, 2$\times$508\,s, 2$\times$600\,s} & {\bf 0.9$\arcsec$} & {\bf 18 Oct. 2009} \\
{\bf NGC\,247} & {\bf NSC} & {\bf 17 Oct. 2009} & \bf {2$\times$830\,s, 2$\times$808\,s, 2$\times$900\,s}  & {\bf 0.8$\arcsec$} & {\bf 18 Oct. 2009} \\
{\bf NGC\,300} & {\bf NSC} & {\bf 13 Jan. 2010} & \bf {2$\times$830\,s, 2$\times$808\,s, 2$\times$900\,s}  & {\bf 0.7$\arcsec$} & {\bf 18 Jan. 2010} \\
HD12642\tablefootmark{1} & K5I & 13 Jan. 2010 & 2$\times$0.7\,s, 2$\times$0.7\,s, 2$\times$2.0\,s & $0.9\arcsec$ & 18 Jan. 2010 \\
HD14802\tablefootmark{1} & G2V & 13 Jan. 2010 & 2$\times$0.7\,s, 2$\times$0.7\,s, 2$\times$1.0\,s & $0.7\arcsec$ & 18 Jan. 2010 \\
{\bf NGC\,5206} & {\bf NSC} & {\bf 14 Feb. 2010} & \bf {2$\times$530\,s, 2$\times$508\,s, 2$\times$600\,s} & {\bf 1.6$\arcsec$} & {\bf 14 Feb. 2010} \\
                \hline
	&      &      & P86       &          &                  \\
HD198357\tablefootmark{1} & K3III & 13 Oct. 2010 & 2$\times$0.7\,s, 2$\times$0.7\,s, 2$\times$1.0\,s & $1.6\arcsec$ & 12 Oct. 2010 \\
HD68758\tablefootmark{1} & A1V & 15 Oct. 2010 & 2$\times$0.7\,s, 2$\times$0.7\,s, 2$\times$1.0\,s & $0.6\arcsec$ & 15 Oct. 2010 \\ 
HD40136\tablefootmark{1} & F1V & 17 Oct. 2010 & 2$\times$0.5\,s, 2$\times$0.5\,s, 2$\times$1.0\,s & $1.6\arcsec$ & 17 Oct. 2010 \\ 
{\bf NGC\,3621} & {\bf NSC} & {\bf 14 Dec. 2010} & \bf {2$\times$530\,s, 2$\times$508\,s, 2$\times$600\,s} & {\bf 3.7$\arcsec$} & {\bf 14 Dec. 2010} \\ 
HD92305\tablefootmark{1} & M0III & 17 Dec. 2010 & 2$\times$0.6\,s, 2$\times$0.2\,s, 2$\times$0.7\,s & $1.5\arcsec$ & 14 Dec. 2010 \\ 
{\bf NGC\,3621} & {\bf NSC} & {\bf 17 Dec. 2010} & {\bf 2$\times$530\,s, 2$\times$508\,s, 2$\times$600\,s} & {\bf 2.2$\arcsec$} & {\bf 14 Dec. 2010} \\ 
{\bf NGC\,5206} & {\bf NSC} & {\bf 18 Feb. 2011} & {\bf 2$\times$530\,s, 2$\times$508\,s, 2$\times$600\,s} & {\bf 0.8$\arcsec$} & {\bf 1 Feb. 2011} \\ 
{\bf NGC\,5102} & {\bf NSC} & {\bf 02 Mar. 2011} & {\bf 2$\times$530\,s, 2$\times$508\,s, 2$\times$600\,s} & {\bf 1.4$\arcsec$} & {\bf 1 Feb. 2011} \\
HD130328\tablefootmark{1} & M3III & 02 Mar. 2011 & 2$\times$0.2\,s, 2$\times$0.6\,s, 2$\times$0.2\,s & $1.5\arcsec$ & 1 Feb. 2011 \\
		\hline
	\end{tabular}
\tablefoot{
\tablefoottext{1}{Velocity calibration stars}
}
\end{table*}

The spectra were reduced using the ESO {\sc reflex} X-Shooter pipeline (v.\,2.6.8).
This included bias subtraction (estimated from the over-scan regions of the CCD frames); flat field correction; order tracing; wavelength solution and spatial resampling; flexure compensation; sky subtraction using the offset sky frames; image combination; and flux calibration using dedicated flux standard observations.
The final products are rectified, 2-dimensional spectra in {\sc fits} format with flux units of erg\,s$^{-1}$\,cm$^{-2}$\,\AA$^{-1}$ and three extensions: flux, flux error, and a bad pixel map.

\begin{figure*}
	\includegraphics[width=18cm]{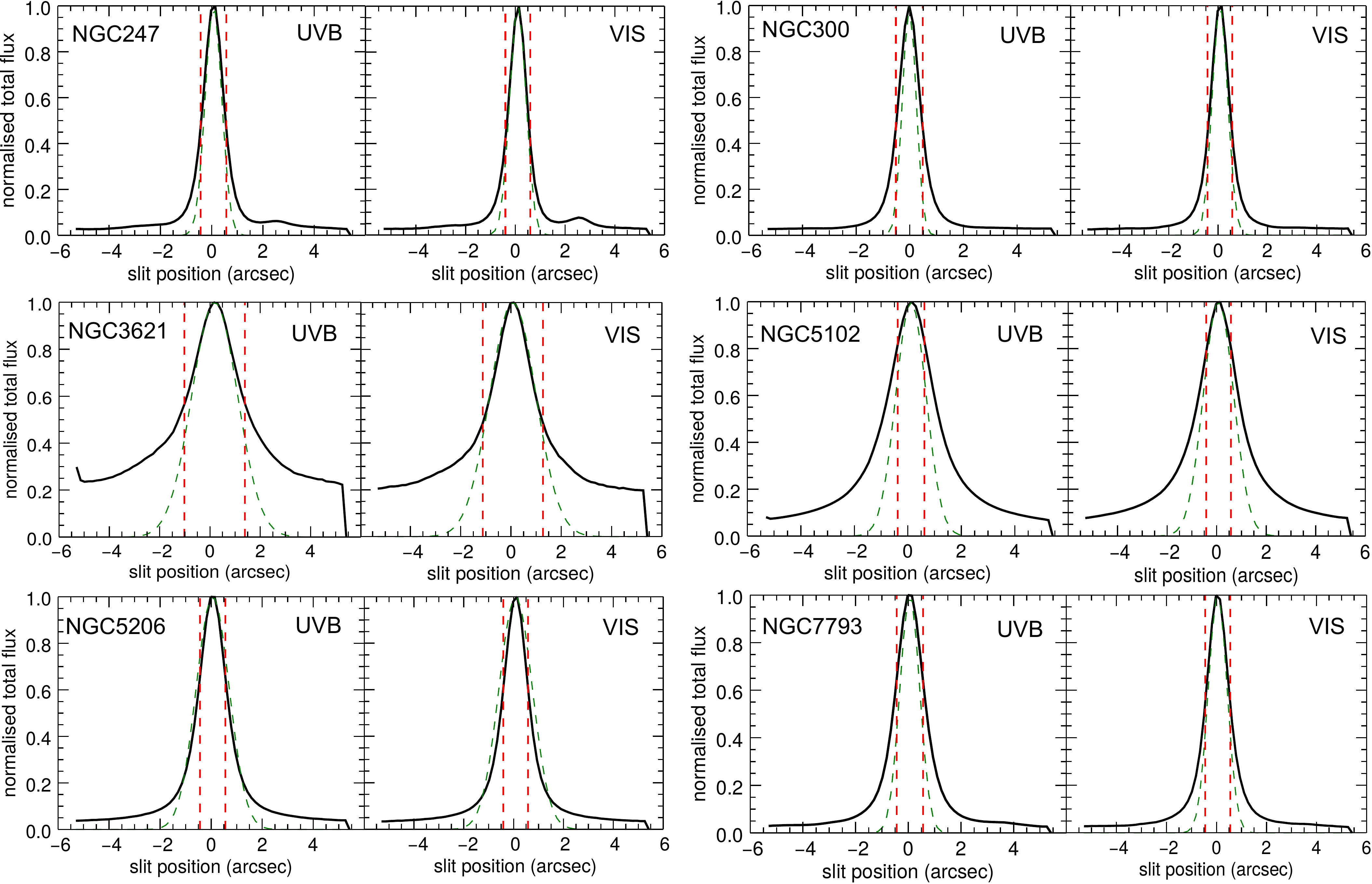}
    \caption{Slit profiles of the light in the UVB and VIS arms for the six galaxies collapsed over all wavelengths and normalised. The green, dashed line is a Gaussian with a FWHM of the seeing as reported by the auto-guiding system of the telescope. The red, dashed, vertical lines indicate the aperture of extraction of the 1D spectrum.}
    \label{fig:profiles}
\end{figure*}

The extraction of the 1D science spectra was done separately in an aperture of $1$\arcsec~by a direct summation of the pixels along the spatial axis.
The resulting rectangular aperture ($1\arcsec\times$the slit width) covers typically between $10$ and $20$\,pc around the centre of the galaxy (see Table \ref{tab:extracted}).
The only exception is the nucleus of NGC\,3621, which is very extended and faint, so to ensure a sufficient signal-to-noise ratio (SNR), we extracted an aperture of $2.4$\arcsec~that covers a diameter of $78$\,pc along the spatial axis of the slit.
The SNR, estimated as the median flux divided by the $1\sigma$ model residuals, varies between $\sim50$ and $\sim85$ per pixel for the different NSCs.

\begin{table*}
	\centering
	\caption{Extracted spectra.}
	\resizebox{\textwidth}{!}{
	\label{tab:extracted}
	\begin{tabular}{ccccccccc} 
		\hline
Galaxy     &  Aperture & Size  & SNR         & $V_{\rm helio}$    & $\sigma_V$  & $E(B-V)_{\rm intr}$ & $\rm{L_B}$\tablefootmark{1} & $\rm{L_V}$\tablefootmark{1} \\
           & [\arcsec] & [pc]  & [px$^{-1}$] & [\kms]             & [\kms]      & [mag] 	      & [$10^6\,\Lsun$]             & [$10^6\,\Lsun$]             \\  
		\hline
NGC\,247   & $1.0$     & $17.4\pm3.5$  & $58$  & $171.6 \pm 0.3$ & $15.1 \pm 0.6$  & $0.084$ & $2.1\pm0.8$  & $1.5\pm0.6$ \\
NGC\,300   & $1.0$     & $9.5\pm1.9$   & $56$  & $146.0 \pm 0.2$ & $13.3 \pm 0.3$  & $0.079$ & $0.7\pm0.3$  & $0.6\pm0.3$ \\
NGC\,3621  & $2.4$     & $78.3\pm15.7$ & $48$  & $720.3 \pm 0.4$ & $41.1 \pm 0.5$  & $0.000$ & $4.9\pm2.0$  & $4.7\pm1.9$ \\
NGC\,5102  & $1.0$     & $18.1\pm3.6$  & $85$  & $464.9 \pm 0.2$ & $46.1 \pm 0.3$  & $0.111$ & $32\pm13$    & $23\pm9$    \\
NGC\,5206  & $1.0$     & $14.5\pm2.9$  & $71$  & $568.4 \pm 0.2$ & $32.2 \pm 0.2$  & $0.038$ & $3.4\pm1.4$  & $3.7\pm1.5$ \\
NGC\,7793  & $1.0$     & $18.9\pm3.8$  & $74$  & $224.1 \pm 0.2$ & $23.1 \pm 0.4$  & $0.147$ & $10\pm4$     & $7.7\pm3.1$ \\
		\hline
	\end{tabular}}
\tablefoot{
\tablefoottext{1}{Total luminosity within the extracted aperture corrected for extinction and systematic uncertainties assuming $20\%$ uncertainty in distance.}
}
\end{table*}

We show in Figure \ref{fig:profiles} the slit profiles of the UVB and VIS arms collapsed over wavelength for all six galaxies.
The NSCs are barely spatially resolved, compared to the expected profile of a point source, approximated here with a Gaussian function with a FWHM of the seeing, as reported from the telescope auto-guiding system at the time of the observation (see Table \ref{tab:observing-log}).
Figure \ref{fig:profiles} also shows that there is generally no significant contamination from the main galaxy component in the NSC (typically $\lesssim10\%$, with the exception of NGC\,3621, where the contamination reaches $\sim25\%$.
In the present analysis we did not subtract the underlying flux to avoid introducing additional noise in the NSC spectra, but we quantified its properties in Section \ref{sec:field}.

The extracted UVB and VIS arm 1D spectra were additionally convolved along the dispersion axis with a Gaussian kernel of $0.7$\,px and $1.2$\,px, respectively, to ensure a smooth dispersion transition between the two arms (see Section \ref{sec:models} for more information on deriving the LSF), and combined together by matching the overlapping regions.
Some small rescaling was generally needed to bring the two arms into scale.
A note of caution at this step is needed, because the overlapping regions of the UVB and VIS arms suffer from time varying dichroic effects \citep[see][]{schoenebeck+2014}, that cannot be corrected out with a flux calibrator in the general case, and the absolute flux in this region is not reliable.
So, we took extra care to ensure that there is no obvious bias between the two arms during the population modeling.
The overlapping region between the two arms ($5400 - 5800$\,\AA) was excluded from the stellar population fits.
We also masked out the pixels marked as bad during the pipeline reduction process and the rest frame region of the Na doublet at $5890$\,\AA, as it is affected by interstellar absorption.

Finally, we corrected the flux for foreground Galactic extinction using the reddening information in the NED database (see also Table \ref{tab:galaxies}) from \citet{schlafly+finkbeiner2011}, assuming the extinction law of \citet{ccm1989}.

\subsection{Population models}\label{sec:models}

In this study we make use of a {\sc pegase-hr} \citep{leBorgne+2004} simple stellar population (SSP) grid provided with the stellar population analysis code {\sc ulyss}\footnote{\url{http://ulyss.univ-lyon1.fr/}} \citep{koleva+2009}.
The {\sc pegase-hr} grid is based on the Elodie 3.1 stellar library \citep{prugniel+soubiran2001,prugniel+2007}.
The SSP models are computed using the Padova isochrones \citep{bertelli+1994}, assuming a Kroupa initial mass function \citep[IMF;][]{kroupa2001}, and normalised to present day mass of $1\,\Msun$.
They include mass loss from supernovae ejecta \citep{woosley+weaver1995} and stellar winds.
They have a resolution $R = 10^4$ and cover a wavelength range between $3900$ and $6800$\,\AA. 
The grid covers 7 metallicity bins with a range $-2.3 < \feh < +0.7$, assuming a scaled Solar abundance pattern, and 68 irregularly spaced age bins between $1$\,Myr and $20$\,Gyr.

We used {\sc ulyss} to interpolate a regular grid in $\ln$\,age and \feh~with $15$ steps in metallicity and $50$ steps in $\ln$\,age between $10$\,Myr and $14$\,Gyr (there are $53$ irregularly spaced age bins in this interval in the original model grid).
This new, interpolated grid was convolved with the LSF of the spectrograph.

\begin{figure}
	\includegraphics[width=\columnwidth]{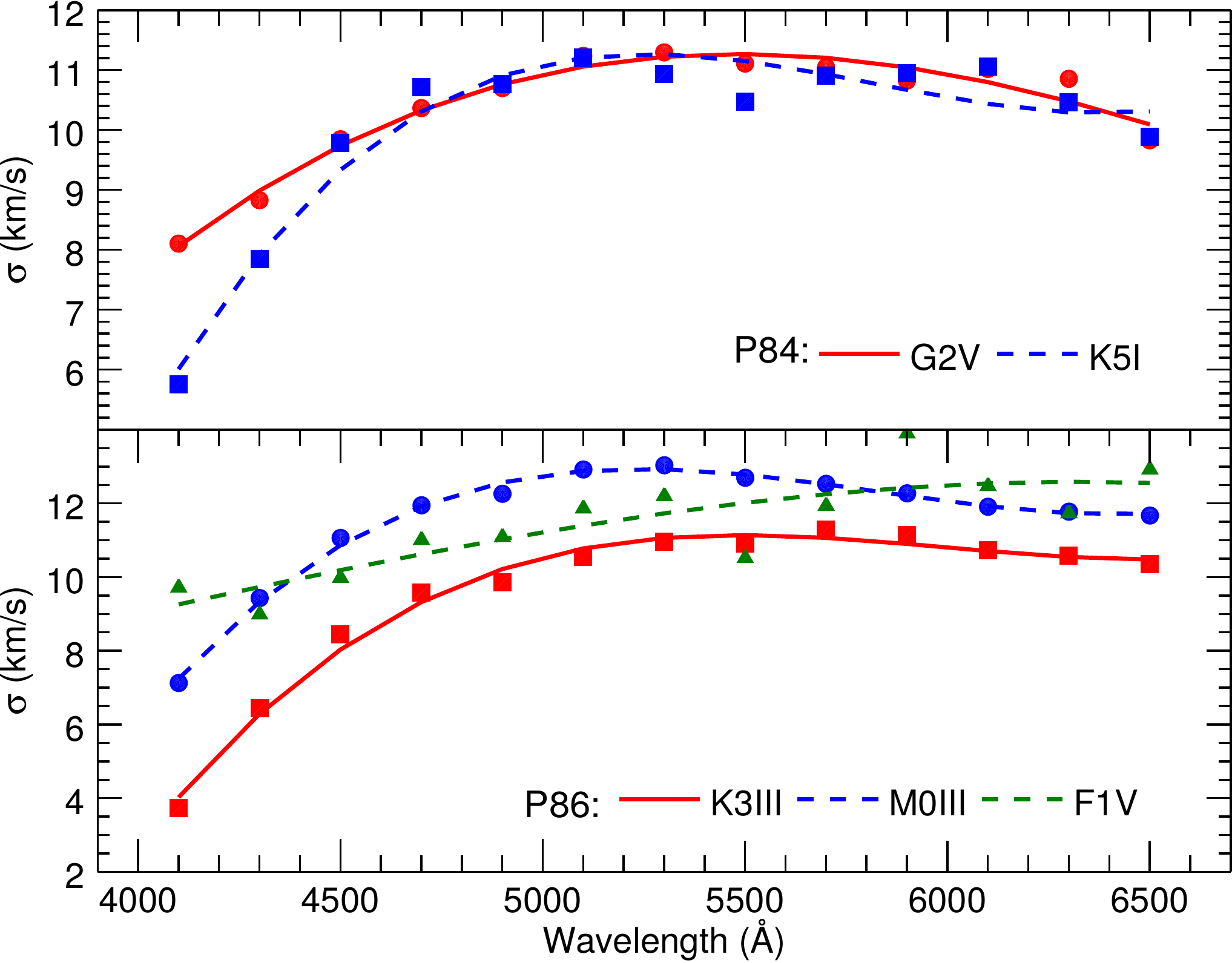}
    \caption{The line broadening function of X-Shooter as derived from two velocity calibration stellar spectra in P84 (top panel) and three spectra in P86 (bottom panel). The adopted LSFs for both observation periods are marked with solid red lines.}
    \label{fig:lsf}
\end{figure}

We assumed a Gaussian LSF with a varying broadening as a function of wavelength.
The broadening function was estimated with {\sc ulyss} by fitting the spectra of the velocity calibration stars (see Table \ref{tab:observing-log} for information about the different type stars) with an empirical stellar grid based on the Elodie 3.1 library, that has the same spectral characteristics as the {\sc pegase-hr} SSP grid.
We measured the necessary broadening of the templates to match the observations in \kms in different parts of the velocity calibration spectra.
We needed to slightly reduce the resolution of the observations by applying a Gaussian kernel with $\sigma = 0.7$\,px in the UVB arm in order to resolve the LSF across the entire wavelength range.
To ensure a smooth LSF transition between the two X-Shooter arms, the VIS arm had to be broadened by a Gaussian kernel with $\sigma = 1.2$\,px.
The derived broadening function was fitted with a 3rd order polynomial, which was subsequently used to correct the SSP model grid.
The broadening functions derived from the different velocity calibration stars are shown in Figure \ref{fig:lsf}.
Although, we observed five velocity calibration stars in P86, only three were suitable for deriving the broadening function and they show a very similar trend also with the observations from P84.
This assures us that the LSF of X-Shooter is temporally stable over sufficiently long periods of time.
The small variance is due to the different spectral types of stars used for each curve.
In the end, we deemed the broadening function derived from the G2V type star for spectra taken in P84 and the K3III type star for spectra taken in P86 as most reliable, because the spectra of these stars have the largest number of narrow absorption lines.
Two of the stars observed in P86 were not suitable for deriving the broadening at all.
The A1V type star is too hot and there are not enough spectral lines in its spectrum, while the M3III type star's spectrum is dominated by broad molecular features.

\subsection{Composite stellar population fits}\label{sec:csp}

While we do provide the age and metallicity of the best fitting SSP to each nucleus (see Section \ref{sec:ssp}), we do not expect that the NSCs have experienced a single burst of star formation.
On the contrary, they have most probably had a prolonged SFH, and have also likely accreted in-falling star clusters of different ages and metallicities \citep[see e.g.][]{walcher+2005}.
In principle, any composite SFH can be described as a linear combination of SSPs.
However, this is a very degenerate problem, as we were convinced by our experiments with {\sc ulyss}, where we tried to reproduce the observed spectra using 2, 3, or more SSP models.
We found that there are solutions with almost identical $\chi^2$ values but very different mix of parameters (ages, metallicities, and weights of the different SSP models), heavily dependent on the initial parameter proposal.
To overcome this problem we decided to use weights regularization, which forces a smooth variation between the weights of neigbouring SSP models in the grid and ensures a unique solution.

Hence, we used the penalized pixel fitting code {\sc ppxf}\footnote{\url{http://purl.org/cappellari/software}} \citep[v.\,5.2.3 for IDL;][]{cappellari+emselem2004,cappellari2017} with the SSP grid described in the previous section.
We used {\sc ppxf} with weights regularization to derive the SFH of the NSCs of the observed galaxies and simultaneously derive their velocity dispersions and line of sight velocities \citep[see also][for other applications of this method]{morelli+2013,norris+2015,mcdermid+2015,morelli+2015,mitzkus+2017}.
The latest major {\sc ppxf} update was specifically designed to properly address situations where the observations and the models have nearly identical spectral resolution and undersampling may become an issue, such is our case \citep{cappellari2017}.
In addition to the SSP grid we added additional templates to model the emission lines from the nuclei if such are present.
The most common emission lines (the Balmer lines, the [S\,II] lines at $6716~ \&~ 6730$\,\AA, the [O\,III] lines at $4959~ \&~ 5007$\,\AA, the  [O\,I] lines at $6300~ \&~ 6364$\,\AA, and the [N\,II] lines at $6548~ \&~ 6583$\,\AA) are modelled as Gaussian functions.
More details on the emission line analysis in the NSCs' spectra are presented in Section \ref{sec:emission}.

\begin{figure}
	\includegraphics[width=\columnwidth]{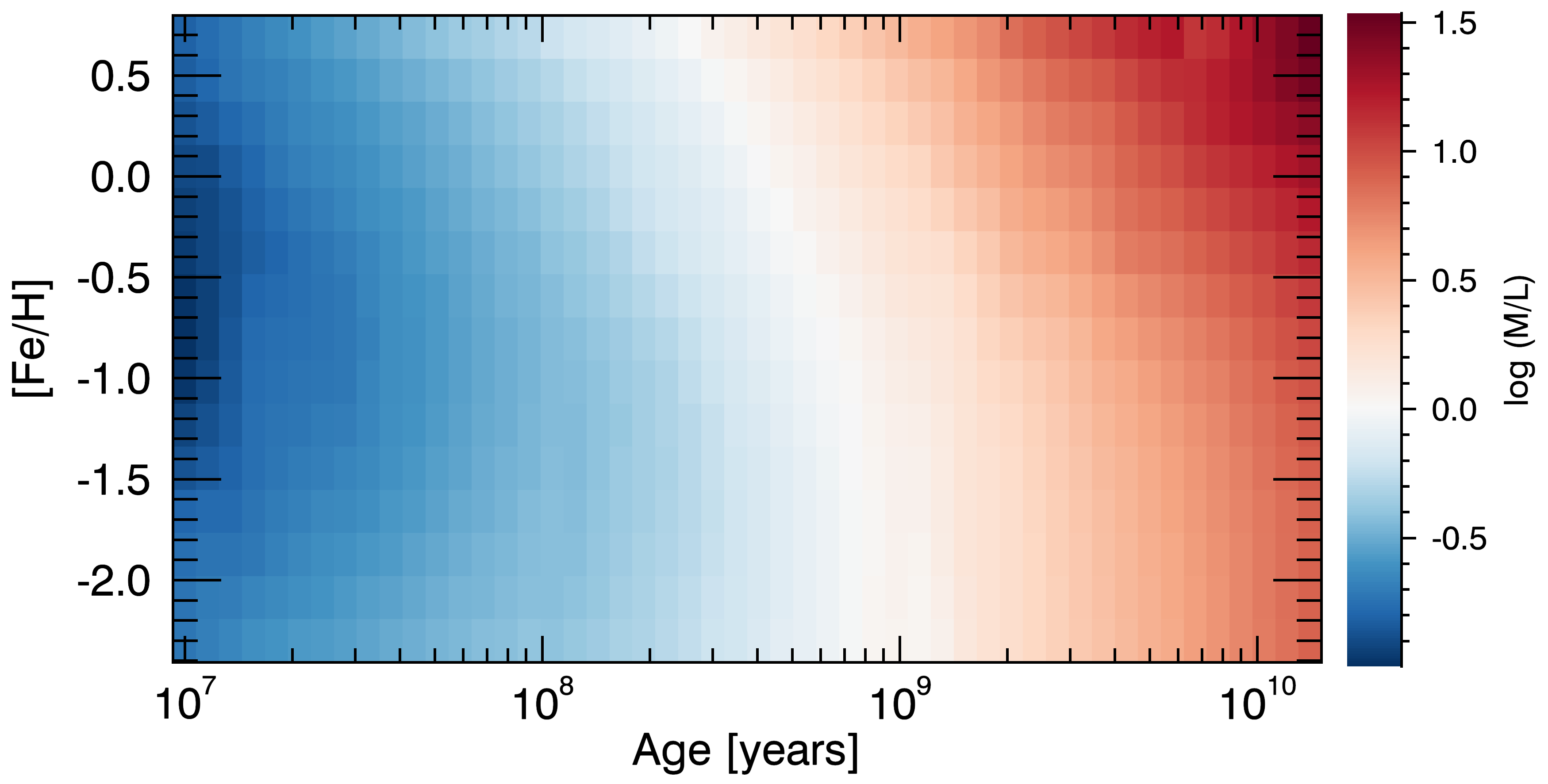}
    \caption{M/L ratios of the individual SSP templates in the model grid. The colour scheme is chosen such that SSP templates with ${\rm M/L}=1$ are indicated with white.}
    \label{fig:ml_templates}
\end{figure}

We scaled the observed spectra and template grid to have a median flux of 1.
This rescaling simplifies our choice of the regularization parameter ($R$; given in Table \ref{tab:age_feh_reg}), which is dependent on the size of the weights.
When the observed and template spectra are more or less on the same scale, we explore regularization factors where $1/R$ is in the order of a few percent.
However, we kept the original templates normalization to unit mass to avoid additional noise introduced from numerical integration.
This means that the median flux of individual SSP templates can be significantly different from $1$ due to the large variation of the M/L ratio ($>2$ orders of magnitude) across the SSP grid.
The M/L ratios of the individual SSP templates are given in Figure \ref{fig:ml_templates}.
In this way, the output {\sc ppxf} weights are proportional to the mass fraction of a given template.
However, {\sc ppxf} fits the light contribution of each SSP, which leads to large range of values for the weights and their variances across the different SSP templates.
For instance, even a tiny light fraction dedicated to old SSPs will result in a large mass weight.
To quantify the weights variance, we performed a bootstrap analysis that is described at the bottom of this section.

We adopted an iterative procedure to fit the SFHs with weights regularization in four steps:
\begin{enumerate}
  \item We estimate the radial velocity and velocity dispersion of the nuclei using a combination of order $10$ additive and multiplicative polynomials along the SSP grid to avoid as much as possible template mismatch. The kinematic results are summarized in Table \ref{tab:extracted}. We keep the derived kinematic parameters fixed in the consequent steps, that pertain to the stellar populations, because we found that even small template mismatch (if no additive polynomial is included in the fit) can lead to an overestimation of the velocity dispersion \citep[see also][]{boardman+2017}. For simplicity, we fit only for the first two velocity moments at this step, keeping higher order velocity moments set to $0$. However, we investigate the effects that ignoring higher order moments have on the SFH recovery in Appendix \ref{sec:h3h4}. We thus confirm that using higher order velocity moments does not noticeably change the population properties of the nuclei, i.e. any changes are well below the quoted uncertainties.
  \item We perform an unregularized fit, which includes no polynomials but internal extinction curve, adopting the reddening law of \citet{calzetti+2000}. The best fit reddening values are shown in Table \ref{tab:extracted}. As noted by \citet{walcher+2006} younger populations may be more attenuated than older ones, but here we make the assumption that a single reddening law describes the entire NSC reasonably well. In Table \ref{tab:extracted} we also list the extinction corrected B- and V-band luminosities of the NSCs as measured from our spectra. 
  \item We redden the SSP model grid with the best-fit extinction curve and perform a second unregularized fit, this time applying a 10th order multiplicative polynomial to correct for irregularities in the flux calibration. The multiplicative polynomial has a mean value of $1$ and if the reddening correction is adequate, it should not have a preferred trend with wavelength. It significantly improves the $\chi^2$ of the fit. We rescale the error spectrum so that the unregularized fit at this step has a reduced $\chi^2=1$ and correct the templates with the multiplicative polynomial. We point out that using the noise model as is from the ESO pipeline reduction package, we typically get reduced $\chi^2$ values close to but lower than $1$. This means that the noise is slightly overestimated and we need to reduce it.
  \item At this step we apply regularization and perform only a linear optimization of the model weights. With {\sc ppxf} we can apply first or second order linear regularization along all dimensions of the SSP-grid, controlled by a single parameter $R$. In this work we use first order regularization ({\sc reg\_ord$=1$} in {\sc ppxf}), where the numerical first derivative of neigbouring weights is equal to $0$ within an error $\sim1/R$, although we tested second order regularization and found that this choice does not affect the results in a significant level. We increase $R$ until the $\chi^2$ is increased by
\begin{equation}\label{eq:max_chi2}
  \Delta\chi^2 = \sqrt{2N},
\end{equation}
where $N$ is the number of pixels considered for the fit. 
The variance of the $\chi^2$ distribution is $2N$, hence this is the smoothest solution that is still compatible with the data within a $1\sigma$ level.
Note that if the data is not compatible with a smooth solution, {\sc ppxf} with regularization will not enforce one due to the $\chi^2$ constraint.
Our spectra have about $1.4\times10^4$ good pixels, which are treated independently.
\end{enumerate}

As we also fit for extinction along the line of sight, this four-step procedure ensures the proper interpretation of the weights of the different SSP templates as mass fractions, given the assumed IMF. 
The sum of all SSP weights (corrected by a known scaling factor) is an estimate of the total mass (including remnants) of the NSC enclosed within the extracted aperture.
We also reconstruct the star formation rate (SFR) in the nuclei as a function of time.
To this aim, we need to correct the inferred current mass fractions in each age bin to their initial values, using the \citet{maraston2005} models for mass loss due to stellar evolution.
Then we divided the initial masses by the time span of each age bin.
The distance uncertainty is the main contributing factor to the systematic error on the mass estimates.
We assumed a $20\%$ error on the distance to all galaxies and computed the systematic mass uncertainty presented in Table \ref{tab:age_feh_reg}. 
Additionally, due to the seeing limited observations, the extraction aperture is only an approximation of the spatial extent of the derived mass.

\begin{figure}
	\includegraphics[width=\columnwidth]{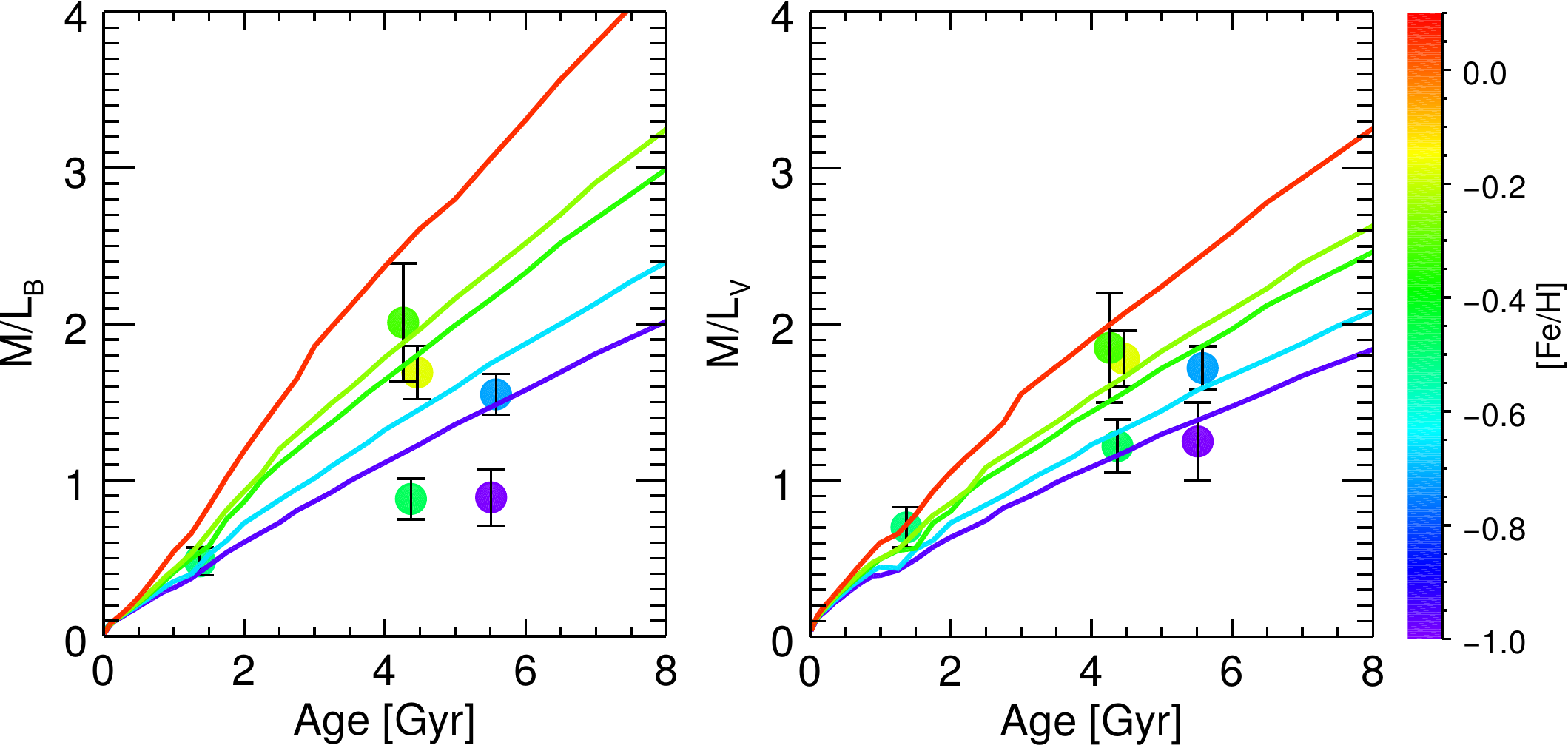}
    \caption{Derived ${\rm M/L_B}$ {\it (left)} and ${\rm M/L_V}$ {\it (right)} ratios of the six NSCs as a function of mass weighted age and metallicity compared to predictions of the BASTI SSP models.}
    \label{fig:ml_basti}
\end{figure}

We present the light- and mass-weighted ages and metallicities computed from the regularized fits, as well as the mass-to-light ratios in the B- and V-bands, for the six NSCs in Table \ref{tab:age_feh_reg}.
Note that the M$/$L ratios are distance independent and thus are not affected by the large distance uncertainties.
In order to compute the luminosities of the extracted spectra, we convolved the contributing SSP models with the B and V Bessel filter transmission curves, integrated the flux, and converted it into Solar luminosities (Table \ref{tab:extracted}).
Since, the presented M$/$L ratios are based on the modelled, rather than on the observed flux, the line-of-sight extinction is automatically taken into account.
Figure \ref{fig:ml_basti} shows that the derived M$/$L ratios in the B- and V-bands are in a good agreement with the predictions of BASTI SSP models \citep{percival+2009}.
The two apparent outliers in the ${\rm M/L_B}$ plot are the NSCs of NGC\,247 and NGC\,7793, which have the most extended SFHs and are hence least compatible with a SSP.

We infer the uncertainties of the individual mass weights of the SSP grid via a bootstrap analysis, where we sampled each science spectrum $100$ times, by resampling the pixels from the best fit model residuals, starting from the residuals of the best fit unregularized solution \citep[see][for the wild bootstrap method]{wu1986}.
Each resampled spectrum is fitted with {\sc ppxf} using a minimum regularization ($R=5$).
This approach allows us to explore the variance of the weights solely due to the variance of the spectrum, without imposing a strong regularization prior, thus not biasing the fit to smooth solutions.
If we were to use maximum regularization in the bootstrap analysis, the solutions would become self-similar.
The derived quantities, such as mean light- and mass-weighted ages and metallicities, mean M$/$L ratios, and masses with their corresponding $1\sigma$ variations are presented in Table \ref{tab:age_feh_mc}.
The corresponding values from the full regularized solutions are generally within these uncertainties.

\begin{table*}
	\centering
	\caption{Mass- and light-weighted ages and metallicities, and extinction corrected M$/$L ratios from the maximum regularized solutions.}
	\label{tab:age_feh_reg}
	\begin{tabular}{ccccccccc} 
		\hline
Galaxy    &  ${\rm AGE_{light}}$   & $\feh_{\rm light}$ & ${\rm AGE_{mass}}$ & $\feh_{\rm mass}$ & Tot. mass\tablefootmark{1} & M$/$L${\rm_B}$            & M$/$L${\rm_V}$            & Max. Regul.  \\
          & [Gyr]                  &                    & [Gyr]              &                   & [${\rm10^6\,M_{\odot}}$]   & [${\rm (M/L_B)_{\odot}}$] & [${\rm (M/L_V)_{\odot}}$] &              \\  
		\hline              
NGC\,247  & $0.60$ & $-0.56$ & $5.51$ & $-0.99$ & $1.88\pm0.75$ & $0.90$ & $1.22$ & $70$  \\
NGC\,300  & $2.04$ & $-0.70$ & $5.58$ & $-0.72$ & $1.10\pm0.44$ & $1.56$ & $1.71$ & $115$ \\
NGC\,3621 & $1.60$ & $-0.23$ & $4.46$ & $-0.17$ & $8.40\pm3.36$ & $1.70$ & $1.78$ & $245$ \\
NGC\,5102 & $0.69$ & $-0.88$ & $1.37$ & $-0.50$ & $15.8\pm6.33$ & $0.49$ & $0.68$ & $15$  \\
NGC\,5206 & $2.91$ & $-0.32$ & $4.26$ & $-0.32$ & $6.89\pm2.75$ & $2.01$ & $1.87$ & $130$ \\
NGC\,7793 & $0.53$ & $-0.39$ & $4.37$ & $-0.47$ & $9.16\pm3.67$ & $0.89$ & $1.19$ & $60$  \\
		\hline
	\end{tabular}
\tablefoot{
\tablefoottext{1}{Total mass within the extracted aperture with systematic uncertainties assuming $20\%$ uncertainty in distance.}
}
\end{table*}

\begin{table*}
	\centering
	\renewcommand{\arraystretch}{1.2}
	\caption{Mass- and light-weighted ages and metallicities, and extinction corrected M$/$L ratios from the bootstrap analysis ($N=100,~R=5$).}
	\label{tab:age_feh_mc}
	\begin{tabular}{cccccccc} 
		\hline
Galaxy    &  ${\rm AGE_{light}}$   & $\feh_{\rm light}$ & ${\rm AGE_{mass}}$ & $\feh_{\rm mass}$ & Tot. mass\tablefootmark{1} & M$/$L${\rm_B}$            & M$/$L${\rm_V}$            \\
          & [Gyr]                  &                    & [Gyr]              &                   & [${\rm10^6\,M_{\odot}}$]  & [${\rm (M/L_B)_{\odot}}$] & [${\rm (M/L_V)_{\odot}}$] \\  
		\hline              
NGC\,247  & $0.61^{+0.11}_{-0.09}$ & $-0.71\pm0.10$ & $4.33^{+2.23}_{-1.47}$ & $-0.94\pm0.22$ & $1.77\pm0.37\pm0.71$ & $0.85\pm0.18$ & $1.15\pm0.24$ \\
NGC\,300  & $1.92^{+0.15}_{-0.14}$ & $-0.72\pm0.06$ & $4.69^{+0.93}_{-0.78}$ & $-0.70\pm0.12$ & $0.99\pm0.09\pm0.40$ & $1.40\pm0.12$ & $1.55\pm0.14$ \\
NGC\,3621 & $1.75^{+0.17}_{-0.16}$ & $-0.42\pm0.08$ & $4.79^{+0.91}_{-0.76}$ & $-0.19\pm0.13$ & $8.52\pm0.84\pm3.64$ & $1.72\pm0.17$ & $1.80\pm0.18$ \\
NGC\,5102 & $0.62^{+0.16}_{-0.13}$ & $-0.89\pm0.34$ & $1.19^{+0.62}_{-0.41}$ & $-0.70\pm0.26$ & $14.4\pm2.8 \pm5.8 $ & $0.44\pm0.09$ & $0.62\pm0.12$ \\
NGC\,5206 & $3.22^{+0.47}_{-0.41}$ & $-0.37\pm0.07$ & $5.28^{+1.62}_{-1.24}$ & $-0.26\pm0.08$ & $8.22\pm1.31\pm3.29$ & $2.38\pm0.37$ & $2.24\pm0.35$ \\
NGC\,7793 & $0.55^{+0.04}_{-0.03}$ & $-0.49\pm0.14$ & $3.89^{+1.10}_{-0.86}$ & $-0.23\pm0.18$ & $9.04\pm1.31\pm3.62$ & $0.88\pm0.13$ & $1.18\pm0.17$ \\
		\hline
	\end{tabular}
\tablefoot{
\tablefoottext{1}{Total mass within the extracted aperture with random and systematic uncertainties assuming $20\%$ uncertainty in distance.}
}
\end{table*}

\subsection{Simple stellar population fits}\label{sec:ssp}

\begin{table*}
	\centering
	\caption{SSP fits.}
	\label{tab:age_feh_ssp}
	\begin{tabular}{ccccccc} 
		\hline
Galaxy  & ${\rm AGE_{SSP}}$ & $\feh_{\rm SSP}$  & Tot. mass\tablefootmark{(1)} & M$/$L${\rm_B}$            & M$/$L${\rm_V}$            & $\chi^2$\tablefootmark{(2)} \\
        & [Gyr]             &                   & [${\rm 10^6\,M_{\odot}}$]  & [${\rm (M/L_B)_{\odot}}$] & [${\rm (M/L_V)_{\odot}}$] & 	        \\  
		\hline              
NGC\,247  & $0.22$ & $+0.26$  & $0.53\pm0.21$ & $0.27$ & $0.31$ & $1.72$ \\
NGC\,300  & $1.53$ & $-0.59$  & $0.47\pm0.19$ & $0.68$ & $0.73$ & $1.16$ \\
NGC\,3621 & $1.77$ & $-0.38$  & $4.72\pm1.89$ & $0.97$ & $0.96$ & $1.06$ \\
NGC\,5102 & $0.54$ & $-0.38$  & $10.2\pm4.1 $ & $0.31$ & $0.44$ & $1.00$ \\
NGC\,5206 & $2.05$ & $-0.16$  & $4.33\pm1.73$ & $1.44$ & $1.36$ & $1.00$ \\
NGC\,7793 & $0.98$ & $-0.38$  & $5.38\pm2.15$ & $0.52$ & $0.63$ & $1.94$ \\
		\hline
	\end{tabular}
\tablefoot{
\tablefoottext{1}{Total mass within the extracted aperture with systematic uncertainties assuming $20\%$ uncertainty in distance.}
\tablefoottext{2}{The unregularized, composite {\sc ppxf} solutions with a grid of $750$ SSP models have a reduced $\chi^2=1$ by construction.}
}
\end{table*}

With the aim to statistically quantify how significant a composite stellar population fit is with respect to a SSP fit, we incorporated a SSP fitting procedure that is maximally similar to the procedure described in the above section.
We used the observed spectrum with rescaled noise properties, where the full unregularized solution gives a reduced $\chi^2=1$, and the same SSP model grid with included emission lines models.
We kept the kinematic solution fixed and estimated the $\chi^2$ at each SSP of the grid, allowing for a multiplicative polynomial of the same order, as in the composite population fit, to vary freely at each grid point.
We give the age and metallicity of the SSP grid point with minimum $\chi^2$ in Table \ref{tab:age_feh_ssp}.

The light weighted ages from the composite fits are in a reasonable agreement with the single population estimates, while older stellar populations (several Gyr) seem to dominate the mass of the nuclei.

We further discuss the quality of the SSP fits and how representative they are of the actual observations in Section \ref{sec:discussion_ssp}.

\section{Results} \label{sec:results}

\subsection{Mock data tests}

\begin{figure*}
	\includegraphics[width=15cm]{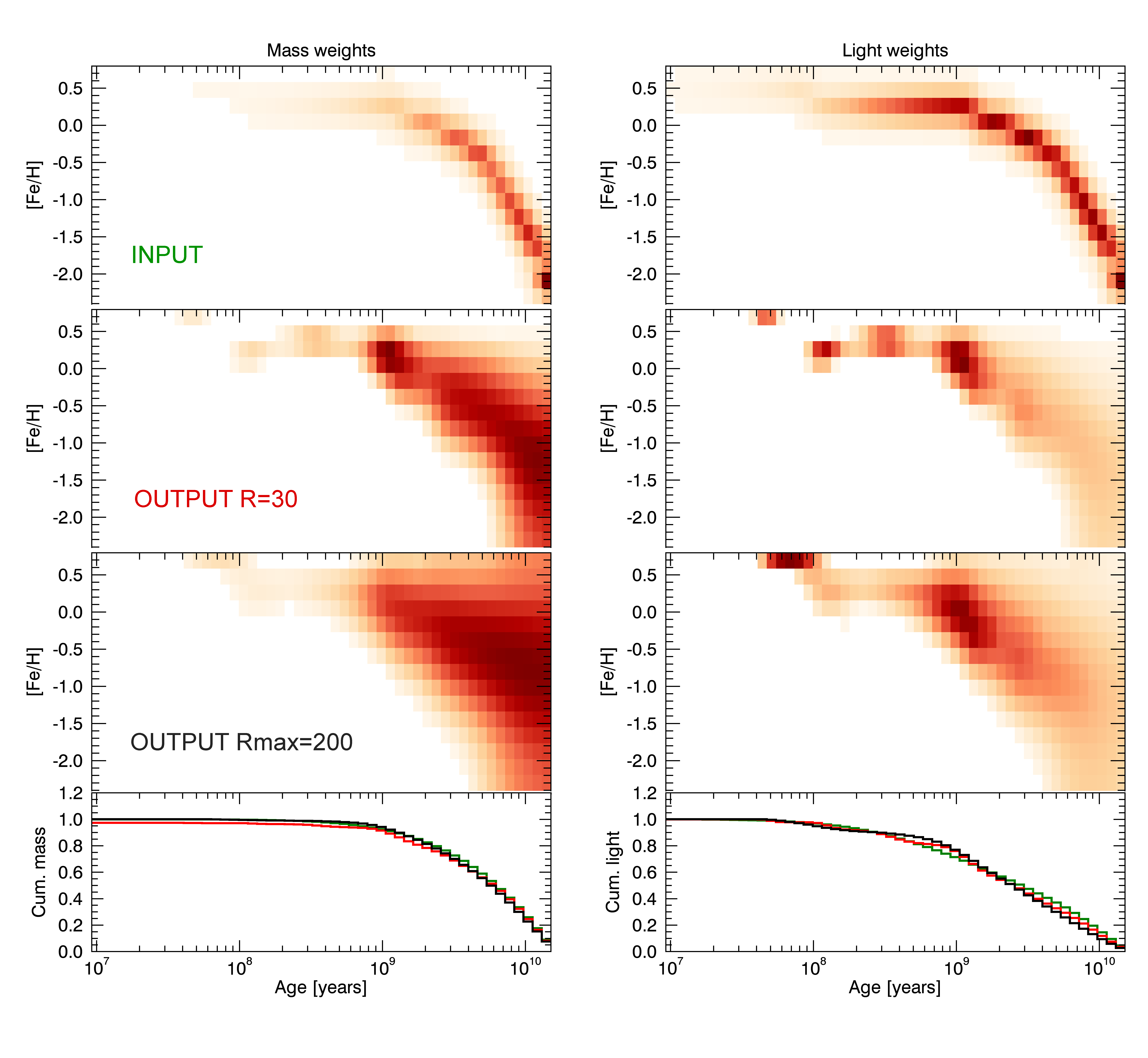}
    \caption{Continuous mock SFH with linear increase of [Fe/H] and slow decline of the SFR. {\it Top panels} show the input weights and corresponding light fractions. {\it Second row panels} show the output weights and corresponding light fractions if low regularization ($R=30$) is applied. {\it Third row panels} show the output weights and corresponding light fractions if maximum regularization is applied. {\it Fourth row panels} show the cumulative mass and light weights of the input (green), low regularization (red) and maximum regularization (black) weights.}
    \label{fig:mock1_ord1}
\end{figure*}

\begin{figure*}
	\includegraphics[width=15cm]{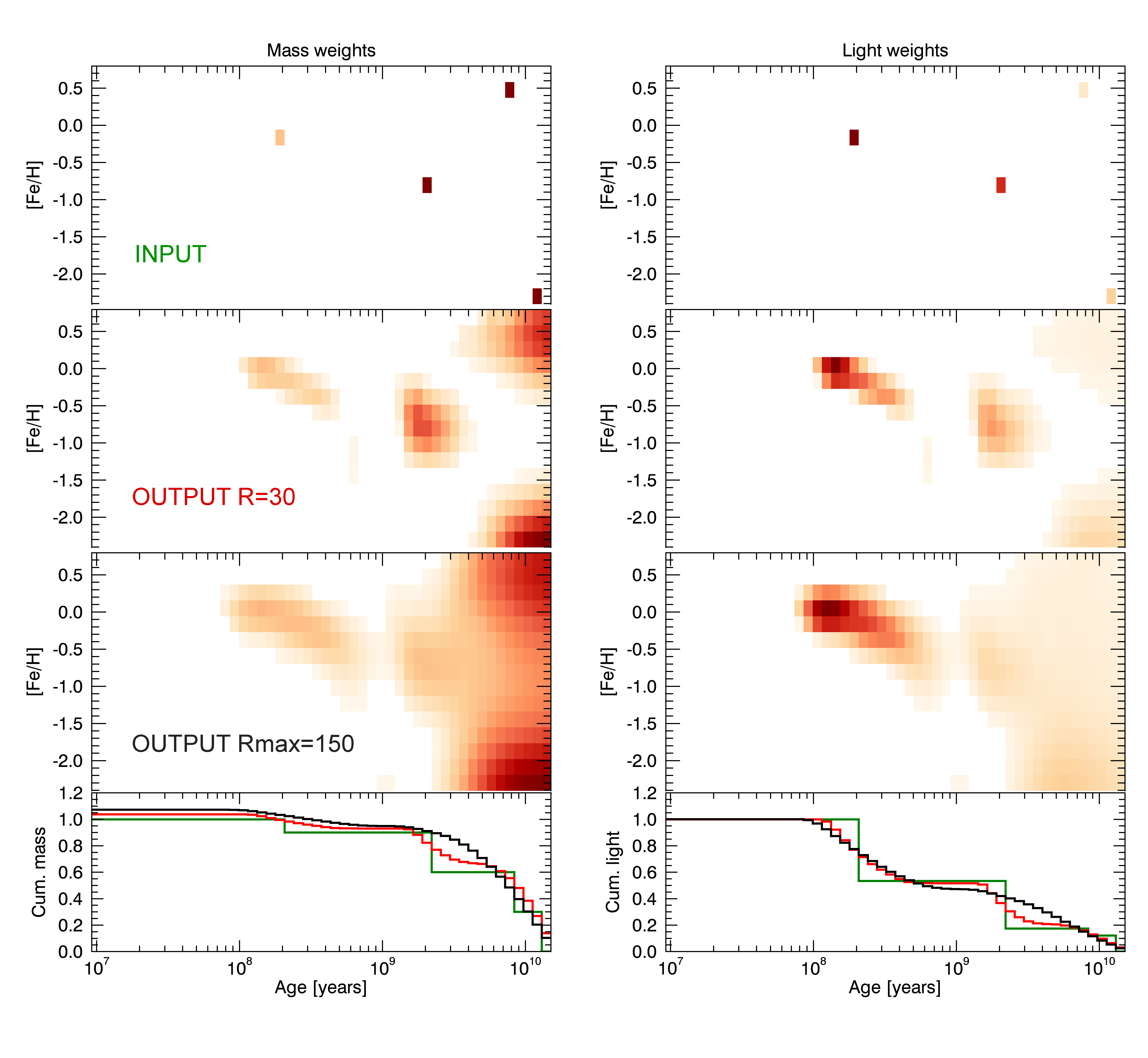}
    \caption{Merger SFH. The notations are the same as in Figure \ref{fig:mock1_ord1}.}
    \label{fig:mock3_ord1}
\end{figure*}

\begin{figure*}
	\includegraphics[width=15cm]{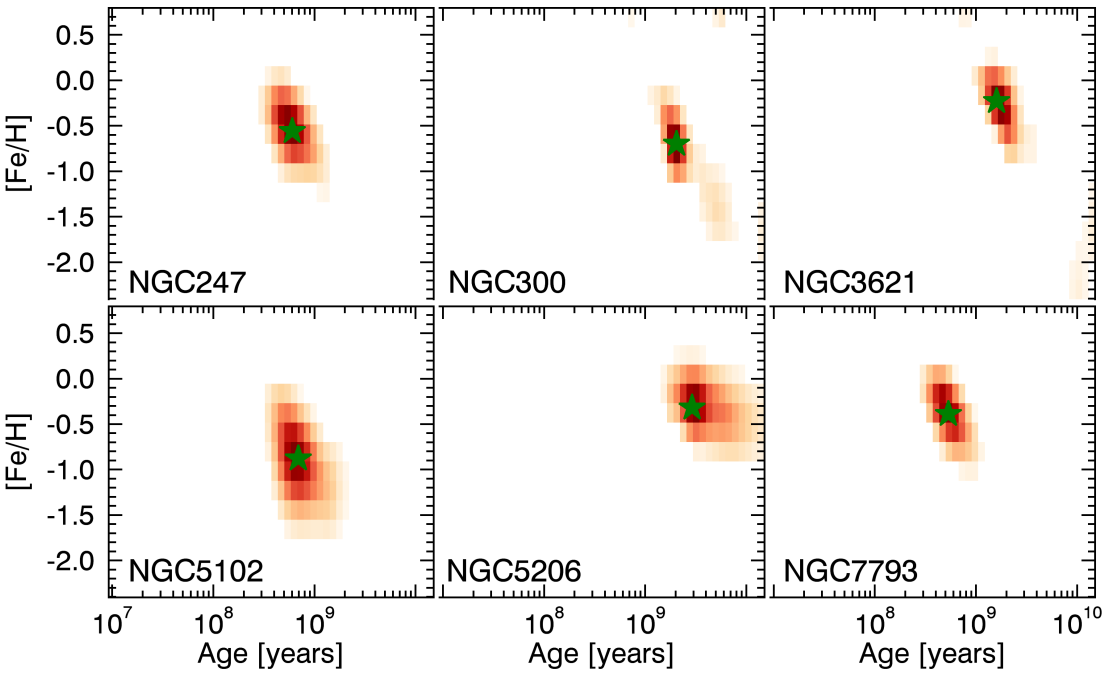}
    \caption{{\sc ppxf} recovery of SSP mock spectra at the light weighted age and metallicity (green asterisk) of the actual observations of the six nuclei and with the same SNR velocity dispersions.}
    \label{fig:all_SSP_max_regul_light}
\end{figure*}

Before turning to the actual observations we performed series of mock data tests to test how well we can recover different SFHs.
We created mock spectra by summing model templates from our SSP grid with different weights, adding noise, and broadening the spectrum to a certain velocity dispersion.
Each mock simulation has a Gaussian SNR\,$=70$\,px$^{-1}$ and a velocity dispersion $\sigma=30$\kms, similar to our observations.
The mock spectra also have the same LSF and the same pixel scale as the observed spectra.
We fit the mock spectra with {\sc ppxf} using the same SSP templates as for the observations.

In essence, regularization is equivalent to adding a prior to the model weights and the only time we will recover a SFH perfectly is if it follows the imposed prior.
We demonstrate this in Figure \ref{fig:mock_prior_sfh} in an appendix where the mock SFH input weights are pre-regularized in accordance with the prior.
In {\sc ppxf}, the prior is a constant in the case of first order regularization and a plane in the case of second order regularization.
In reality, of course, star formation does not need to obey linear regularization priors and we will never be able to perfectly reconstruct a model.
We test the performance of {\sc ppxf} on arbitrary SFHs by testing the common theories of NSC formation.
Figure \ref{fig:mock1_ord1} mimics the {\it in situ} formation scenario, which includes gradual self-enrichment (linear increase of \feh) and slowly declining SFR, highest in the early Universe.
We present two solutions, using low regularization ($R=30$) and the maximum regularization, that is still compatible with the input spectrum (see Eq. \ref{eq:max_chi2}).
One can see that the chemical enrichment history is well reproduced by the {\sc ppxf} fit.
It is especially worth noting, that this method recovers very well the metallicity of the young populations with a very small uncertainty, while there is a large spread in the metallicity estimates at old ages.
Given that young stellar populations are generally not very sensitive to [Fe/H] variations because they lack strong metal lines, this is a somewhat counter-intuitive result.
The young populations, however, dominate the spectrum flux, while the old populations contribute relatively little light and thus have a high variance.

We need to point out also, that the estimated total mass of the mock system is very well matched to the input mass for the maximum regularized solution.
This is not a requirement by construction, but rather a success of the model fits in reproducing the total M$/$L.
This is also the main reason that we decided to use first order regularization rather than second order, where we found that the solution with maximum regularization generally has a slightly biased M/L ratio.
We show the results of the same experiments but using second order regularization in an appendix (see Figure \ref{fig:mock1_ord2}).

The second mock test that we present here assumes a merger of randomly distributed (in age and metallicity) clusters for the origin of the NSC.
In Figure \ref{fig:mock3_ord1} we input four SSPs, three of them contain $30\%$ of the total mass each, and the youngest one contains $10\%$ of the mass, but almost $50\%$ of the total light.
Here low regularization represents better the input weights, but the mock spectrum is still compatible with higher regularization according to our $\chi^2$ criterion.
Figure \ref{fig:mock3_ord2} shows the same test but using second order regularization.

Additionally, in Figure \ref{fig:all_SSP_max_regul_light} we show a {\sc ppxf} fit to mock SSPs, derived at the light weighted age and metallicity of each NSC (Table \ref{tab:age_feh_reg}), that have the same SNR and velocity dispersions as the observed spectra.
These fits have the maximum regularization that a SSP can take before the $\chi^2$ becomes inconsistent with the input data at a $1\sigma$ level according to Eq. \ref{eq:max_chi2}.
They give a good indication about the resolution in age and \feh~of our SFHs and illustrate how different from SSPs the actual spectra are (see Figures \ref{fig:ngc247} -- \ref{fig:ngc7793}).

In summary, our tests demonstrate that we have the ability to accurately recover smooth star formation histories and metallicity evolution, but we have poorer constraints on recovering how bursty a SFH is.

\subsection{NGC\,247}

\begin{figure*}
	\includegraphics[width=\columnwidth]{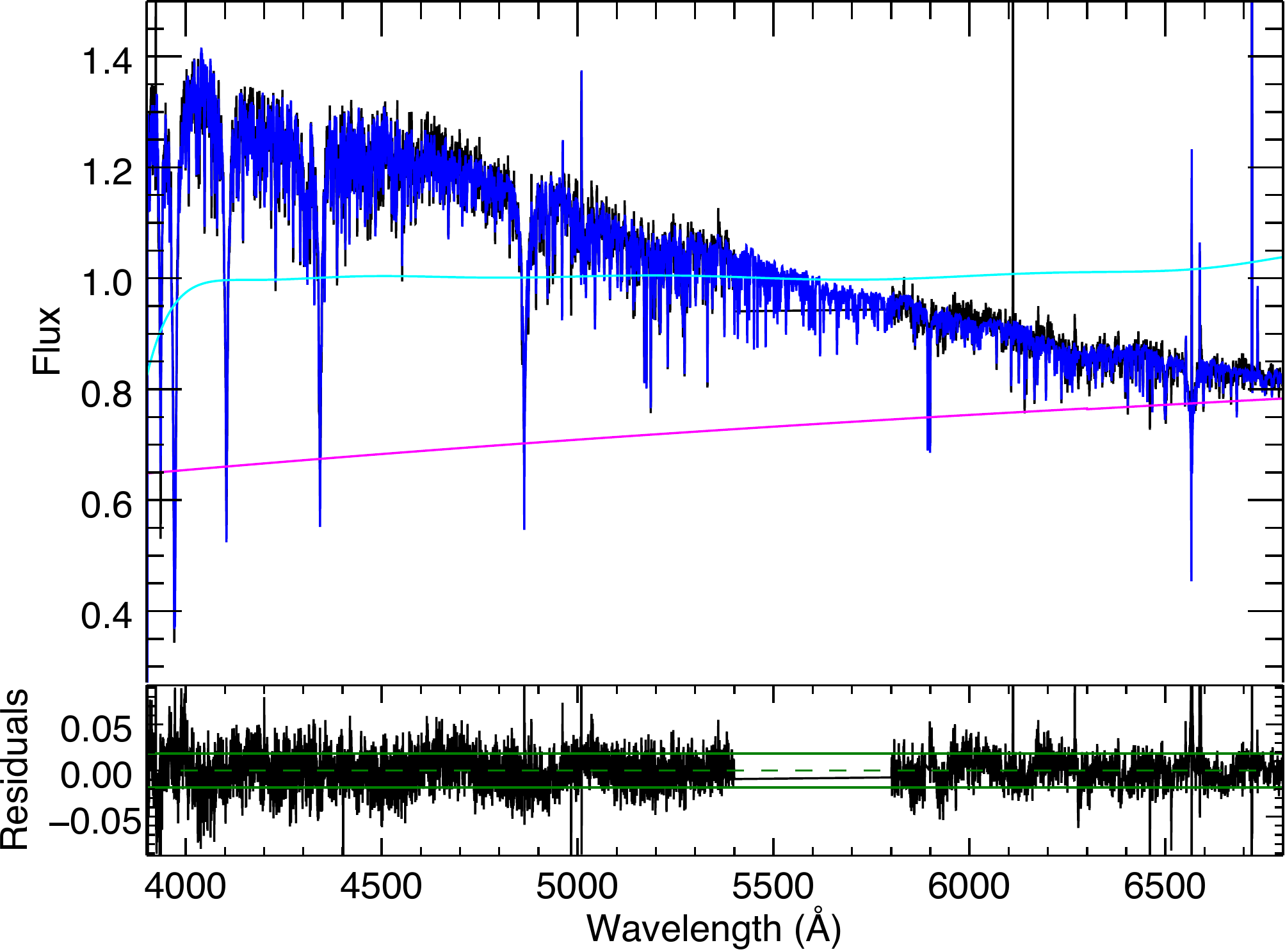}
	\includegraphics[width=\columnwidth]{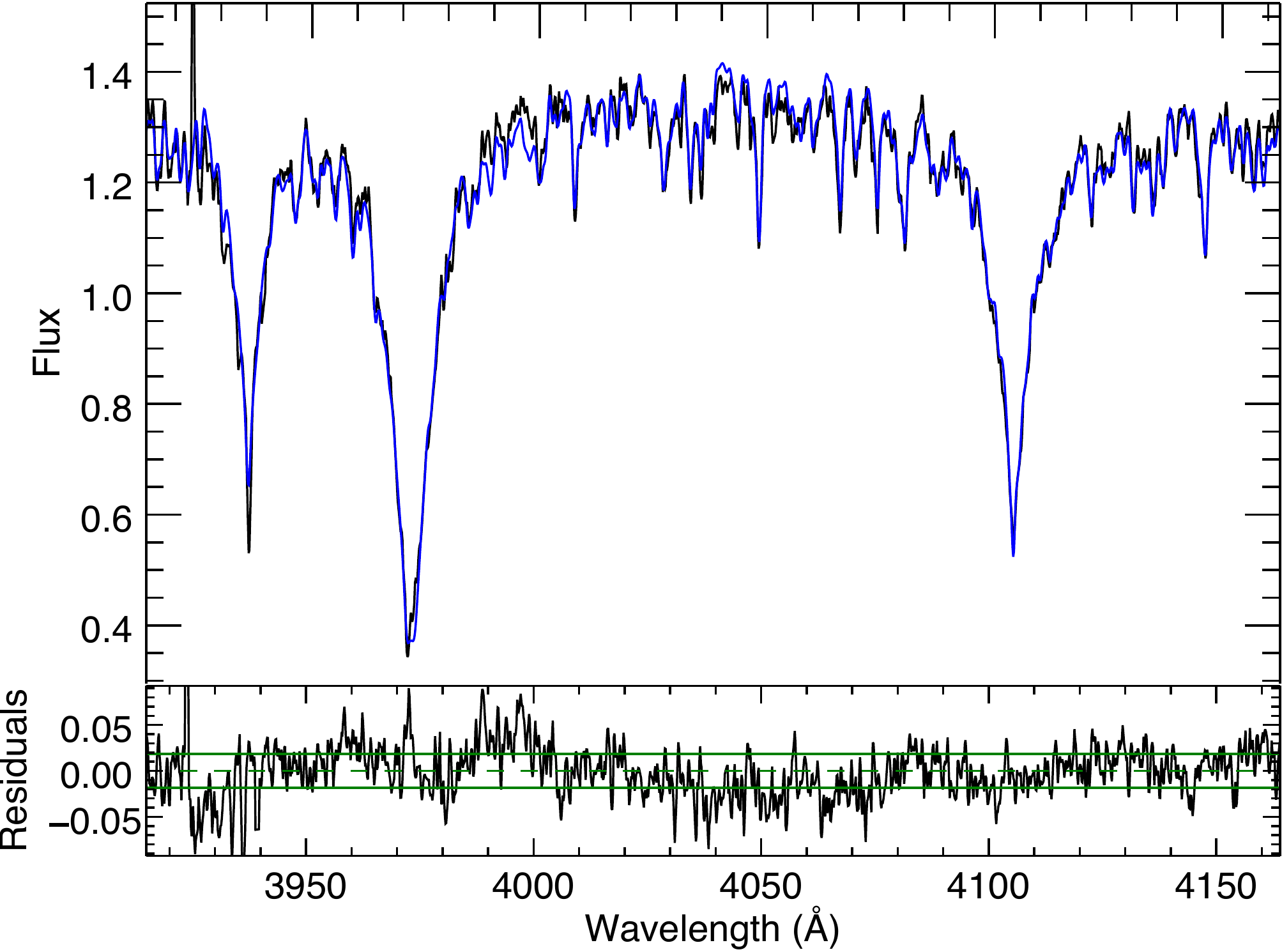}
	\includegraphics[width=\columnwidth]{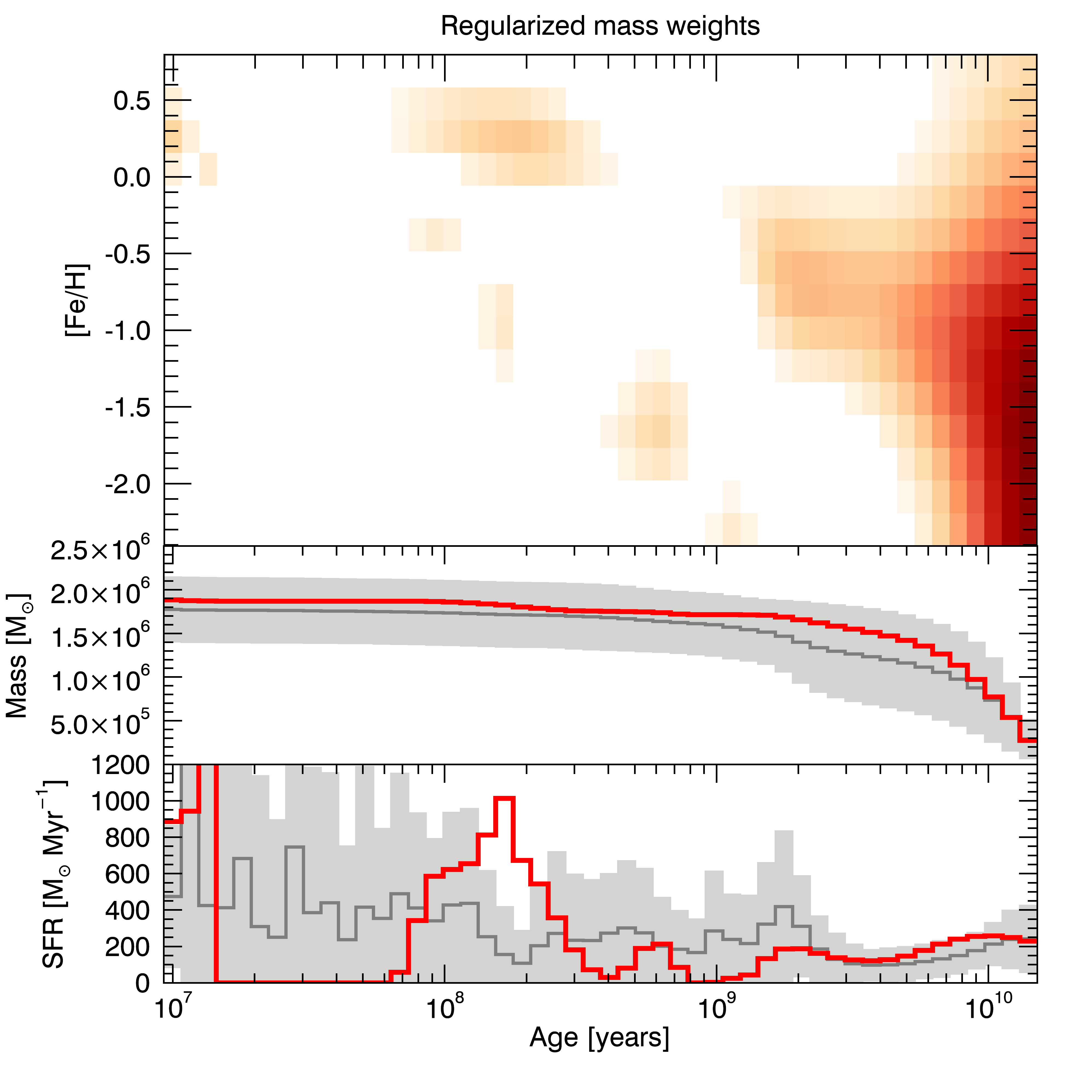}
	\includegraphics[width=\columnwidth]{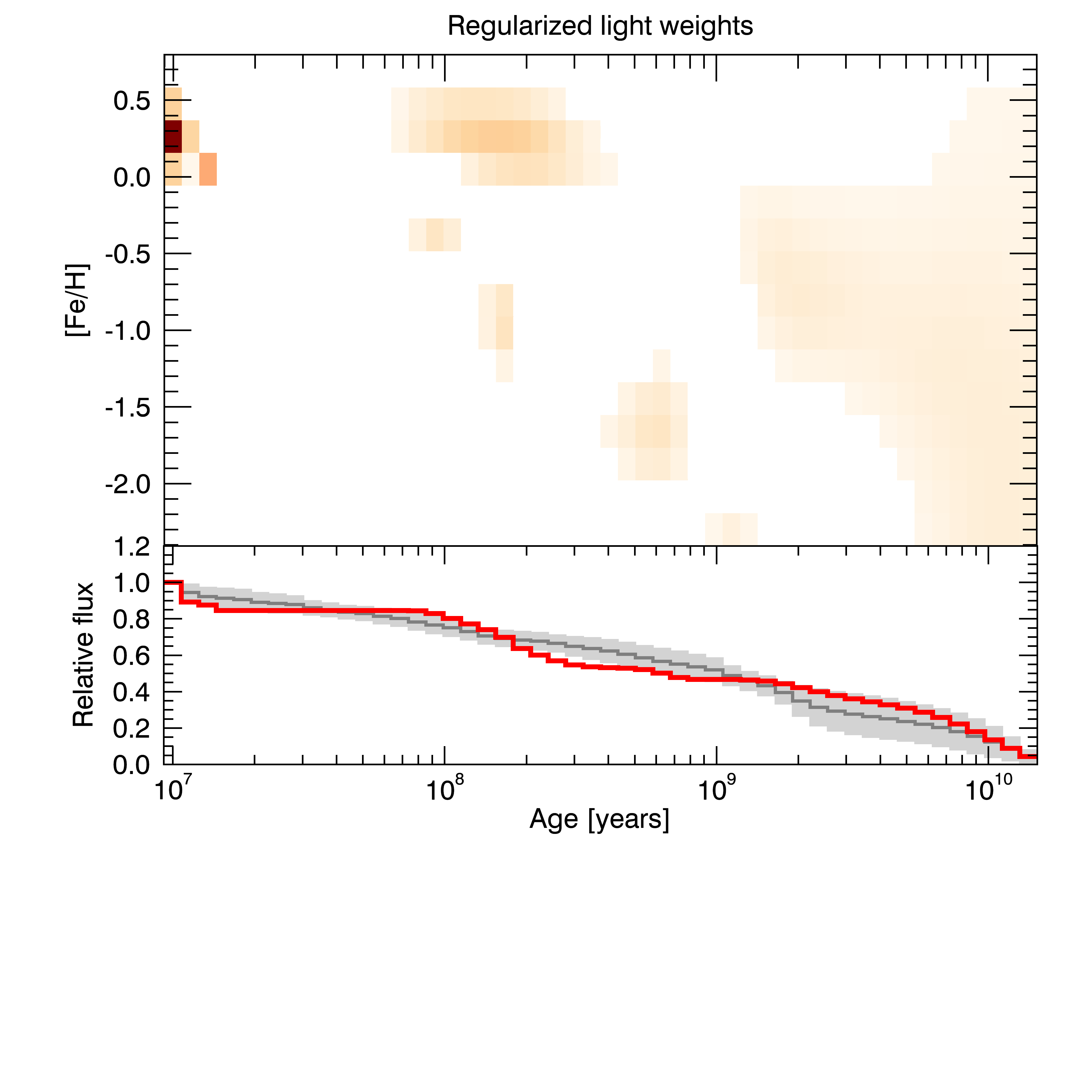}
    \caption{SFH of NGC\,247 NSC.
            {\it Top left panel:} the observed spectrum (black) and the best fit spectrum (regularized solution; blue) with the residuals shown in a sub-panel (the green lines shows the mean of the residuals and $\pm1\sigma$ intervals). The magenta line is the best fit reddening law \citep{calzetti+2000} and the cyan line is the best fit multiplicative polynomial.
            {\it Top right panel:} zoom-in to the region of the H\,\&\,K Ca lines and the H$\delta$ line.
            {\it Bottom left panel, top:} colour-coded, the mass weights of the SSP grid as a function of age and metallicity of the maximum regularized solution shows the enrichment history; {\it middle:} cumulative mass as a function of time - the red line is the best fit maximum regularized model and the gray line and shaded areas show the mean and $\pm1\sigma$ spread of the MC simulations; {\it bottom}: SFR.
            {\it Bottom right panel:} colour-coded, the luminosity weights of the maximum regularization solution and cumulative luminosity distribution.
            }
    \label{fig:ngc247}
\end{figure*}

The results of the best fit SFH solution for the nucleus of NGC\,247 are presented in Figure \ref{fig:ngc247}.
The top panels show the full spectrum fit (left) and a zoom-in to the region of the H$\delta$ and the Calcium H\,\&\,K lines, which are good age and metallicity indicators.
The regularized mass weights of the SSP grid amounting to the observed spectrum of the nucleus of NGC\,247 are shown in the bottom left panel of Figure \ref{fig:ngc247}, while the bottom right panel shows the light contribution of each SSP.
The spectrum is dominated by the light of younger populations, which, however, carry only a small fraction of the total mass.
We also show cumulative mass and light plots, and the estimated SFR in the NSC as a function of time.
The grey line and shaded areas depict the mean and $\pm1\sigma$ spread of the bootstrap solutions, while the red lines show the regularized solution.
Note that, owing to the logarithmic age scale of the SSP grid, the SFR at old ages is averaged over much longer periods of time than in the younger age bins, and hence the latter appear much more stochastic.

We measure an average velocity dispersion of $15.1\pm0.6$\kms, which is similar to the velocity dispersion values of the most massive Galactic GCs (\citealt[][2010 catalogue edition]{harris1996}\footnote{\url{http://www.physics.mcmaster.ca/resources/globular.html}}).

NGC\,247's NSC seems to have experienced a very prolonged star formation that continues until the present day.
It built most of its mass in the early Universe ($\sim50\%$ already present $10$\,Gyr ago) at a relatively low metallicity ($\feh \sim -1.5$\,dex but with a significant spread), with younger generations being more and more metal rich on average, reaching $\feh\sim-0.5$\,dex roughly $1$\,Gyr ago and $\sim90\%$ of the present day mass.
A significant increase of the SFR has occurred $\sim100-500$\,Myr ago giving birth to a population of super-Solar metallicity and contributing $\sim10\%$ of the total mass.
There is also an indication of a very young component ($\sim15$\,Myr but contributing a negligible fraction of the total mass $<1\%$), which is further supported by the presence of strong emission lines in the nucleus and the surrounding region, indicative of ongoing star formation.
We discuss the current SFR of the nuclei based on the emission line analysis in Section \ref{sec:emission}.


\subsection{NGC\,300}

\begin{figure*}
	\includegraphics[width=\columnwidth]{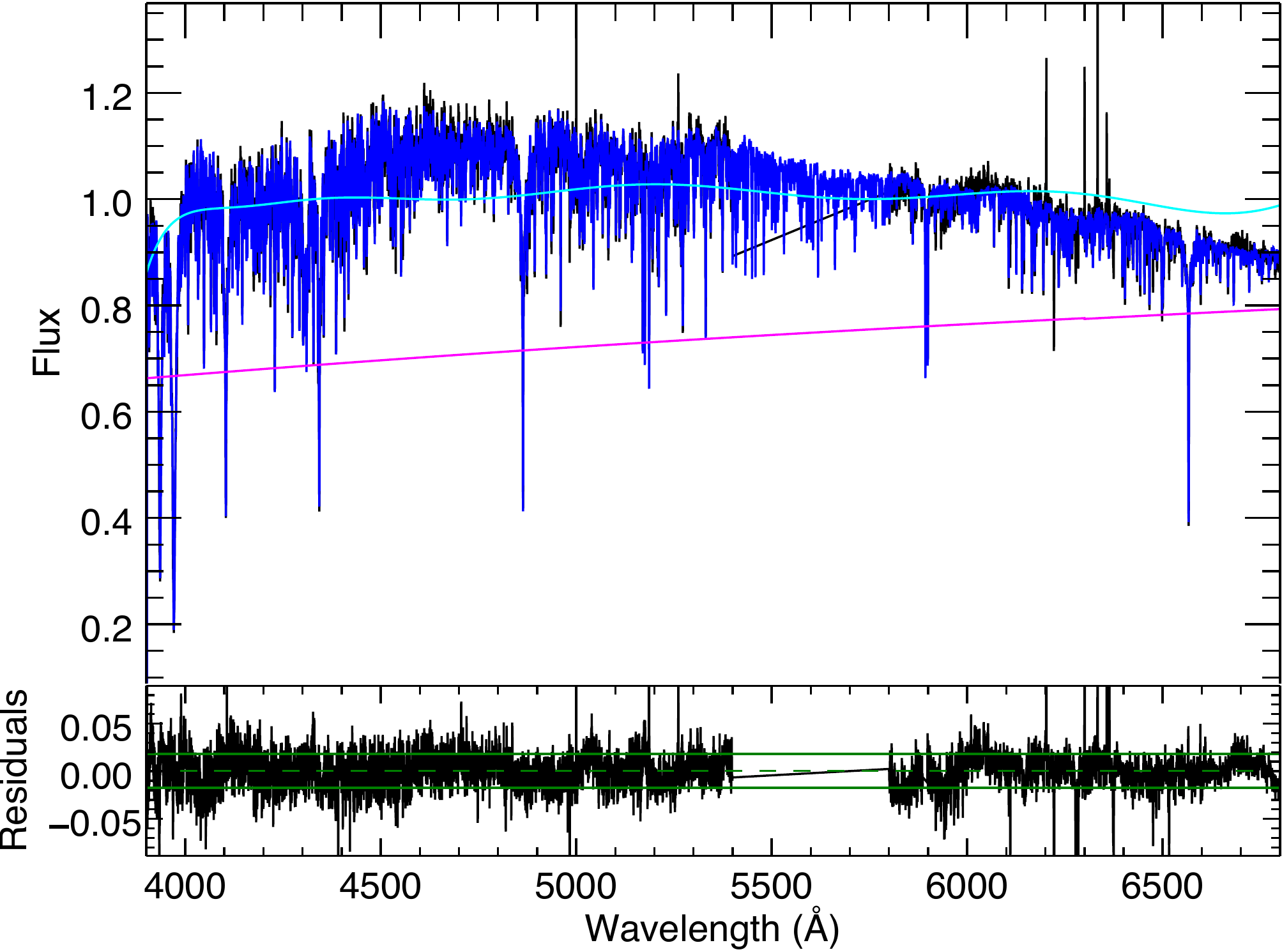}
	\includegraphics[width=\columnwidth]{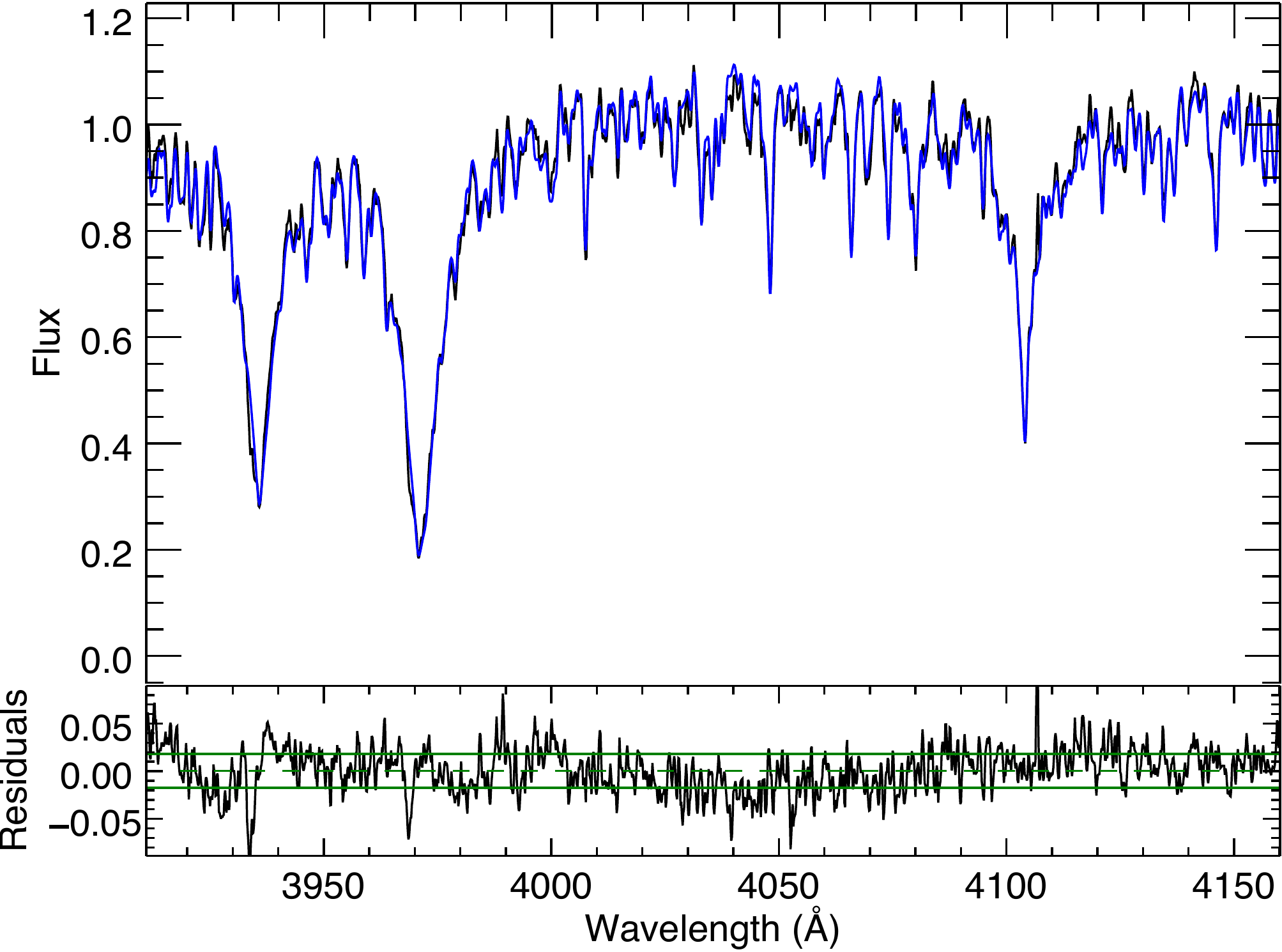}
	\includegraphics[width=\columnwidth]{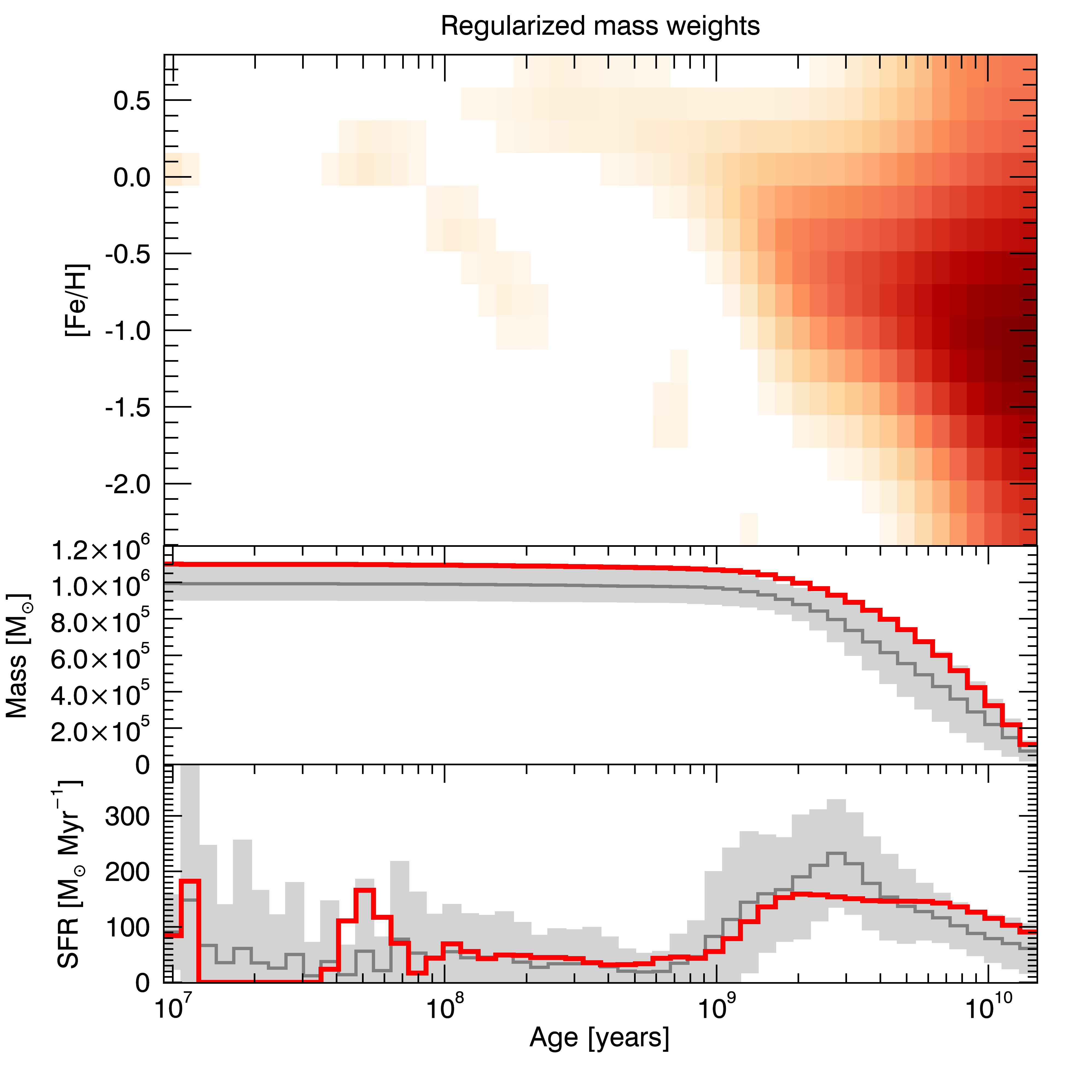}
	\includegraphics[width=\columnwidth]{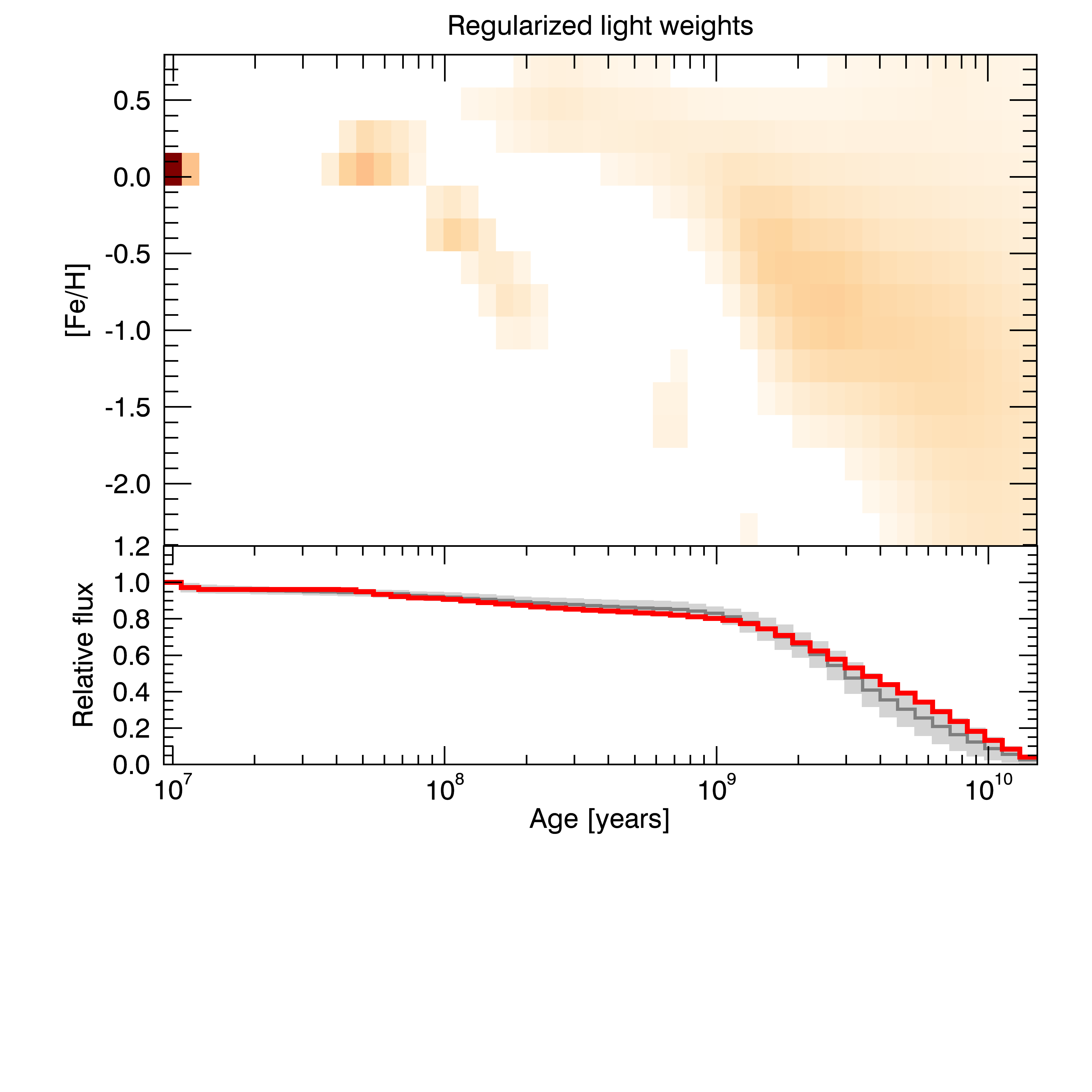}
    \caption{SFH of NGC\,300 NSC - same as Figure \ref{fig:ngc247}.}
    \label{fig:ngc300}
\end{figure*}

Figure \ref{fig:ngc300} shows the best fit SFH of the nucleus of NGC\,300.
We measure an average velocity dispersion of $13.3 \pm 0.3$\kms, similar to NGC\,247.
This value is in an excellent agreement with the results by \citet{walcher+2005}, who find $\sigma_V = 13.3\pm2$\kms from high resolution UVES/VLT spectroscopy.


Similarly to NGC\,247's NSC, NGC\,300's nucleus formed most of its stars at early times ($50\%$ more than $10$\,Gyr ago) and low metallicity ($\feh\sim-1$\,dex with a significant spread) and continued forming stars up until few hundred Myr ago, with increasingly higher metallicity, reaching Solar values about a Gyr ago.
However, at difference with NGC\,247, NGC\,300's NSC did not experience a significant increase of the SFR several hundred Myr ago and its younger population is considerably less massive.
It reached $90\%$ of its mass $\sim1.3$\,Gyr ago and $99\%$ - roughly $200$\,Myr ago.
About $10\%$ of the light in the spectrum is attributed to populations younger than $100$\,Myr, which, however, contribute less than $1\%$ to the total mass.

Overall, this result is in a reasonable agreement with the composite stellar population fit performed by \citet{walcher+2006}, who find the youngest population in this NSC to be $\sim100$\,Myr old and contributing $\sim2.7\%$ of the mass.
They find that $96\%$ of the total mass was formed between $3$ and $6$\,Gyr ago, and do not find evidence for a very old ($>10$\,Gyr) population.
Note that we can also reproduce the spectrum of the nucleus restricting the SSP grid to include only ages younger than $\sim4$\,Gyr (see Section \ref{sec:yng_old}).
In addition, \citet{walcher+2006} use a constant metallicity of $\feh\sim-0.65$\,dex for their fit, as well as a different spectral range, and no regularization, which may also explain the differences between the two studies.

Finally, there is a contribution from a very young population in the NSC of NGC\,300 ($\sim10$\,Myr), although we do not see emission lines in the spectrum, indicative of ongoing star formation, as we do in the case of NGC\,247.
Such a young component might instead be caused by the presence of an extended horizontal branch \citep[see e.g.][]{conroy+2018} or blue straggler stars \citep{schiavon2007} in the nucleus.
However, the presence of a genuine young population is the simplest and most likely explanation.

\subsection{NGC\,3621}

\begin{figure*}
	\includegraphics[width=\columnwidth]{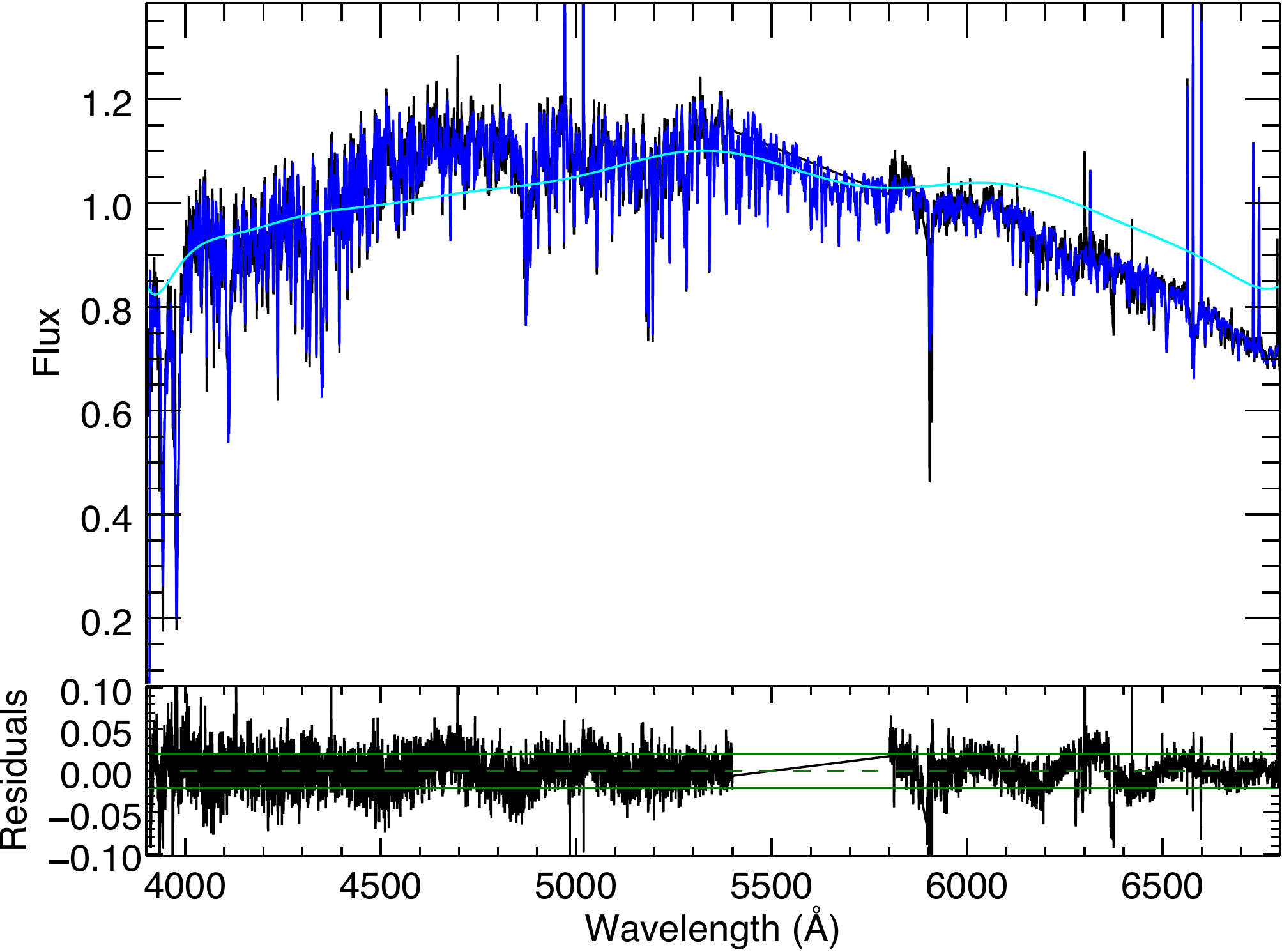}
	\includegraphics[width=\columnwidth]{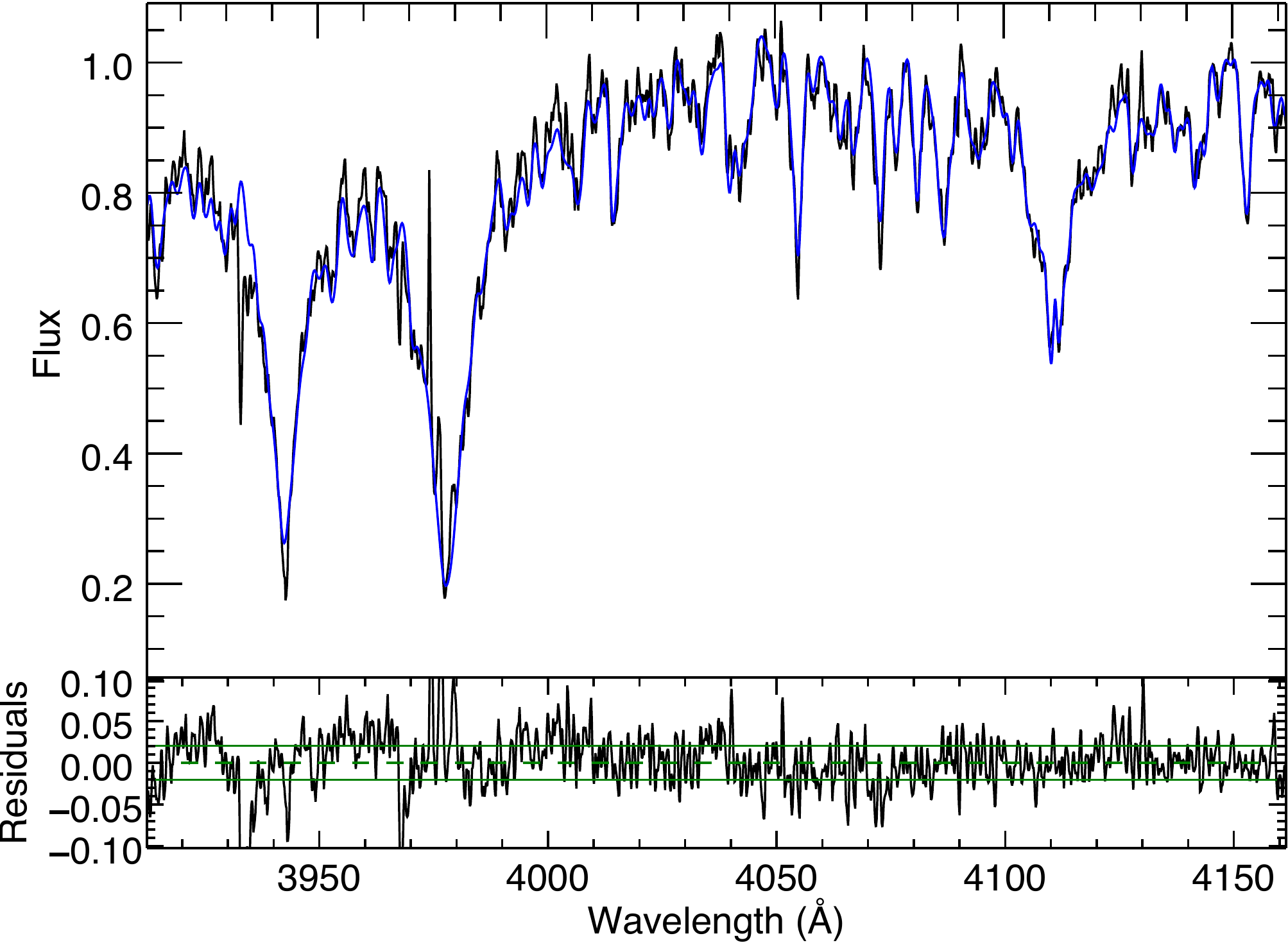}
	\includegraphics[width=\columnwidth]{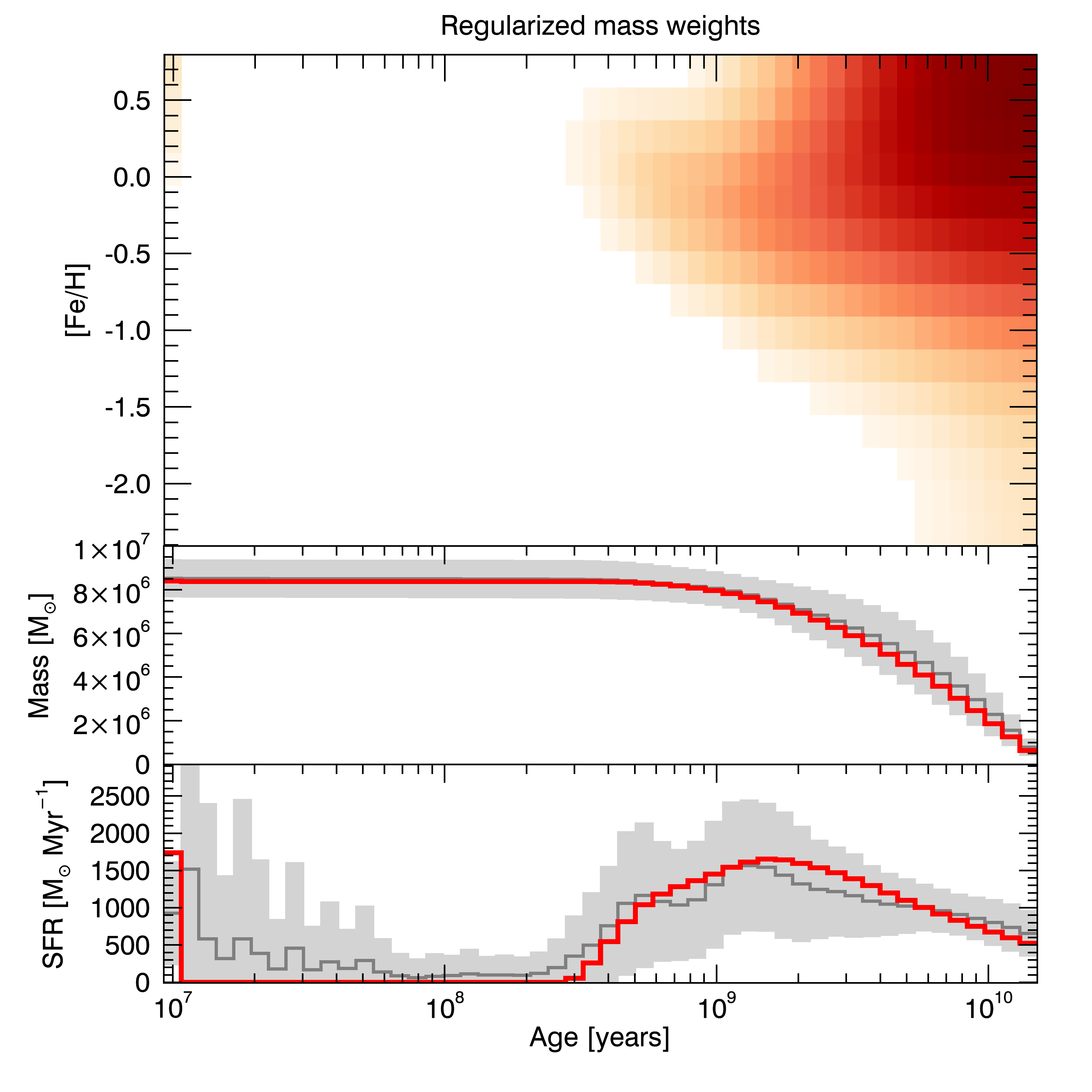}
	\includegraphics[width=\columnwidth]{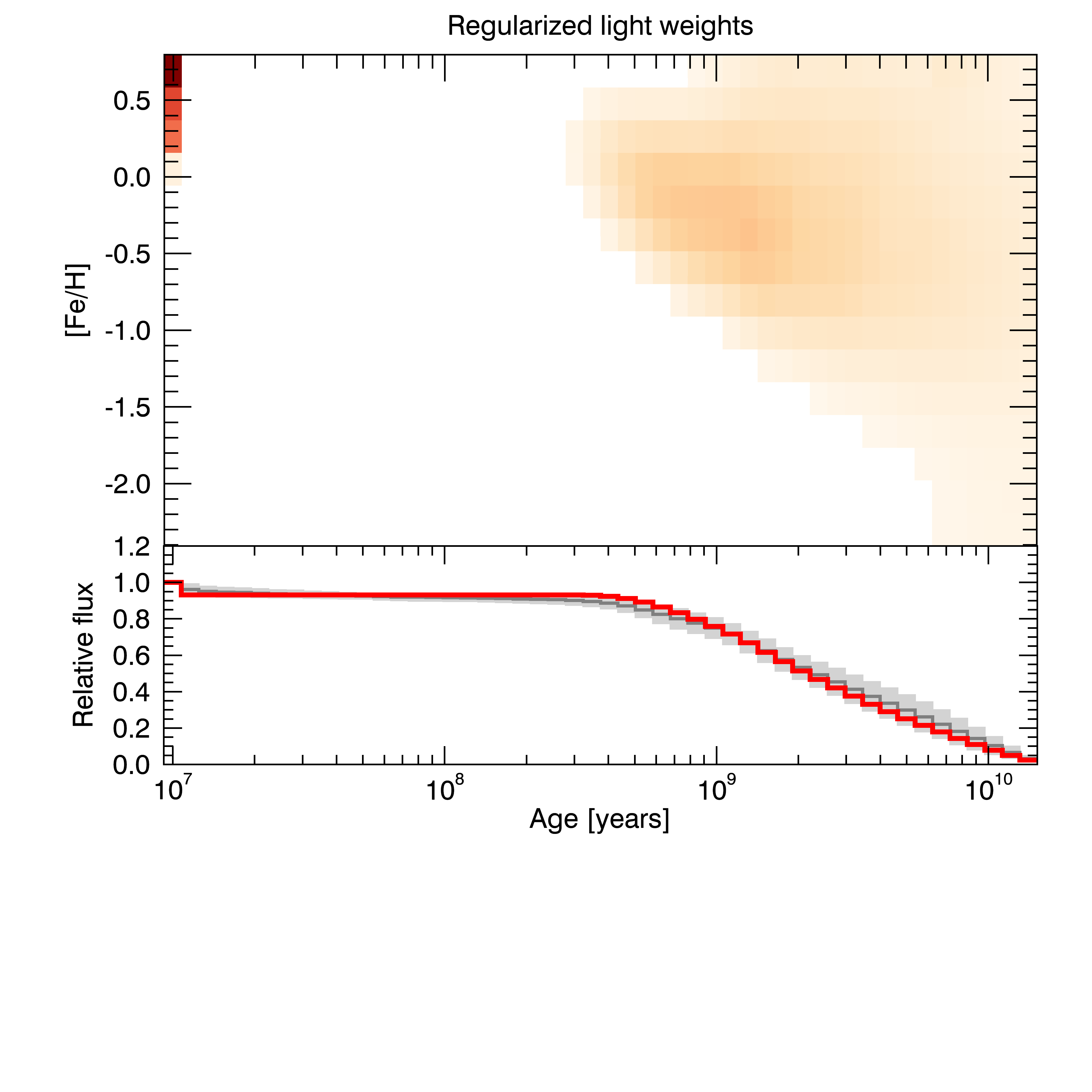}
    \caption{SFH of NGC\,3621 NSC - same as Figure \ref{fig:ngc247}.}
    \label{fig:ngc3621}
\end{figure*}

NGC\,3621 (Figure \ref{fig:ngc3621}) is the most distant galaxy in our sample and was also observed under sub-optimal conditions - the seeing reported from the auto-guiding system of the telescope is $3.7\arcsec$ during the first run of observations and $2.2\arcsec$ during the second.
We have combined the exposures from the two nights using a weighted average method but analyzing them separately yields the same result. 
In order to achieve a desirable SNR ($48$\,px$^{-1}$; the lowest in the sample), we had to integrate the light in a $2.4\arcsec$ wide aperture, which covers $\sim80$\,pc at the distance of NGC\,3621.
Hence, we may have also included a non-negligible fraction of the galaxy's disk component.
Within this aperture we measured an average velocity dispersion $\sigma = 42.2 \pm 0.6$\kms.
This is in excellent agreement with the measurement by \citet[][$43\pm3$\kms]{barth+2009} from an ESI/Keck similar quality spectrum.

The best fit multiplicative polynomial appears to have different trends for the UVB and VIS arms of X-Shooter (see Figure \ref{fig:ngc3621}).
This suggests that there may be issues with the flux calibration of this source.
Nevertheless, the composite stellar population fit suggests that the NSC of NGC\,3621 built most of its mass in the early Universe and continued forming stars until several hundred Myr ago, similarly to the other late-type NSCs in this sample.

NGC\,3621's NSC likely experienced an extremely rapid enrichment because even the oldest SSPs contributing to the fit appear to be very metal rich.
We note that \citet{boardman+2017} analysed a sample of early-type galaxies using the same code (but with different models) and found a similar effect that most of the mass is situated in an old, metal rich population.
This is due to the very high M$/$L ratios of these SSP models (see Figure \ref{fig:ml_templates}), which leads to significant mass weights even if only a very small fraction of the total light is attributed to these templates.

The spectrum has very strong emission lines and although it is not excluded that there has been some recent star formation, NGC\,3621 is classified as a Seyfert 2 type galaxy with an X-ray source at its centre \citep{barth+2009}, which is the main source of emission.

\subsection{NGC\,5102}

\begin{figure*}
	\includegraphics[width=\columnwidth]{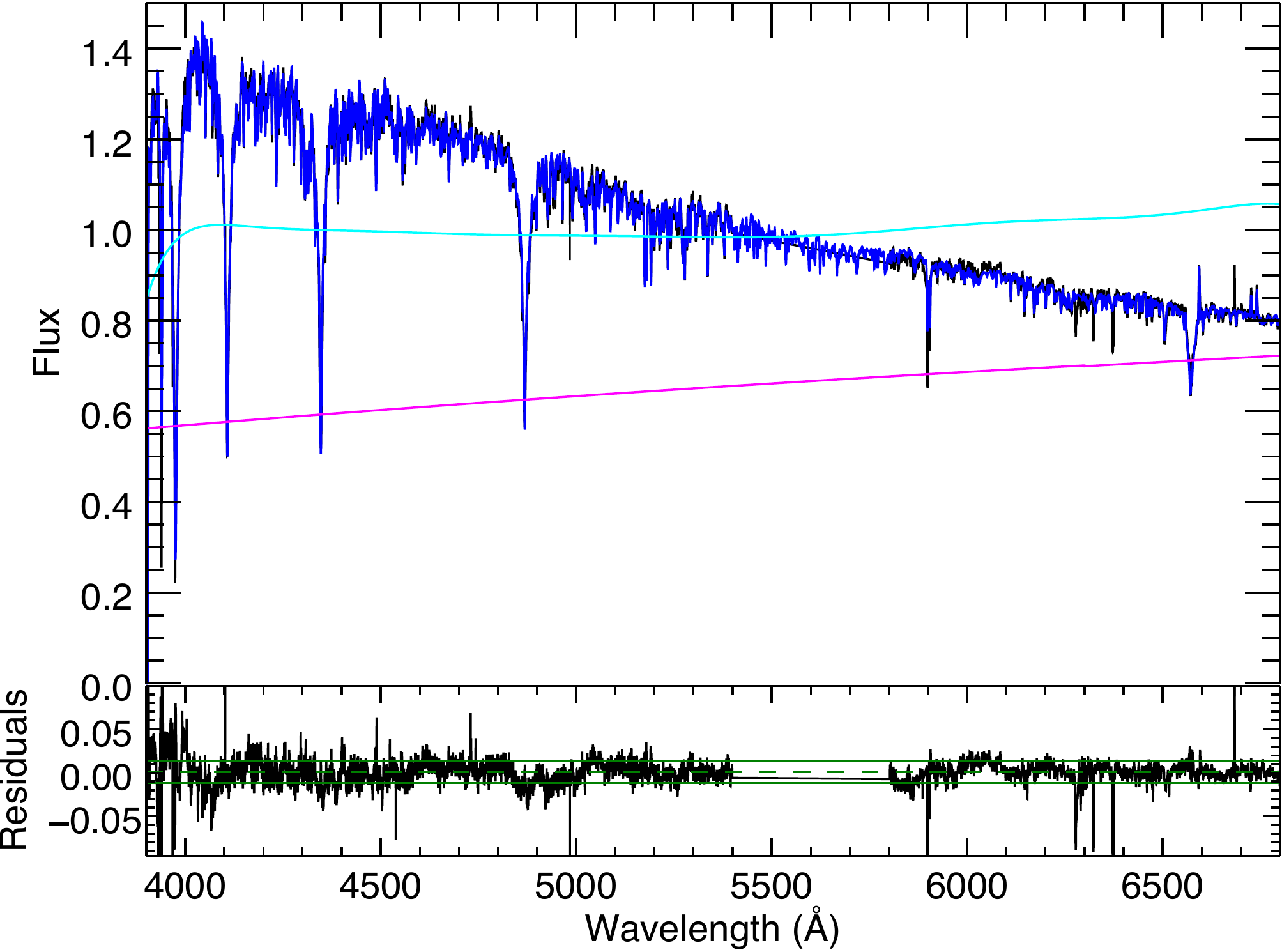}
	\includegraphics[width=\columnwidth]{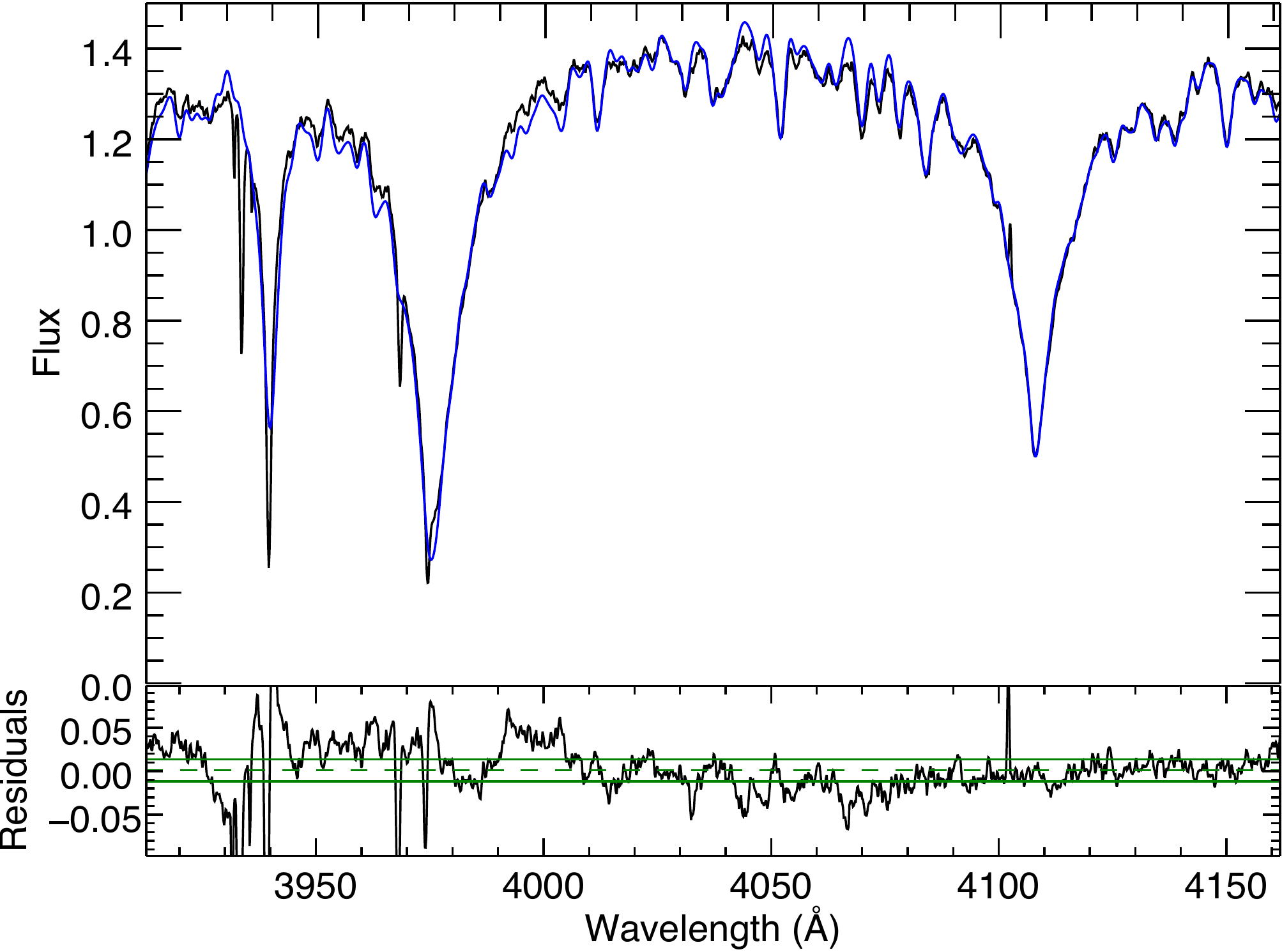}
	\includegraphics[width=\columnwidth]{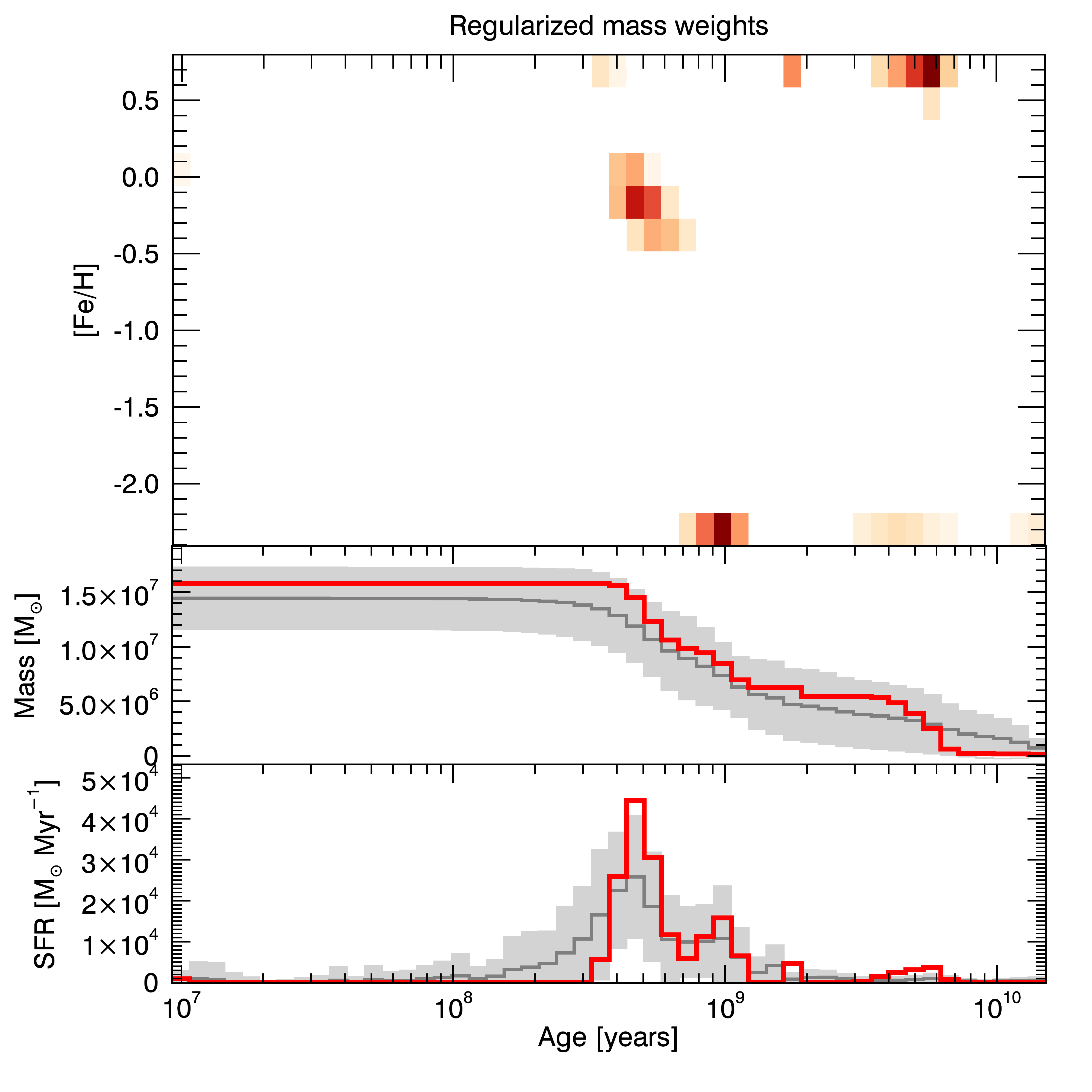}
	\includegraphics[width=\columnwidth]{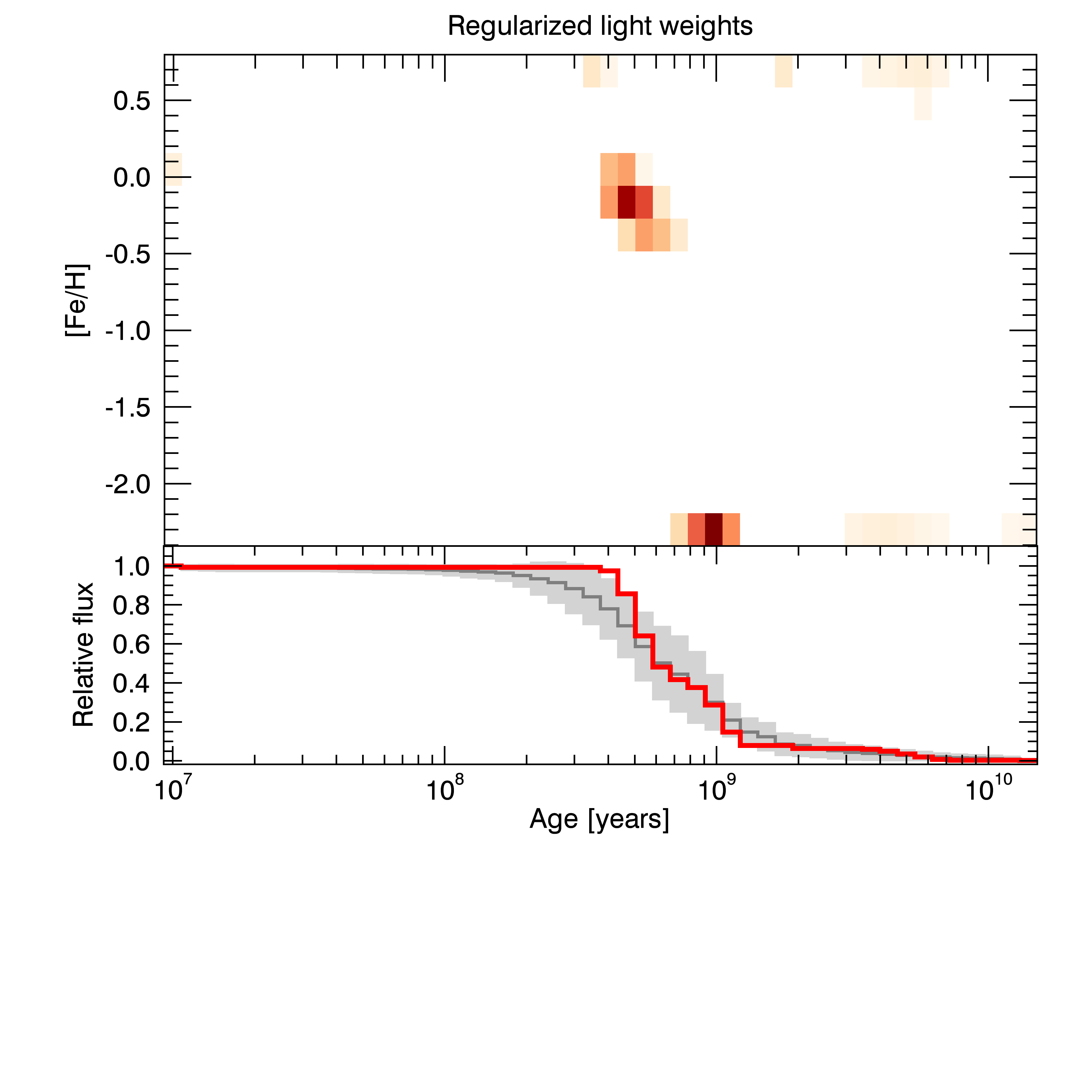}
    \caption{SFH of NGC\,5102 NSC - same as Figure \ref{fig:ngc247}.}
    \label{fig:ngc5102}
\end{figure*}

NGC\,5102 is one of two early-type galaxies in our sample. The spectrum of its NSC has the highest SNR among the six nuclei ($85$\,px$^{-1}$). We measured a velocity dispersion $\sigma = 45.2 \pm 0.4$\kms.
\citet{mitzkus+2017,nguyen+2018} report a flat dispersion profile at $44$\kms at radii larger than $0.3\arcsec$ that peaks at $60$\kms at the centre of the NSC.
It is possible that our seeing-limited observations are not sensitive to the high dispersion peak in the very centre of the nucleus.

The spectrum of NGC\,5102's nucleus, however, does not broaden with regularization as is seen in the other galaxies (Figure \ref{fig:ngc5102}).
The maximum regularization that would still yield a fit within $1\sigma$ of the unregularized solution is $R=15$, compared to $R\geq60$ for the other galaxies.
On one hand, this might be an effect of the high SNR, which does not leave tolerance for smoothing the solution, but may also indicate a very bursty and stochastic SFH.
This is the only NSC, in which we do not find an evidence for a population older than $10$\,Gyr and the most significant star formation happened less than $1$\,Gyr ago.
We find a mass weighted age of $1.4$\,Gyr and light weighted age of $0.7$\,Gyr, while \citet{mitzkus+2017} estimate a mass weighted age of $0.8$\,Gyr.
Comparing it to our SSP resolution (Figure \ref{fig:all_SSP_max_regul_light}), it is consistent with being a SSP.
The best fit SSP solution also has the same $\chi^2$ value as the composite fit (Table \ref{tab:age_feh_ssp}).
However, an older population may simply be completely outshined by a massive young population, making NGC\,5102 a post-starburst galaxy.
This actually seems to be the case if we compare our result with the SFH derived by \citet{mitzkus+2017}, who use {\sc ppxf} to model a MUSE spectrum and detect the same massive, high metallicity, young component ($<1$\,Gyr) as we do, but also an old, metal poor component (age $> 10$\,Gyr, $\feh<-1.5$\,dex), perhaps benefiting from the redder wavelengths ($4800-9000$\,\AA) used in their fit.

There are also emission lines in the spectrum of this NSC indicative of the presence of possible ongoing star formation.

\subsection{NGC\,5206}

\begin{figure*}
	\includegraphics[width=\columnwidth]{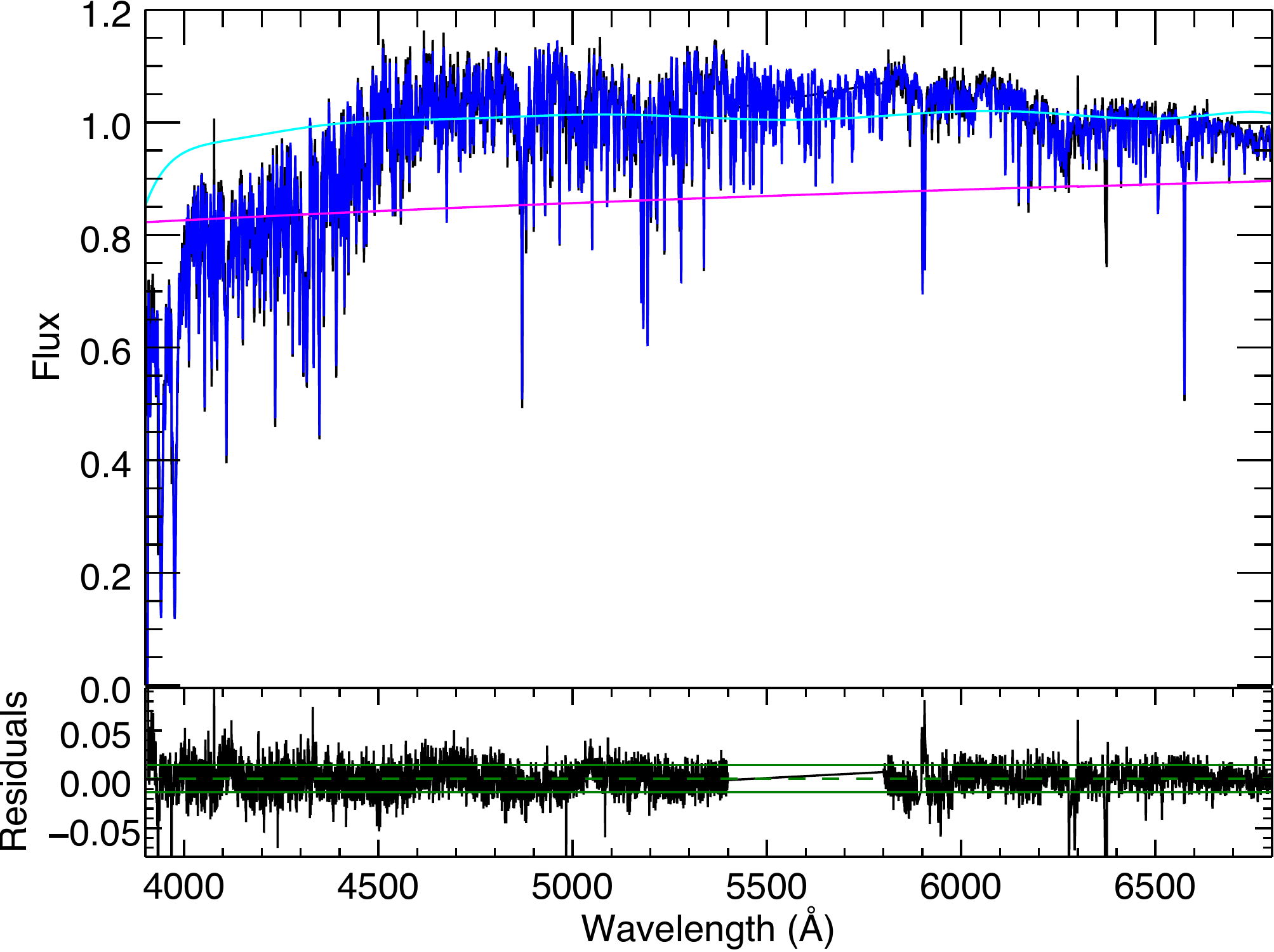}
	\includegraphics[width=\columnwidth]{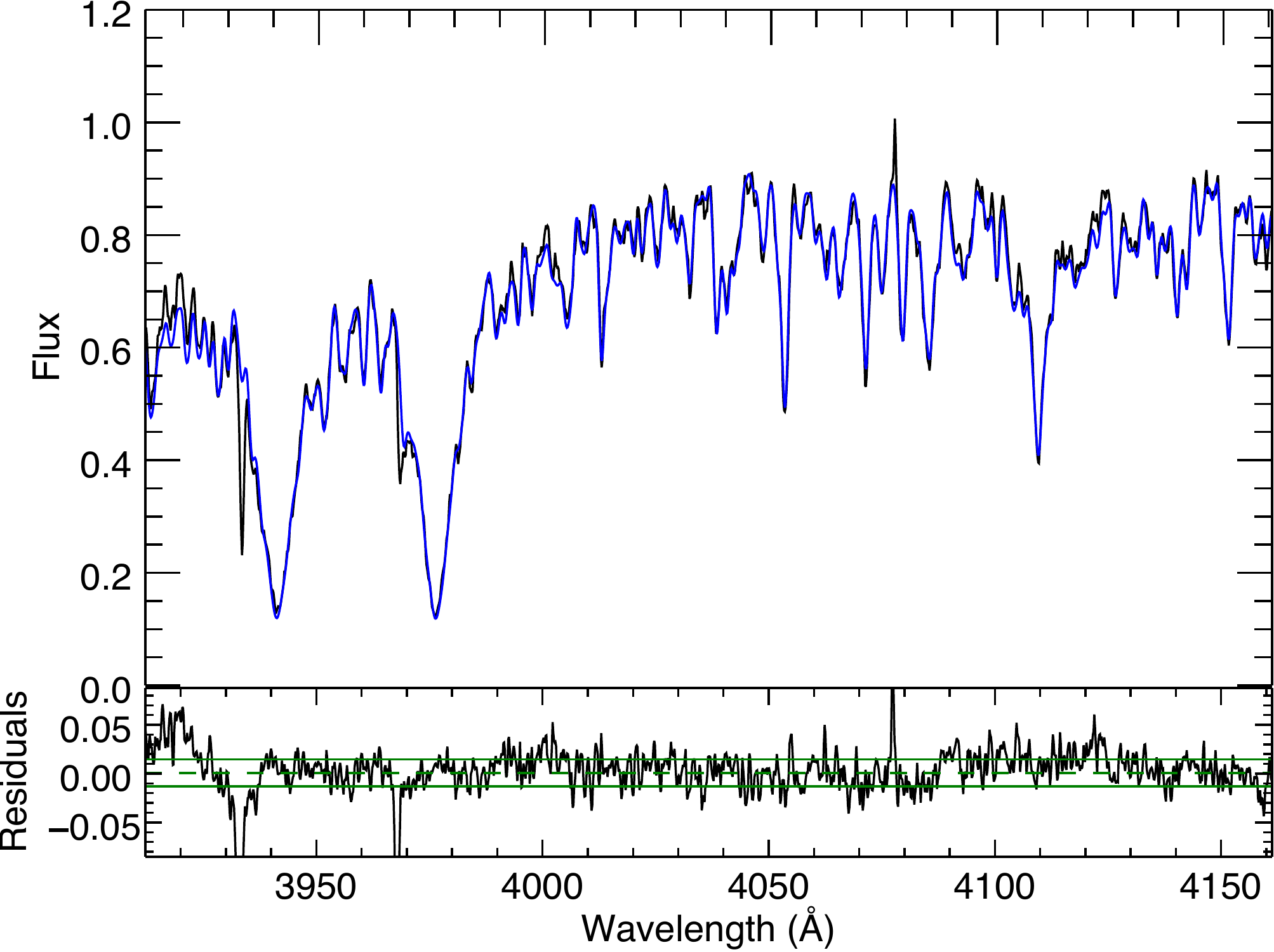}
	\includegraphics[width=\columnwidth]{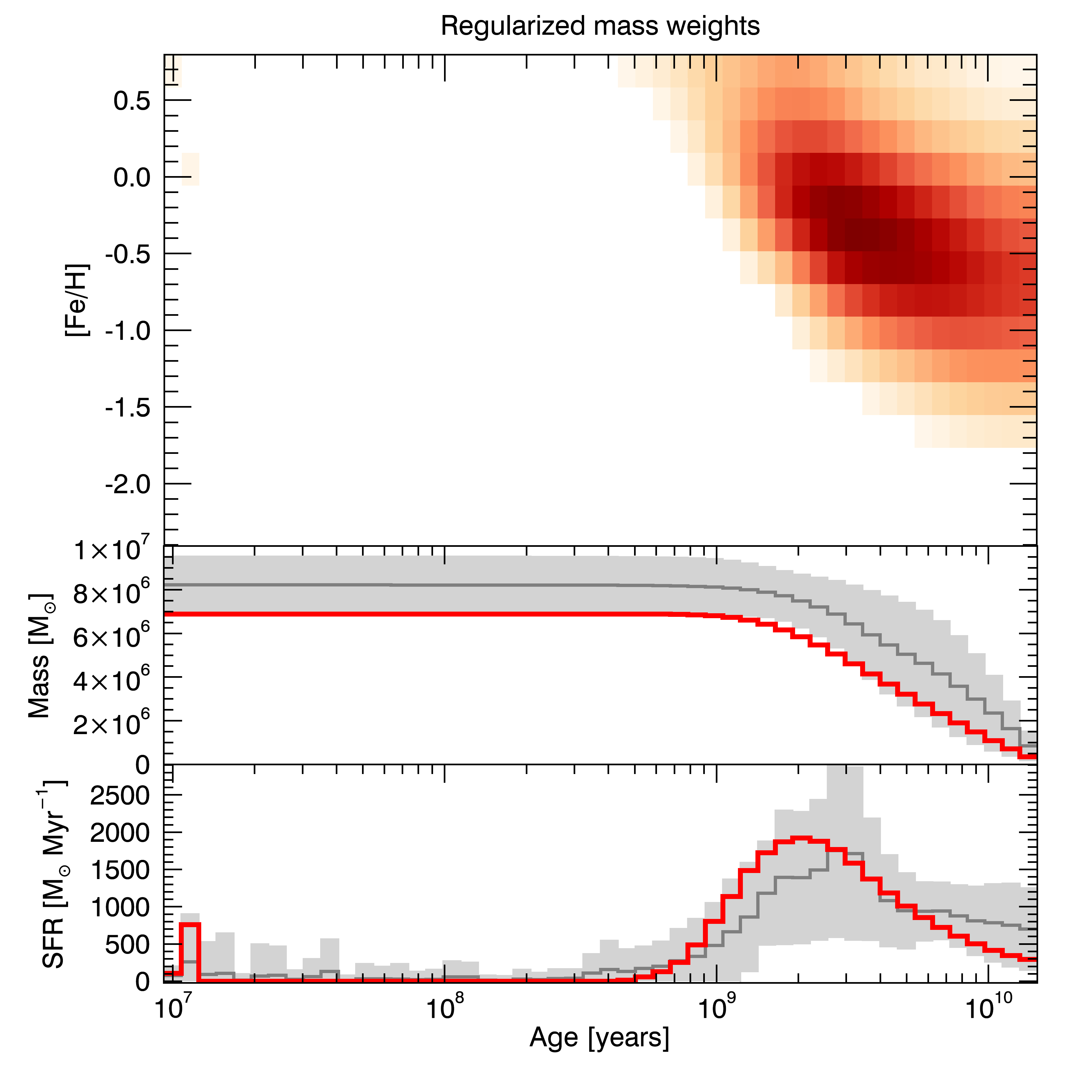}
	\includegraphics[width=\columnwidth]{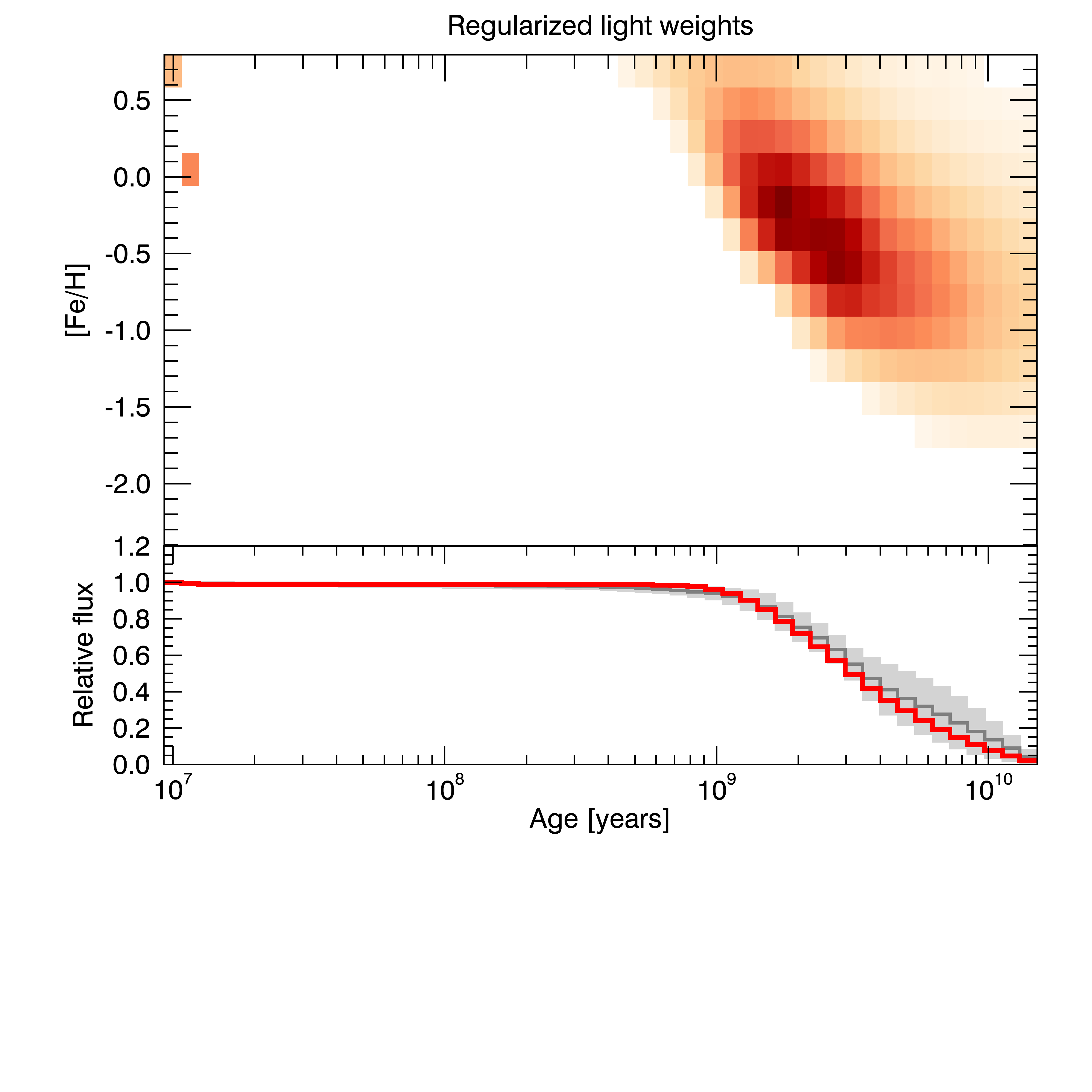}
    \caption{SFH of NGC\,5206 NSC - same as Figure \ref{fig:ngc247}.}
    \label{fig:ngc5206}
\end{figure*}

NGC\,5206 is the other early-type galaxy in the sample.
We combined the spectra from both periods of observation (P84 and P86) using a weighted average method.
We measured an average velocity dispersion $\sigma = 32.2 \pm 0.3$\kms.

At a difference with the nucleus of NGC\,5102, this NSC seems to be compatible with continuous SFH and a gradual enrichment from $\feh\sim-1.0$\,dex to $+0.5$\,dex with a peak of the SFR $\sim2$\,Gyr ago at a first glance (Figure \ref{fig:ngc5206}).
But NGC\,5206's NSC also formed most of its stars less than $10$\,Gyr ago, similarly to the other early-type galaxy nucleus NGC\,5102, with only $\sim15\%$ of the total mass being present $>10$\,Gyr ago.
Both early-type galaxies are different in this respect to all four spirals in the sample.
The observations of the latter are compatible with at least $50\%$ of their total masses being formed more than $10$\,Gyr ago.
NGC\,5206's NSC reached $50\%$ of its present total mass $\sim4.5$\,Gyr ago and $99\%$ $\sim1$\,Gyr ago.
However, considering the resolution of our fit (Figure \ref{fig:all_SSP_max_regul_light}), NGC\,5206 is marginally consistent with also being a SSP.
The quality of the SSP fit to this spectrum is also as good as the composite fit (Table \ref{tab:age_feh_ssp}).

There are no emission lines in the spectrum of this nucleus.

\subsection{NGC\,7793}

\begin{figure*}
	\includegraphics[width=\columnwidth]{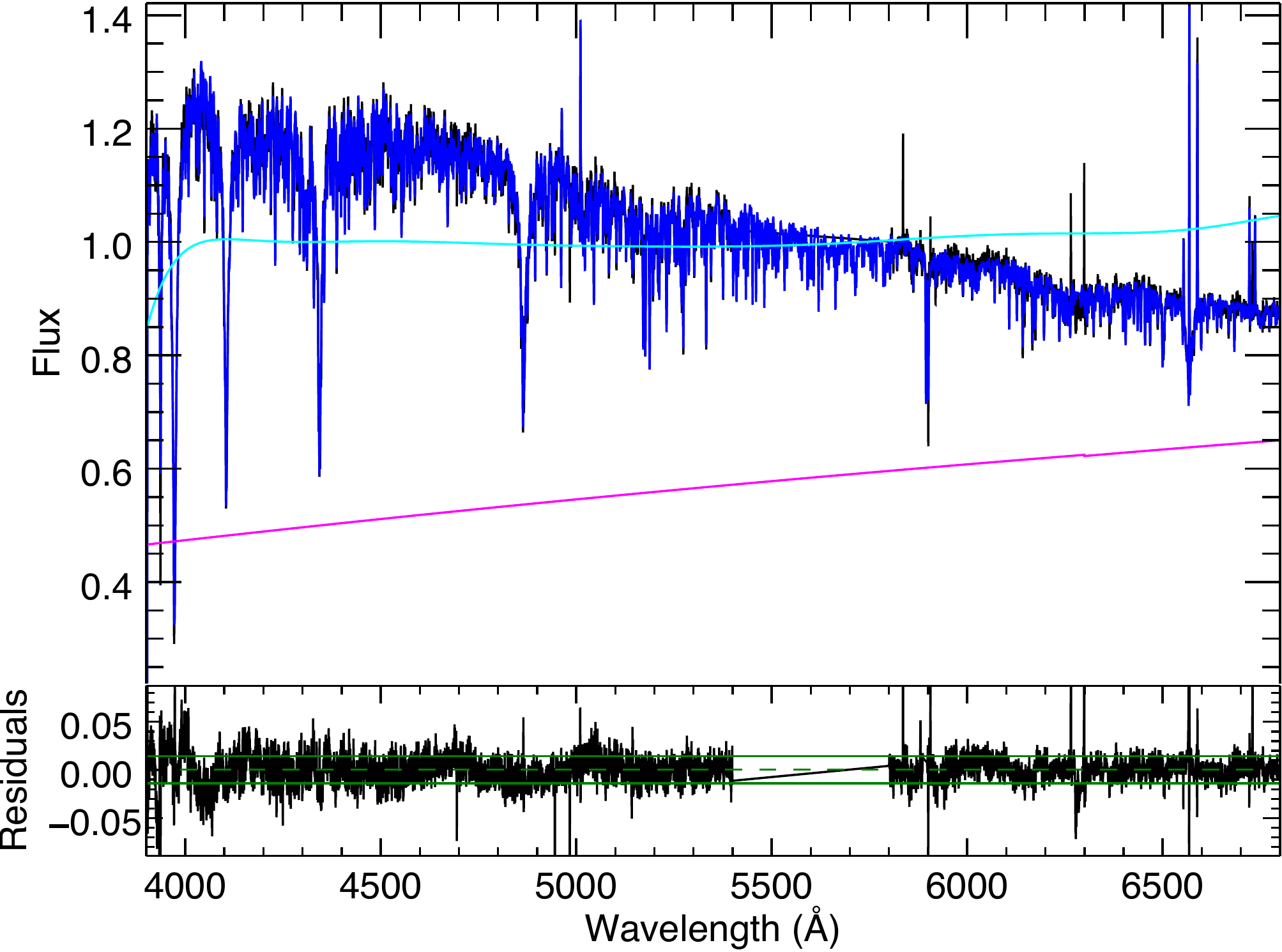}
	\includegraphics[width=\columnwidth]{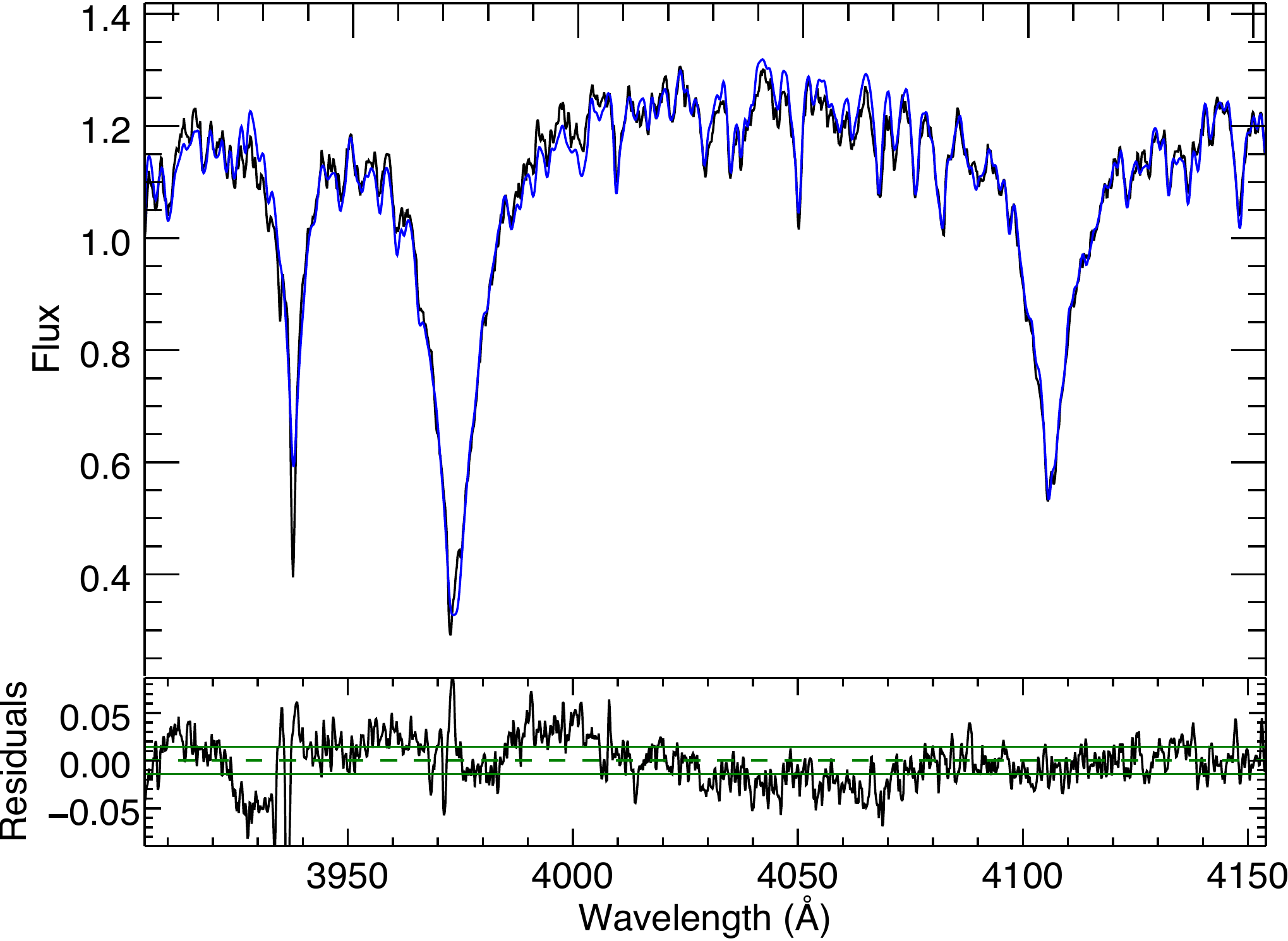}
	\includegraphics[width=\columnwidth]{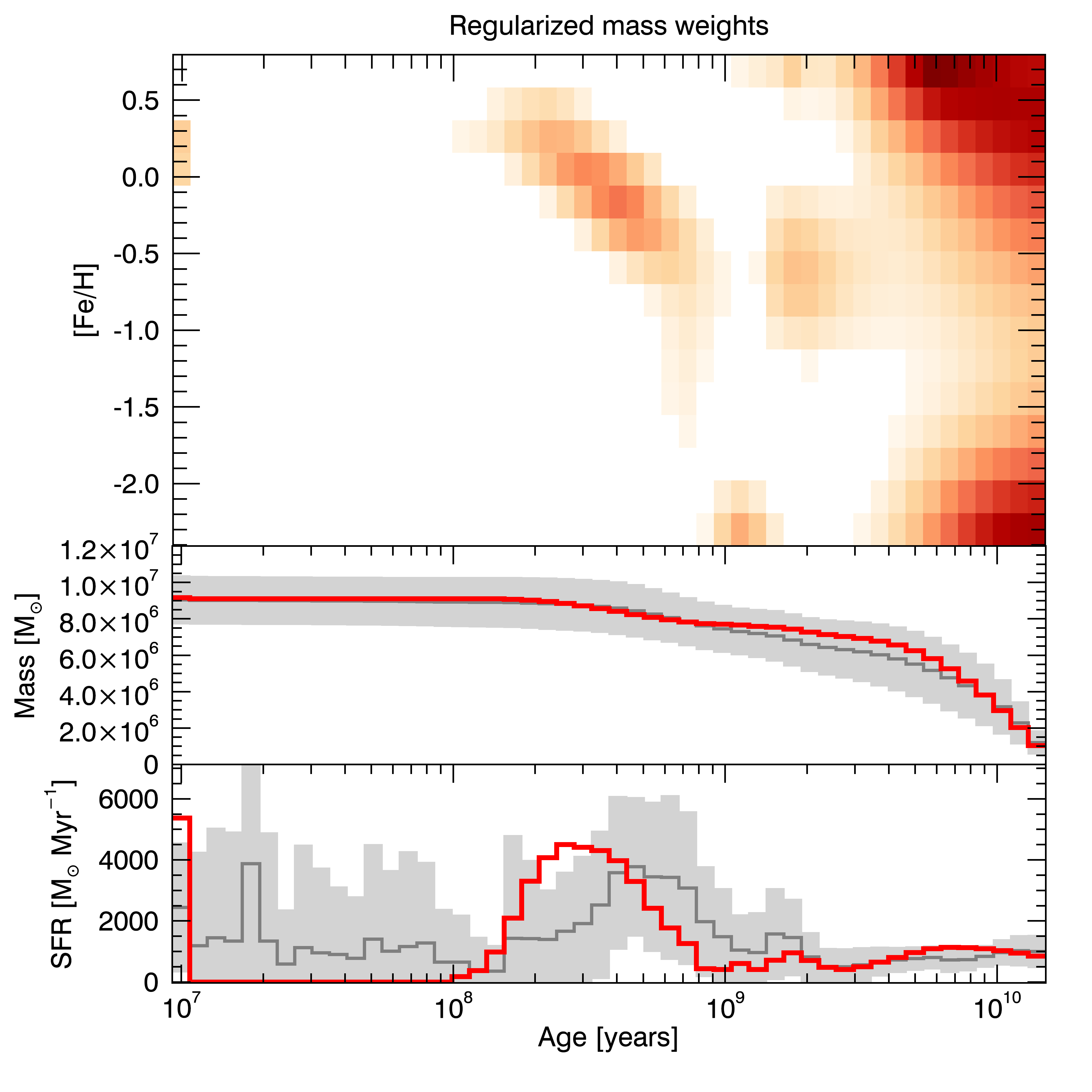}
	\includegraphics[width=\columnwidth]{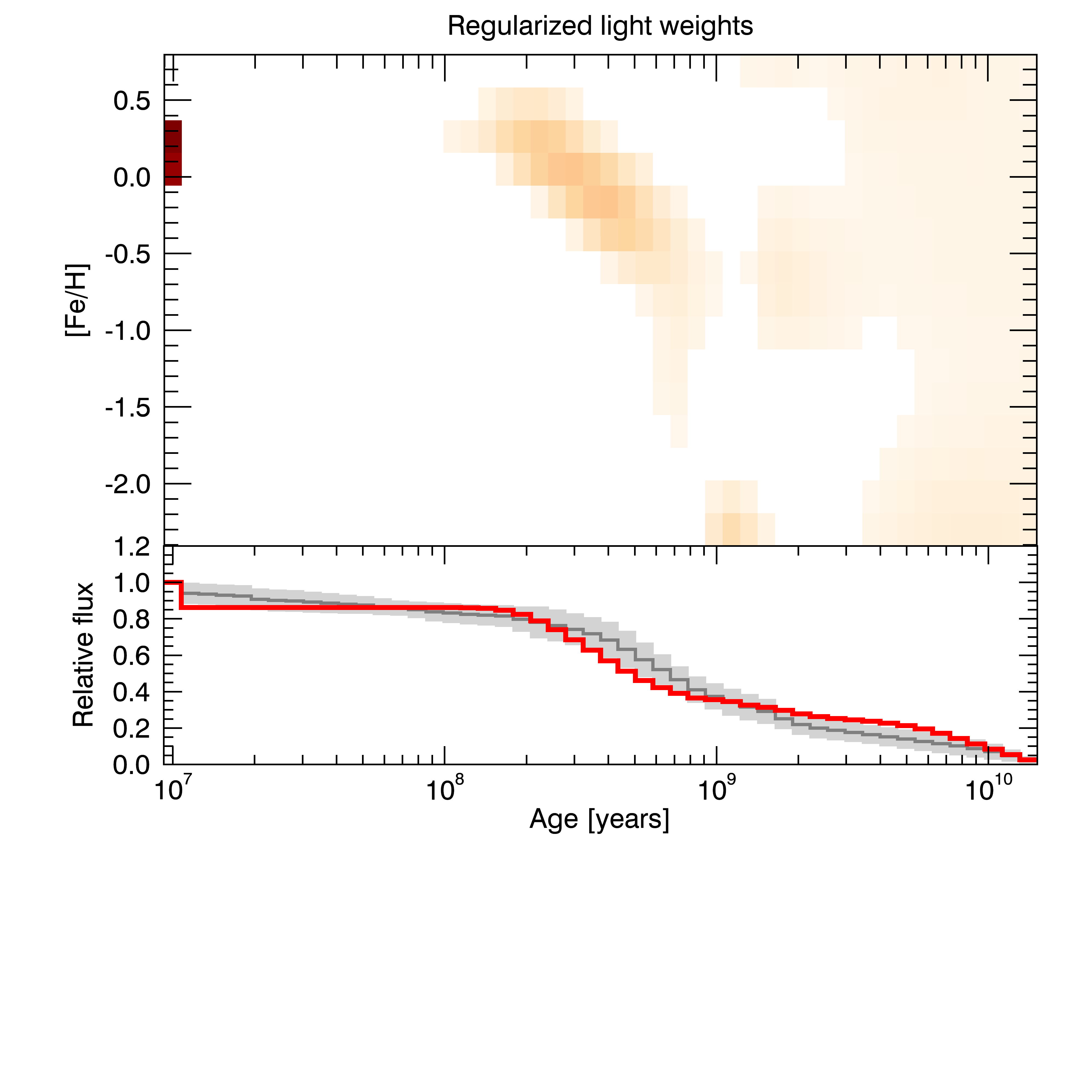}
    \caption{SFH of NGC\,7793 NSC - same as Figure \ref{fig:ngc247}.}
    \label{fig:ngc7793}
\end{figure*}

We measured an average velocity dispersion $\sigma = 23.1 \pm 0.4$\kms \citep[in excellent agreement with the result from][$24.6\pm3.7$\kms]{walcher+2005} in the nucleus of the spiral galaxy NGC\,7793.

The best fit composite stellar population model is quite complex (Figure \ref{fig:ngc7793}) and might be interpreted as a merger of multiple star clusters of different ages and metallicities.
The major components are two old ($>10$\,Gyr) populations with very different metallicities, essentially bordering the limits of our SSP grid - $\feh = -2.3$ and $+0.6$\,dex, respectively.
Together they amount to $75\%$ of the total mass formed in this NSC.
The very metal rich component is similar to what we find in the nucleus of NGC\,3621.
Other components include a population with $\feh\sim-0.5$\,dex of about $2$\,Gyr of age ($\sim6\%$ of the total mass) and a population of Solar metallicity and an age estimate of $200-600$\,Myr ($\sim15\%$ of the total mass).
There is also an indication of a very young ($\sim10$\,Myr, $6.6\times10^4\,\Msun$), metal rich component, and the nucleus has strong emission lines, indicative of ongoing star formation. 
These findings are in a reasonable agreement with the result of \citet{walcher+2006} for this NSC, who found $1\%$ of the mass in a $30$\,Myr old population, $23\%$ aged between $100$ and $600$\,Myr, and the rest $76\%$ to be about $6$\,Gyr.

Overall, the metallicity marginalized SFR rate of NGC\,7793 seems to be very similar to the estimated SFR of NGC\,247.
Both galaxies are also very similar morphologically.
However, the striking difference in the metallicity enrichment processes in both nuclei (stochastic in the case of NGC\,7793 and smooth in NGC\,247) might be illustrative for their different origin.
\citet{carson+2015} found from HST photometry that the young stars are situated in the outskirts of NGC\,7793's NSC, which points towards accretion origin, while in NGC\,247's nucleus, the young stars are more centrally concentrated, pointing towards an {\it in situ} origin.

\section{Discussion} \label{sec:discussion}

\subsection{Comparison of the population M$/$L ratios to dynamical estimates}

\begin{figure}
	\includegraphics[width=\columnwidth]{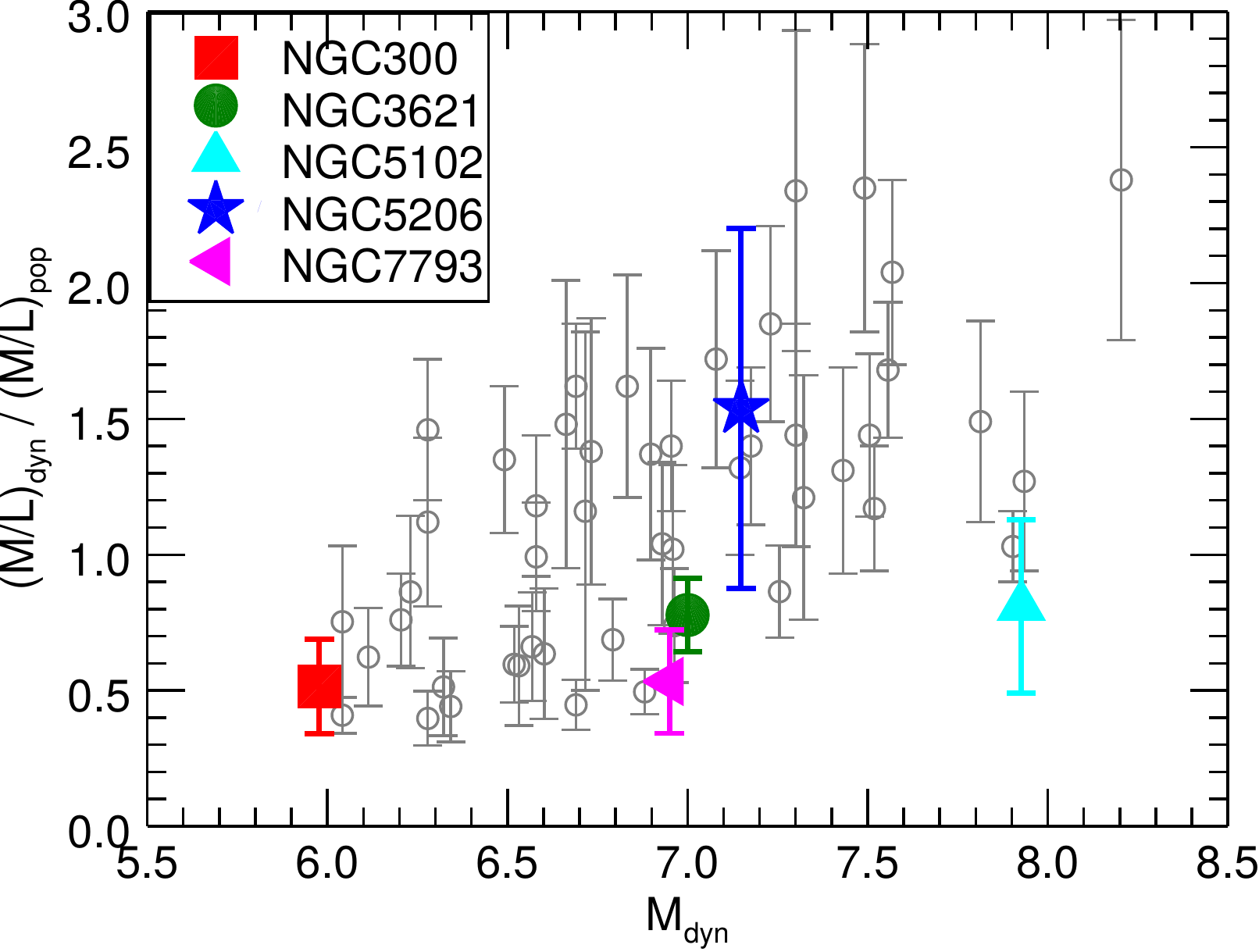}
    \caption{Dynamical vs. Population M$/$L ratios of the NSCs in this study (blue asterisks), compared to UCDs from \citet[][grey circles]{mieske+2013}. The dynamical masses and M$/$L ratios come from \citet{walcher+2005} for the nuclei of NGC\,300 and NGC\,7793; \citet{nguyen+2018} for the nuclei of NGC\,5102 and NGC\,5206; and \citet{barth+2009} for the nucleus of NGC\,3621. The population M$/$L ratios are from this study.}
    \label{fig:ml_ratio}
\end{figure}

Here we compare the M$/$L ratios, which we estimate based on the stellar population analysis of the nuclei (Tables \ref{tab:age_feh_reg} and \ref{tab:age_feh_mc}), to literature results based on dynamical modelling. 
In Figure \ref{fig:ml_ratio}, we present the ${\rm (M/L)_{dyn}}\, /\, {\rm (M/L)_{pop}}$ as a function of the dynamical mass of the system for a compilation of UCDs \citep{mieske+2013} and five of the nuclei in this work.
We did not find dynamical results only for the NSC of NGC\,247 in the literature.
Dynamical stellar M$/$L ratios have frequently been found previously to fall below their population expectations in metal-rich globular clusters and UCDs \citep{strader+2011,mieske+2013,seth+2014,ahn+2017,baumgardt2017}, but the cause of this is not yet well understood.
The trend of increasing ${\rm (M/L)_{dyn}}\, /\, {\rm (M/L)_{pop}}$ with the dynamical mass may be indicative of the increasing importance of the central SMBH to the dynamics of the system.

The dynamical ${\rm (M/L)_{dyn}}\propto\frac{1}{d}$, where $d$ is the distance to the galaxy, due to a linear scaling of the dynamical mass with the effective radius in the virial theorem, while the ${\rm (M/L)_{pop}}$ is distance independent.
Thus, we brought the literature ${\rm (M/L)_{dyn}}$ values to the distances adopted in this study (Table \ref{tab:galaxies}).
\citet{walcher+2005} estimate the dynamical mass of NGC\,300's and NGC\,7793 nuclei to be $0.95^{+0.77}_{-0.45}\times10^6$ and $8.9^{+3.3}_{-2.5}\times10^6\,\Msun$ (corrected to the adopted in this study distances of $2.0$ and $3.8$\,Mpc, respectively).
They have originally adopted a slightly larger distance to NGC\,300 ($d = 2.2$\,Mpc) and lower to NGC\,7793 ($d = 3.3$\,Mpc).
Thus, their published ${\rm (M/L_I)_{dyn}}$ corrected for distance differences become $0.72\pm0.22$ and $0.56\pm0.17$ for the nuclei of NGC\,300 and NGC\,7793.
We estimate ${\rm (M/L_I)_{pop}} = 1.4\pm0.2$ and ${\rm (M/L_I)_{pop}} = 1.05\pm0.2$ from the X-Shooter spectra for the two nuclei, respectively.
The ${\rm (M/L)_{dyn} / (M/L)_{pop}}$ is therefore $\sim0.5$ for both of them, suggesting SMBHs at the lower mass limit in the centres of these galaxies, if present at all.
However, our ${\rm (M/L)_{pop}}$ estimates are considerably higher than the composite population fits by \citet{walcher+2006}, who find ${\rm (M/L_I)_{pop}} = 0.64$ for NGC\,300's NSC and ${\rm (M/L_I)_{pop}} = 0.47$, which would bring the ${\rm (M/L)_{dyn} / (M/L)_{pop}}$ ratio closer to 1.

\citet{barth+2009} infer the dynamical mass of NGC\,3621's NSC to be $1\times10^7\,\Msun$.
They put an upper limit of the ${\rm (M/L_V)_{dyn}} = 1.4\pm0.2$ for a SMBH with a mass of $10^4\Msun$.
Our estimate ${\rm (M/L_V)_{pop}} = 1.8\pm0.2$ is in line with the population analysis by \citet[][${\rm M/L_V} = 2$]{barth+2009}.
This yields a ${\rm (M/L)_{dyn}}\, /\, {\rm (M/L)_{pop}} = 0.8$, which is very similar to e.g. M60-UCD1 and other metal-rich, old systems \citep{mieske+2013,seth+2014,ahn+2017,baumgardt2017}.

\citet{nguyen+2018} infer the dynamical masses of the NSCs of the early type galaxies NGC\,5102 and NGC\,5206.
Corrected to the adopted distances of this study, the masses of the nuclei and dynamical M$/$L ratios are $8.4\pm2.7\times10^7\,\Msun$~and $1.4\pm0.5\times10^7\,\Msun$ for NGC\,5102 and NGC\,5206 NSCs, and ${\rm (M/L_V)_{dyn}} = 0.5\pm0.2$ and ${\rm (M/L_I)_{dyn}} = 2.3\pm0.9$, respectively.
We get ${\rm (M/L_V)_{pop}} = 0.6\pm0.1$ and ${\rm (M/L_I)_{pop}} = 1.5\pm0.25$, respectively.
The ${\rm (M/L)_{dyn}}\, /\, {\rm (M/L)_{pop}}$ ratios are, therefore, $0.8\pm0.3$ and $1.5\pm0.7$.
\citet{nguyen+2018} detected SMBHs in both of them with masses $8.8^{+4.2}_{-6.6}\times10^5\,\Msun$ in NGC\,5102 and $4.7^{+2.3}_{-3.4}\times10^5\,\Msun$ in NGC\,5206.

\subsection{Comparison of the SFHs of the nuclei in the sample}

\begin{figure}
	\includegraphics[width=\columnwidth]{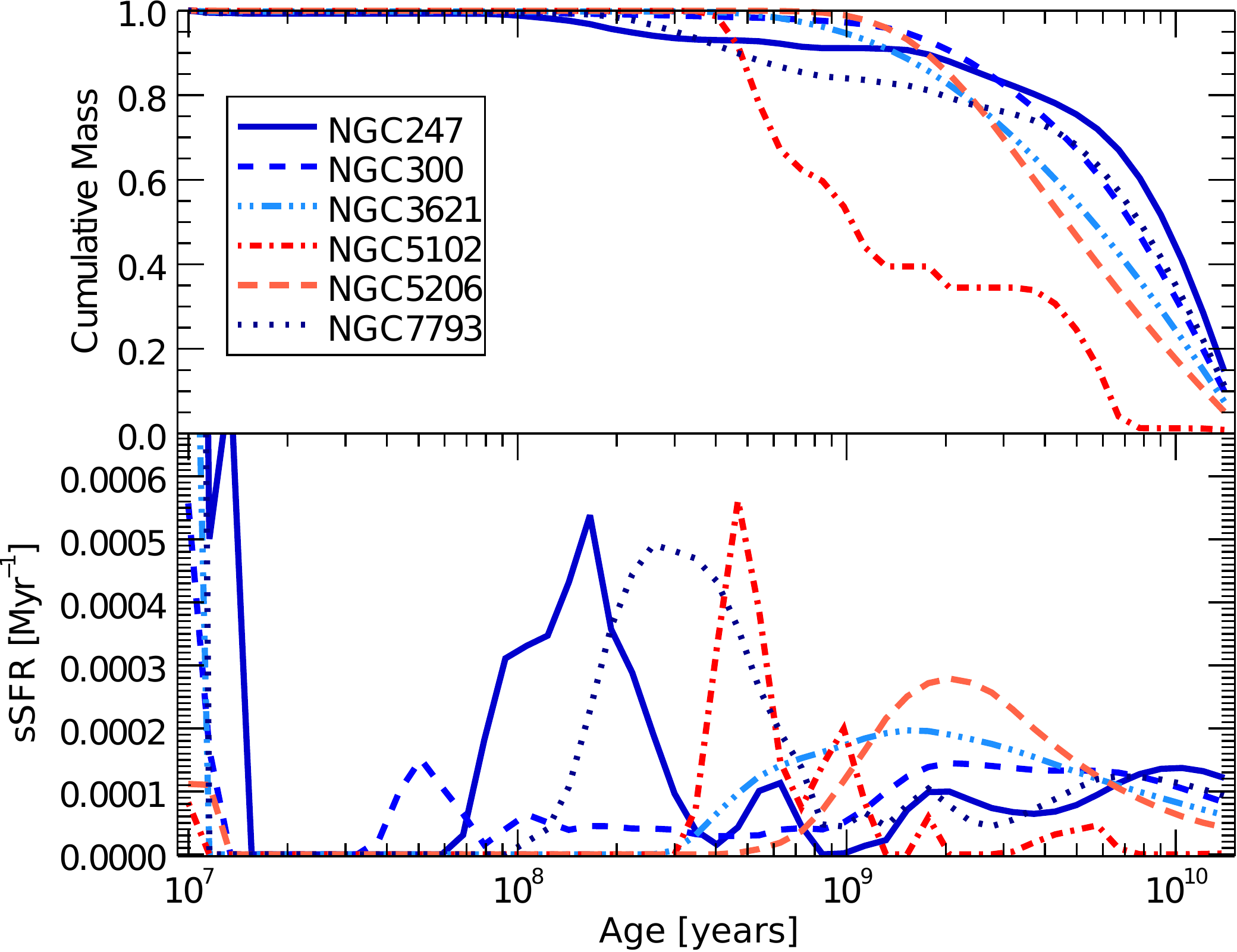}
    \caption{Normalised cumulative mass and specific SFR as a function of age for all six NSCs. Late-type galaxies are shaded in blue and early-type galaxies are shaded in red.}
    \label{fig:sfe}
\end{figure}

As discussed in Section \ref{sec:results}, there are a number of similarities and differences among the stellar content of the six NSCs analysed in this work.
In Figure \ref{fig:sfe} we compare the derived SFHs by plotting the normalised cumulative mass of each nucleus and the specific star formation rate defined as ${\rm sSFR} = {\rm SFR}/{\rm M_{tot}}$.
The two early-types stand out as the only nuclei that do not appear to have a significant fraction of old population ($>10$\,Gyr) and are more consistent with being SSPs than the spirals.
The dominant populations of the two nuclei are both young, with NGC\,5206 having a mass-weighted age of $4.3$\,Gyr, while NGC\,5102 has a mass-weighted age of $1.4$\,Gyr (Table \ref{tab:age_feh_reg}).
Their spectra are also compatible with SSP models.

The NSCs of the four spirals, on the other hand, were all assembled $>10$\,Gyr ago and continued forming stars at a roughly constant rate up until $\sim1$\,Gyr ago.
With the exception of NGC\,3621, which does not show evidence for recent star formation, the other three are consistent with experiencing a star forming burst of different intensity $\lesssim1$\,Gyr ago and there are strong indications that there is ongoing star formation even today in the NSCs of NGC\,247 and NGC\,7793.
Note, however, that the apparent low SFR recovered for ages $>1$\,Gyr may simply be due to the lack of age resolution to detect individual strong bursts as we do for populations younger than $1$\,Gyr and the detailed SFH is smeared out in time.

Three of the four late type spirals NSCs seem to be consistent with {\it in situ} star formation and gradual chemical enrichment, while the nucleus of NGC\,7793 is more consistent with a merger origin - a result also supported by the inverted colour gradient in this object \citep{carson+2015}.

\subsection{Comparison to SSP fits}\label{sec:discussion_ssp}

\begin{figure*}
	\includegraphics[width=15cm]{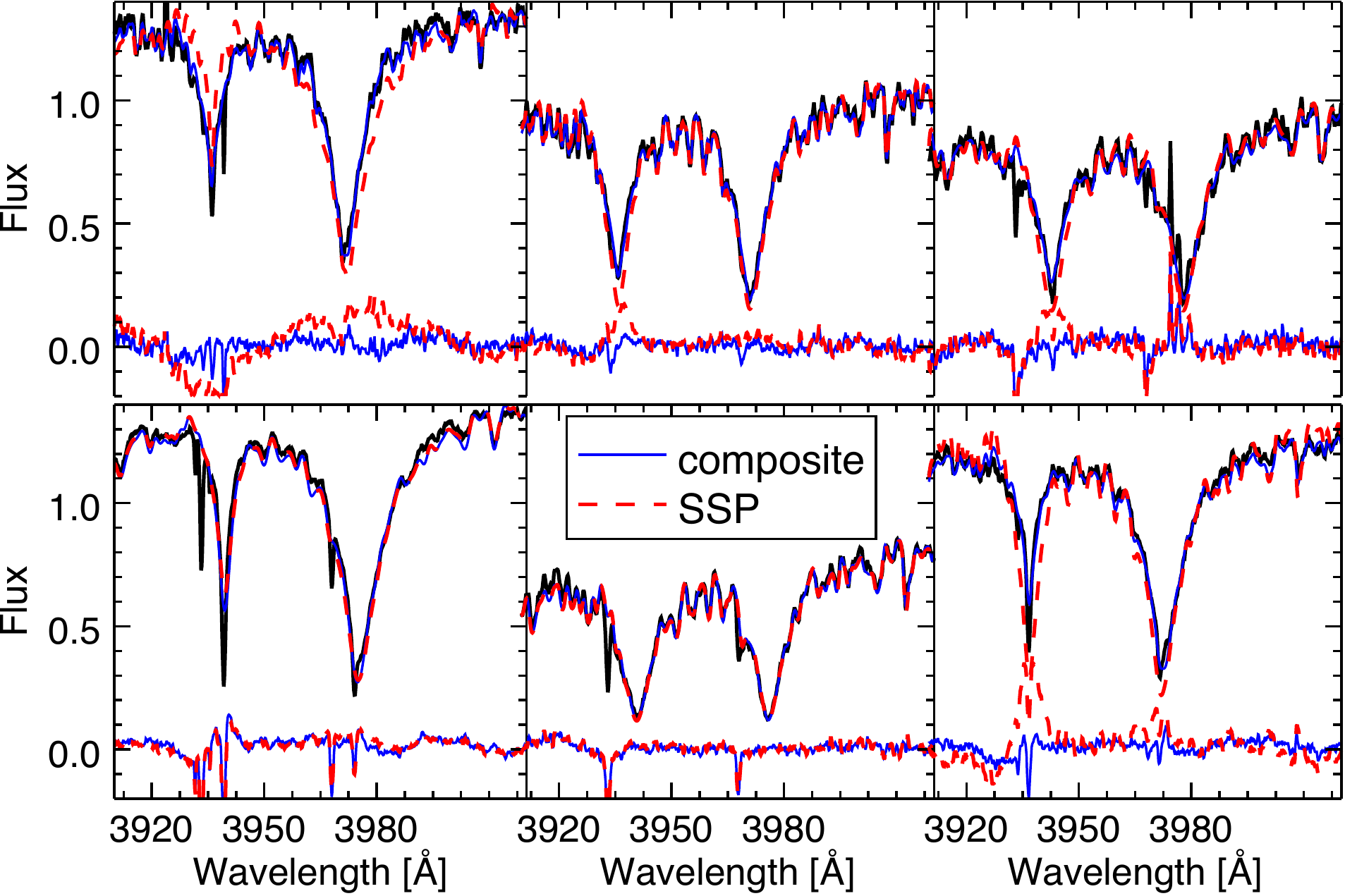}
    \caption{Composite (blue) vs SSP (red) fit models in the region of the Ca H\&K lines with residuals; {\it top row:} NGC\,247, NGC\,300, NGC\,3621; {\it bottom row:} NGC\,5102, NGC\,5206, NGC\,7793. The SSP fits to the spectra of the NSCs of the early-type galaxies NGC\,5102 and NGC\,5206 are of the same quality as the composite fits, while the SSP fits to the spectra of the late-type galaxies' nuclei fail to reproduce well this region.}
    \label{fig:cahk}
\end{figure*}

Here we compare the goodness of fit of the composite vs. SSP models to see if our NSCs are consistent with an SSP model alone.
The best fit SSP solutions together with the minimum $\chi^2$ for each NSC are presented in Table \ref{tab:age_feh_ssp}.
Throughout the text, we adopted a $\chi^2$ criterion (Eq. \ref{eq:max_chi2}), for which we no longer consider solution compatible with the data at the $1\sigma$ level.
This value is typically $\chi^2 = 1.013$, varying slightly due to the different number of pixels flagged as bad in the spectra.
One could see from Table \ref{tab:age_feh_ssp} that two NSCs are fully compatible with being SSP, namely the nuclei of NGC\,5102 and NGC\,5206, where an SSP and an order $10$ multiplicative polynomial can provide a fit equivalent to the best possible solution using the full SSP grid.
On the other hand, none of the late-type spirals is consistent with a SSP within the adopted goodness of fit criterion.
Overall, the largest difference between SSP and composite models is visible in our spectra in the region of the H\&K Ca II lines, which are generally not reproduced well by SSP models only.
This is demonstrated in Figure \ref{fig:cahk}, where we show this spectral region together with the best fit SSP and composite stellar population solutions.

\subsection{Constraints on the youngest and oldest population}\label{sec:yng_old}

\begin{figure}
	\includegraphics[width=\columnwidth]{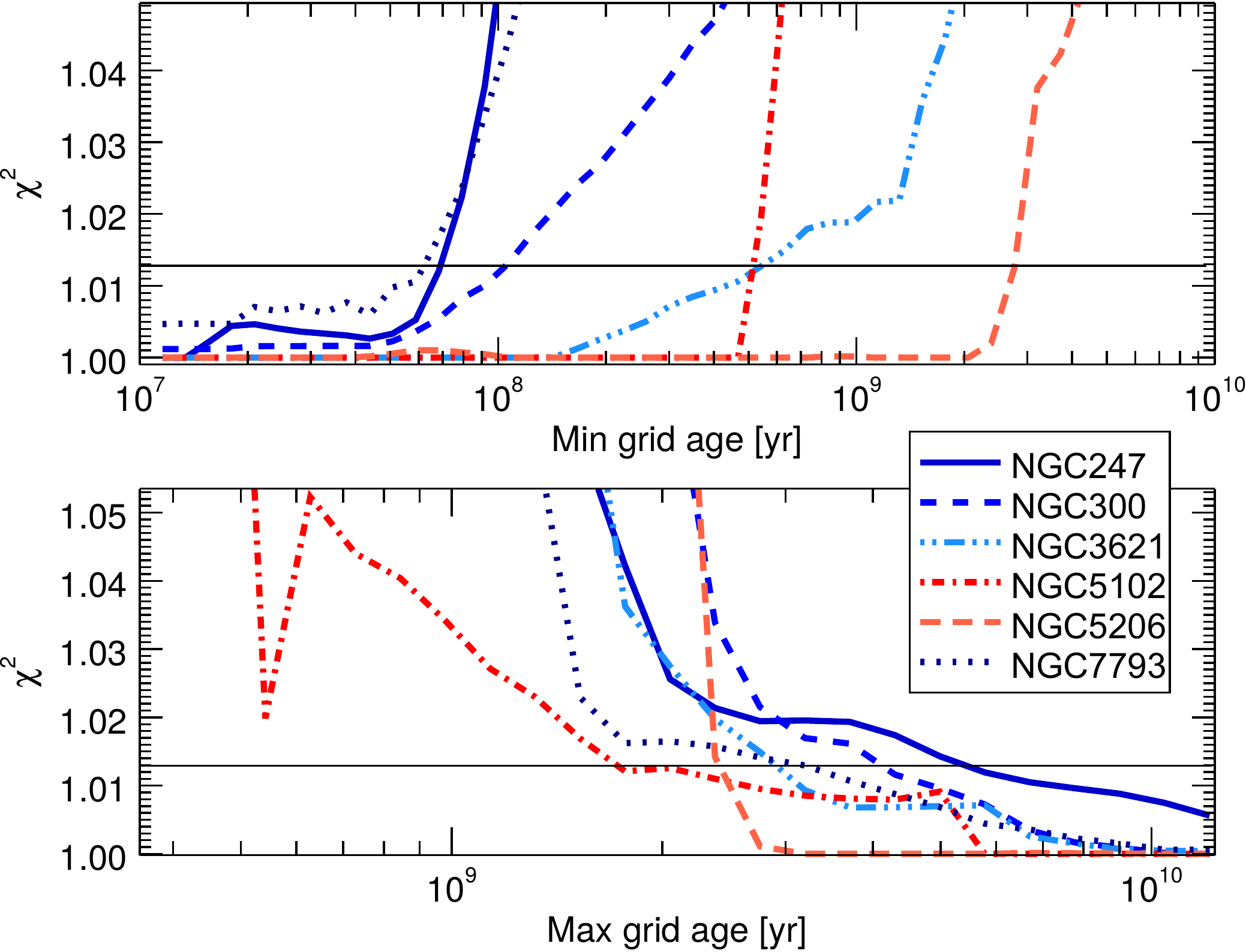}
    \caption{$\chi^2$ as a function of the minimum ({\it top panel}) and maximum ({\it bottom panel}) age in the model grid for the six NSCs. The $1\sigma$ threshold of the $\chi^2$ distribution is indicated with a horizontal black line. Late-type galaxies are shaded in blue and early-type galaxies are shaded in red.}
    \label{fig:remove_old_young}
\end{figure}

\begin{table}
	\centering
	\caption{Minimum and maximum age required in the SSP grid by the data.}
	\renewcommand{\arraystretch}{1.2}
	\label{tab:min_max_age}
	\begin{tabular}{ccc} 
		\hline
NSC	& Min grid age & Max grid age \\
        & [Myr]    &   [Myr]    \\  
		\hline
NGC\,247  & $70$   &  $5800$  \\
NGC\,300  & $90$   &  $4300$  \\
NGC\,3621 & $500$  &  $3200$  \\
NGC\,5102 & $500$  &  $1700$  \\
NGC\,5206 & $2700$ &  $2700$  \\
NGC\,7793 & $60$   &  $3700$  \\
		\hline
	\end{tabular}
\end{table}

An interesting question to address would be to determine how young the youngest population in the NSCs actually is.
As one can see from Figure \ref{fig:sfe}, most of our composite population fits require a small contribution from a very young population $\sim10-20$\,Myr.
To test the plausibility of such a young population, we performed unregularized {\sc ppxf} fits to all NSCs by gradually shrinking the age grid from the young side.
We show in the upper panel of Figure \ref{fig:remove_old_young} how the reduced $\chi^2$ changes as a function of the minimum age in the grid.
The results of this analysis are summarized in Table \ref{tab:min_max_age}, where we show the minimum age required to ensure a good fit for each NSC.
The values in the first column give the age of the youngest population, which is required in the SSP grid to get a reduced $\chi^2=1.013$, which is the $1\sigma$ deviation (Eq. \ref{eq:max_chi2}) from the unregularized solution using the full grid.
Not surprisingly, the three nuclei that must contain stars younger than $\sim100$\,Myr are those of NGC\,247, NGC\,7793, and NGC\,300.
The nuclei of NGC\,5102 and NGC\,3621 do not necessarily contain stars younger than $500$\,Myr, and NGC\,5206's NSC does not necessarily contain stars younger than $\sim2.7$\,Gyr.

In a similar fashion we tried to constrain the oldest age in the grid required by the data (bottom panel of Figure \ref{fig:remove_old_young}).
We achieve good fits limiting the grid size to ages significantly younger than $10$\,Gyr (second column in Table \ref{tab:min_max_age}).
This result highlights the lack of sensitivity at old populations, especially in the cases where they contribute very little light in the composite spectrum, dominated by the flux of the younger stars.
The age-metallicity degeneracy likely also plays a role, as the spectra of old, metal poor SSPs are similar to those of younger, more metal rich SSPs.
In general, we are able to get satisfactory fits using only an age grid going up to $4-6$\,Gyr for most NSCs.
Looking at the minimum and maximum grid age required by the spectra of NGC\,5102, we can confirm our previous result that NGC\,5102's nucleus experienced a very brief SFH and NGC\,5206 is compatible with being an SSP.

\subsection{Insights from the surrounding galaxy light}\label{sec:field}

\begin{table*}
	\centering
	\renewcommand{\arraystretch}{1.2}
	\caption{Mass- and light-weighted ages and metallicities of the NSC surroundings.}
	\label{tab:age_feh_disc}
	\begin{tabular}{cccccccc} 
		\hline
Galaxy   & ${\rm AGE_{SSP}}$ & $\feh_{\rm SSP}$  & ${\rm AGE_{light}}$ & $\feh_{\rm light}$ & ${\rm AGE_{mass}}$ & $\feh_{\rm mass}$ & SNR         \\
         & [Gyr]             &                   & [Gyr]               &                    & [Gyr]              &                   & [px$^{-1}$] \\  
		\hline              
NGC\,247  & $2.05$ & $-1.02$ & $0.61^{+0.08}_{-0.07}$ & $-0.63\pm0.16$ & $4.24^{+1.34}_{-1.02}$ & $-0.76\pm0.18$ & $23$ \\
NGC\,300  & $2.76$ & $-0.81$ & $2.24^{+0.45}_{-0.37}$ & $-0.81\pm0.08$ & $6.75^{+1.20}_{-1.02}$ & $-0.78\pm0.09$ & $20$ \\
NGC\,3621 & $1.77$ & $-0.59$ & $1.21^{+0.14}_{-0.12}$ & $-0.25\pm0.09$ & $4.98^{+1.18}_{-0.95}$ & $-0.16\pm0.09$ & $44$ \\
NGC\,5102 & $1.14$ & $-0.38$ & $0.73^{+0.09}_{-0.08}$ & $-0.38\pm0.15$ & $2.25^{+0.64}_{-0.50}$ & $-0.35\pm0.10$ & $73$ \\
NGC\,5206 & $1.53$ & $-0.16$ & $2.29^{+0.32}_{-0.28}$ & $-0.39\pm0.09$ & $5.64^{+1.31}_{-1.06}$ & $-0.31\pm0.09$ & $27$ \\
NGC\,7793 & $1.77$ & $-1.02$ & $0.71^{+0.13}_{-0.11}$ & $-0.97\pm0.13$ & $4.70^{+1.24}_{-0.98}$ & $-1.06\pm0.29$ & $33$ \\
		\hline
	\end{tabular}
\end{table*}

\begin{figure*}
	\includegraphics[width=15cm]{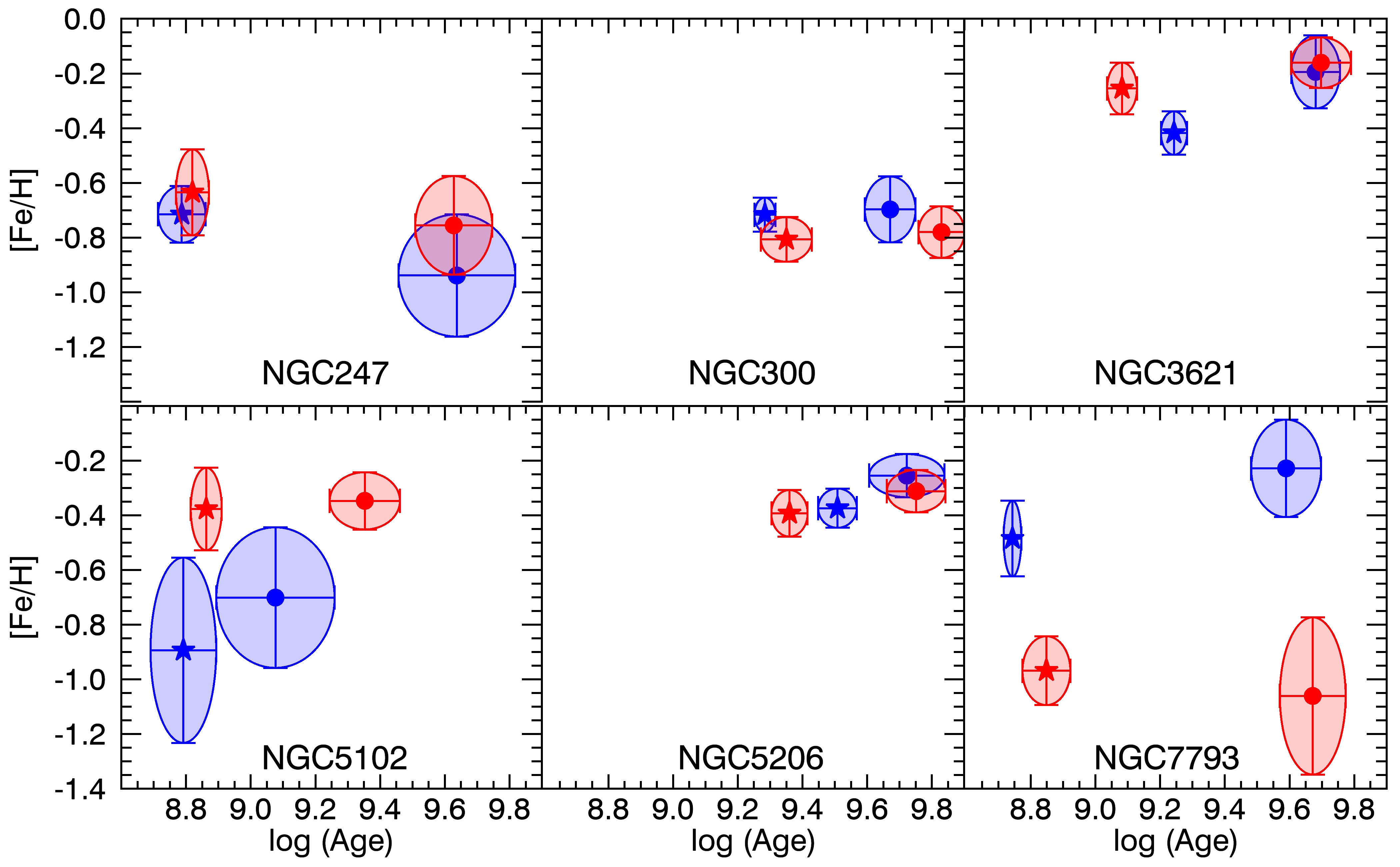}
    \caption{Comparison of the light-weighted (asterisks) and mass-weighted (dots) ages and metallicities in the NSCs (blue ellipses) and the integrated field light (red ellipses) in all six galaxies.}
    \label{fig:nsc_disc_comp}
\end{figure*}

While NSCs have been found to be younger on average than the field stars of their hosts in early-type galaxies \citep[see e.g.][]{paudel+2011}, it is still not determined if that is also the case in spirals.
In disk galaxies, there is generally ample gas to support star formation, both in the disk and the nucleus.
In this case, stellar ages in the disk and NSC should not be systematically different, as starbursts in the disk and the nucleus happen often, but without correlation.
However, once a galaxy is disrupted and morphs into an early-type, the gas is mostly removed, and any remaining gas will be driven into the nucleus.
In this case, any post-merger star formation will occur there (if at all) and so early-type nuclei should appear younger than their host bodies on average.

In an attempt to shed some light to this question, we extracted the light in an aperture with a typical size $3.5\arcsec$ centred $4\arcsec$ away on both sides from the NSC from the X-Shooter long slit spectra and combined them.
The extracted spectra from the nuclear surroundings of the sample galaxies have a SNR between $20$ and $45$ per pixel.
The only exception is NGC\,5102, where we reach a SNR$\sim70$.
We analysed them the same way, we analysed the extracted spectra from the NSCs - presenting both the SSP and composite population fit results based on bootstrap resampling with low regularization ($R=5$) in Table \ref{tab:age_feh_disc}.
We compare these results with the corresponding quantities for the NSCs, presented in Table \ref{tab:age_feh_mc}, in Figure \ref{fig:nsc_disc_comp}.
We plot both, the light-weighted and mass-weighted mean ages and metallicities of the NSC and its surroundings in Figure \ref{fig:nsc_disc_comp}, where the mass-weighted mean age is always higher than the light-weighted counterpart.
Perhaps, somewhat counter intuitively, the age and metallicity estimates of the lower SNR spectra from the surrounding galaxy light appear better constrained in Figure \ref{fig:nsc_disc_comp} than the equivalent quantities derived from the higher SNR spectra of the nuclei.
For both regions (nuclear and field population), the ellipses in the figure represent the mean and $1\sigma$ spread of the light- and mass-weighted ages and metallicities, estimated from $100$ residual bootstrap resamplings of the spectra, fit with regularization parameter $R=5$.
However, the same regularization parameter gives smoother solutions for lower SNR spectra than for higher SNR equivalents.
At the same time, the smoother solutions tend to be more self-similar when bootstrapping the residuals, which leads to the smaller spreads in the estimated ages and metallicities of the field populations.
One could say that the fits to the lower SNR spectra of the galaxies' field population are dominated by the regularization prior to a larger extent than the higher SNR spectra of the nuclei.

In any case, looking at the early-type galaxies, the NSC of NGC\,5102 does appear younger and more metal poor than its galaxy field, while the NSC of NGC\,5206 appears coeval with the field population and with the same metallicity.
The SFH of the field population in NGC\,5206, however, might be a bit more extended than in the NSC as the difference between the mean light- and mass-weighted ages of the stars is slightly larger in the field.

The situation in the late-type galaxies is also very heterogeneous.
The field and NSC populations in NGC\,247 and NGC\,3621 appear very coeval and with the same metallicity.
The NSCs of NGC\,300 and NGC\,7793 might be slightly younger than the surrounding field stars and NGC\,7793's nucleus is considerably more metal rich than the field population.

In general, if we focus on the mass weighted ages, all NSCs have either similar ages or are younger than their surrounding galaxy light.

\subsection{Insights from the emission lines}\label{sec:emission}

\begin{figure}
	\includegraphics[width=\columnwidth]{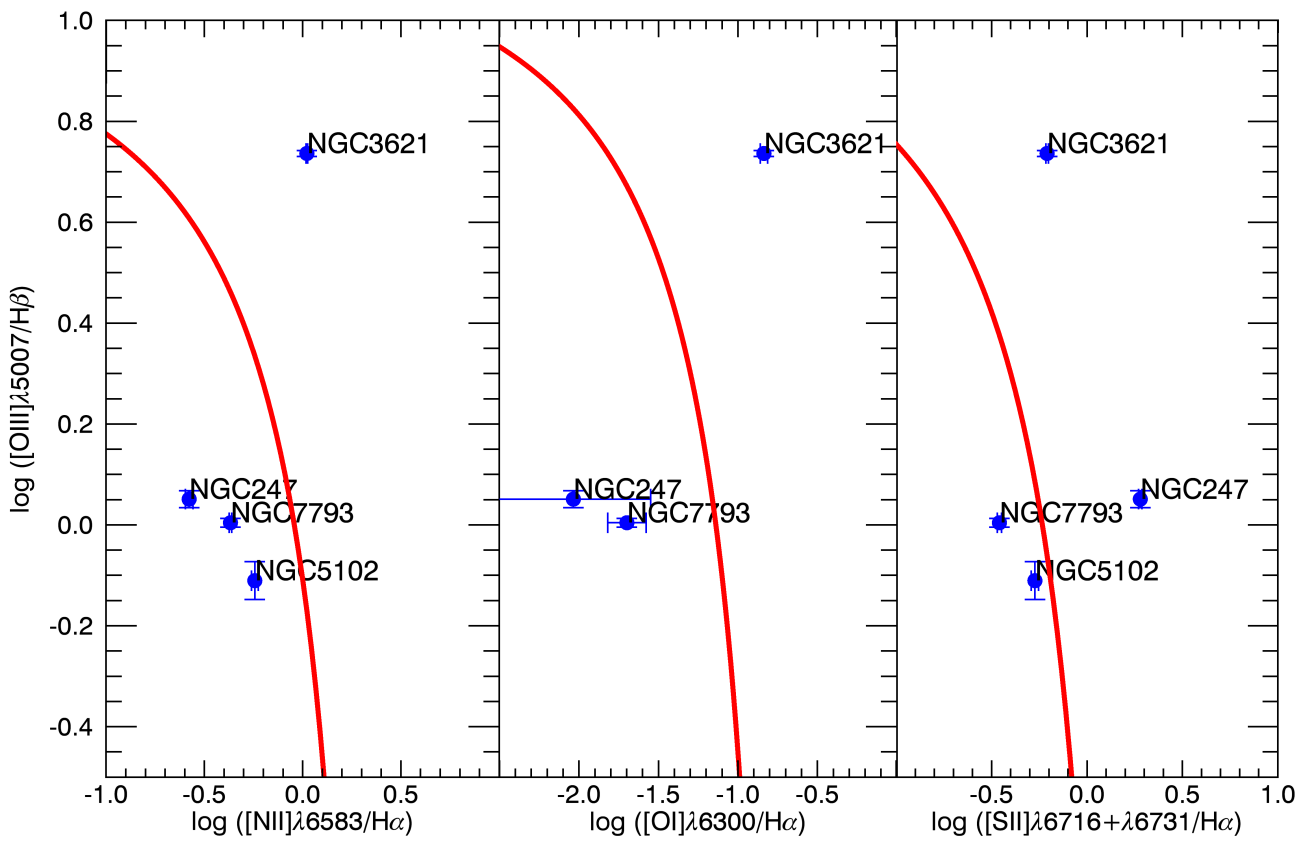}
    \caption{BPT diagram for the NSCs with emission lines. The separation curves \citep[][in red]{kewley+2001} define star forming regions (below) or AGN and LINER type of emission (above).}
    \label{fig:bpt}
\end{figure}

\begin{table*}
	\centering
	\caption{Emission line measurements.}
	\label{tab:emission}
	\begin{tabular}{ccccccc} 
		\hline
Galaxy  & ${\rm L_{H\alpha}}$\tablefootmark{1} & ${\rm SFR_{H\alpha}}$ & ${\rm H_{\alpha}/H_{\beta}}$\tablefootmark{1} & [O/H]\tablefootmark{2}   & [O/H]\tablefootmark{3} & ${\rm SFR_{pop}}$\tablefootmark{4} \\
        & [erg/s]                              & [${\rm\Msun\,Myr^{-1}}$] &                           &                         &                  & [${\rm\Msun\,Myr^{-1}}$] \\  
		\hline
NGC\,247  & $7.7\times10^{35}$ & $4.26\pm0.07$ & $2.93\pm0.10$ & $-0.10\pm0.06$ & $-0.16\pm0.06$ & $475\pm390$ \\
NGC\,3621\tablefootmark{5} & $5.1\times10^{36}$ & $---$         & $2.77\pm0.04$ & $---$          & $---$          & $830\pm640$ \\
NGC\,5102 & $7.8\times10^{36}$ & $42.7\pm0.85$ & $4.18\pm0.22$ & $-0.14\pm0.05$ & $0.00\pm0.06$  & $<2850$\\
NGC\,7793 & $6.2\times10^{36}$ & $33.8\pm0.25$ & $2.78\pm0.05$ & $-0.13\pm0.05$ & $-0.06\pm0.06$ & $2400\pm2100$\\
		\hline
	\end{tabular}
\tablefoot{
\tablefoottext{1}{Extinction corrected.}
\tablefoottext{2}{Based on the [O\,III]\,/\,[N\,II] index calibration.}
\tablefoottext{3}{Based on the [N\,II] index calibration.}
\tablefoottext{4}{SFR according to the bootstrap composite stellar population fits.}
\tablefoottext{5}{Values for NGC\,3621's NSC are not given due to the Seyfert nature of its emission lines.}
}
\end{table*}

As we mentioned in Section \ref{sec:models} we used emission line templates in {\sc ppxf} in addition to the SSP models.
The emission lines were modelled as Gaussian functions with an integral of $1$ and were treated the exact same way as the population models during the composite population fits (e.g. they were extinction corrected, broadened by the LSF, etc.), so their weights correspond directly to the strengths of the observed lines. 
There are visible emission lines in the spectra of four out of six NSCs studied in this paper: NGC\,247, NGC\,3621, NGC\,5102, and NGC\,7793.
There are no visible emission lines in the spectra of the NSC of NGC\,300 and NGC\,5206.
The uncertainties of the emission line strengths are estimated as the $1\sigma$ spreads of the template weights from the residual bootstrap analysis.

It is well known that the source of the emission (star formation or AGN/LINER activity) can be determined by analyzing different emission line ratios through the so called BPT diagrams \citep*{bpt}, after the names of the scholars that first presented them.
We show such a diagram for four different emission line ratios in Figure \ref{fig:bpt}.
We overplot the maximum star formation separation curve by \citet{kewley+2001}.
With the exception of NGC\,3621, which is a well known Seyfert 2 galaxy \citep{satyapal+2007}, all other nuclei appear to be star forming.
NGC\,247's nucleus is a bit strange as it looks like a normal star forming region in two of the diagrams but manifests as LINER if the sulfur lines are considered.

The luminosity of H$_{\alpha}$ is a good SFR estimator in star forming galaxies and we provide the SFR estimates based on the \citet{osterbrock+ferland2006} calibration for the four star forming nuclei in Table \ref{tab:emission}, together with the measured, extinction corrected ${\rm H_{\alpha}\,/\,H_{\beta}}$ ratios (using the best fit continuum-based correction curve).
The nominal ${\rm H_{\alpha}\,/\,H_{\beta}}$ ratio for star forming regions is $2.85$ \citep{osterbrock1989,kewley+2006}, which is very close to what we measure, meaning that the average dust attenuation that we have estimated from the stellar population analysis is also adequate for the gas.
The only exception is NGC\,5102, where we measure considerably higher line ratio, possibly indicative of different extinction towards the gas and the stars.
All nuclei seem to experience relatively low SFR now, much lower than what would be required to reproduce the mass formed in the last $\sim10$\,Myr according to the composite population fits (see the last column of Table \ref{tab:emission}).
As we noted earlier, the large light fractions in the youngest population bins in the composite population fits may also reflect effects, such as extended horizontal branch morphologies \citep{conroy+2018}, blue straggler stars \citep{schiavon2007}, or other complexities on the Hertzsprung-Russell diagram, unaccounted for in the SSP models.
The estimated SFRs today are also lower than their average rate throughout their entire history.
One might argue that this result supports theories where {\it in situ} star formation happens only in rare burst events, or stars are forming outside of the nucleus and migrate inwards.
There is also the possibility that the generally weak H$\alpha$ in the nuclei in this sample is related to stochasticity in the formation not recovered by our models \citep{weisz+2012}.
In addition, there may be other sources of emission, such as active stars.

We also measured the metallicity of the gas based on oxygen abundance using the [O\,III]\,/\,[N\,II] \citep{alloin+1979} and [N\,II] \citep{storchi-bergmann+1994} line indices and the calibrations by \citet{marino+2013}.
We find that the star forming gas in all nuclei has a slightly subsolar metallicity with a typical ${\rm [O/H]} = -0.1$\,dex (Table \ref{tab:emission}).

\section{Conclusions} \label{sec:conclusions}

We analysed unresolved, flux-calibrated, optical spectra of the NSCs of six nearby galaxies (four bulgeless spirals and two dwarf ellipticals) taken with the X-Shooter instrument at the VLT.
We fitted a linear combination of a 2-dimensional grid of SSPs (50 age bins and 15 [Fe/H] bins) to the observed spectra to derive their SFHs, using the spectral fitting code {\sc ppxf} \citep{cappellari+emselem2004,cappellari2017}, and applying weights regularization.
The SSP models were computed with the {\sc pegase-hr} \citep{leBorgne+2004} stellar population synthesis code and are based on the Elodie 3.1 stellar library \citep{prugniel+soubiran2001,prugniel+2007}.
The model grid has a spectral resolution of $10\,000$ and covers the wavelength region from $3900$ to $6800$\,\AA, and is distributed with {\sc ulyss} \citep{koleva+2009}.
In addition to the SSP grid, we also included intrinsic extinction, emission lines, and a multiplicative polynomial to correct for uncertainties in the flux calibration in our pixel by pixel modelling.

We found that the four late-type galaxies experienced a prolonged SFH, with a significant fraction $\sim50\%$ of their stellar mass formed more than $10$\,Gyr.
This is the era where massive globular cluster formation was also most common and at the peak of star formation efficiency in the Universe \citep{madau+dickinson2014}.
However, we note that our tests suggest that our age resolution is fairly limited at ages beyond $\sim5$\,Gyr.

The NSCs in the late-type galaxies have continued forming stars up until few hundred Myr ago.
The nuclei of NGC\,247 and NGC\,7793 are the only ones, where we find strong evidence for the presence of stars younger than $100$\,Myr and possibly still ongoing star formation, confirmed from emission lines in their spectra.
The NSCs of the spirals NGC\,247 and NGC\,300 are consistent with {\it in-situ} star formation showing a gradual metallicity enrichment from $\sim-1.5$\,dex more than $10$\,Gyr ago reaching super-Solar values a few hundred Myr ago.
We found the NSC of the Seyfert 2 galaxy NGC\,3621 to be very metal rich already in the early Universe.
Overall, we conclude that the SFHs of these three spirals is consistent with the {\it in-situ} formation scenario, where they are predominantly formed from self-enriched gas.
This conclusion is further supported by the colour gradient found by \citet{carson+2015}, who noticed that younger (bluer) stars are more centrally concentrated than the older (redder) stars in these NSCs.

NGC\,7793, on the other hand, has a very complex SFH, likely dominated by merging of various massive star clusters coming from different environments.
In support to this finding, \citet{carson+2015} reported that NGC\,7793's NSC has an inverted colour gradient.

The characteristic age of the nuclei of the early-type galaxies appear surprisingly younger than their counterparts in late-type spirals.
While the sample size is small, this may point to a physical difference in NSC formation in these environments.
This is consistent with the stellar populations in the NSCs of a number of elliptical galaxies, which are substantially younger than their hosts field population: NGC\,404 \citep{cid-fernandes+2005,seth+2010,nguyen+2017}, M\,32 \citep{worthey2004, rose+2005}, NGC\,205 \citep{butler+martinez-delgado2005, valluri+2005}, and NGC\,5102 (this work).
The NSCs of the two early-types in our sample are consistent with being SSPs but NGC\,5206 ($\sim2$\,Gyr) is significantly older than NGC\,5102 ($\sim0.5$\,Gyr).
There is considerable evidence that NGC\,404's younger nucleus was predominantly formed during a minor merger \citep{seth+2010}, and given the lack of evidence for older populations in NGC\,5206 and NGC\,5102 it is easy to imagine a similar origin for these NSCs.
We note, however, that it is not clear if the light from an older population may just be hidden by this dominant younger population in these galaxies.
Given the very high nucleation fraction of nearby and cluster dwarf ellipticals similar in luminosity to those studied here \citep[e.g.][]{cote+2006,denBrok+2014}, it seems unlikely that these two galaxies lacked a NSC prior to $0.5$ and $2$\,Gyr ago.
Nevertheless, quenching in NGC\,5102 happened recently and proceeded outside in.
Hence, NSCs may be a good place to look for remnant gas in early-type dwarf galaxies.

The results presented in this paper can be improved considerably if a new high resolution SSP grid that includes the near-IR is made available.
The modelling of stellar populations in the IR, however, still remains a challenge as shown in recent work by \citet{baldwin+2018}

\section*{Acknowledgements}
We thank Mark Norris, Iskren Georgiev and Morgan Fouesneau for insightful discussions.
This research has made use of NASA's Astrophysics Data System.
This research has made use of the NASA/IPAC Extragalactic Database (NED) which is operated by the Jet Propulsion Laboratory, California Institute of Technology, under contract with the National Aeronautics and Space Administration. 




\bibliographystyle{mnras}
\bibliography{NSC} 




\appendix

\section{Higher order kinematic moments}\label{sec:h3h4}

\begin{figure*}
	\includegraphics[width=15cm]{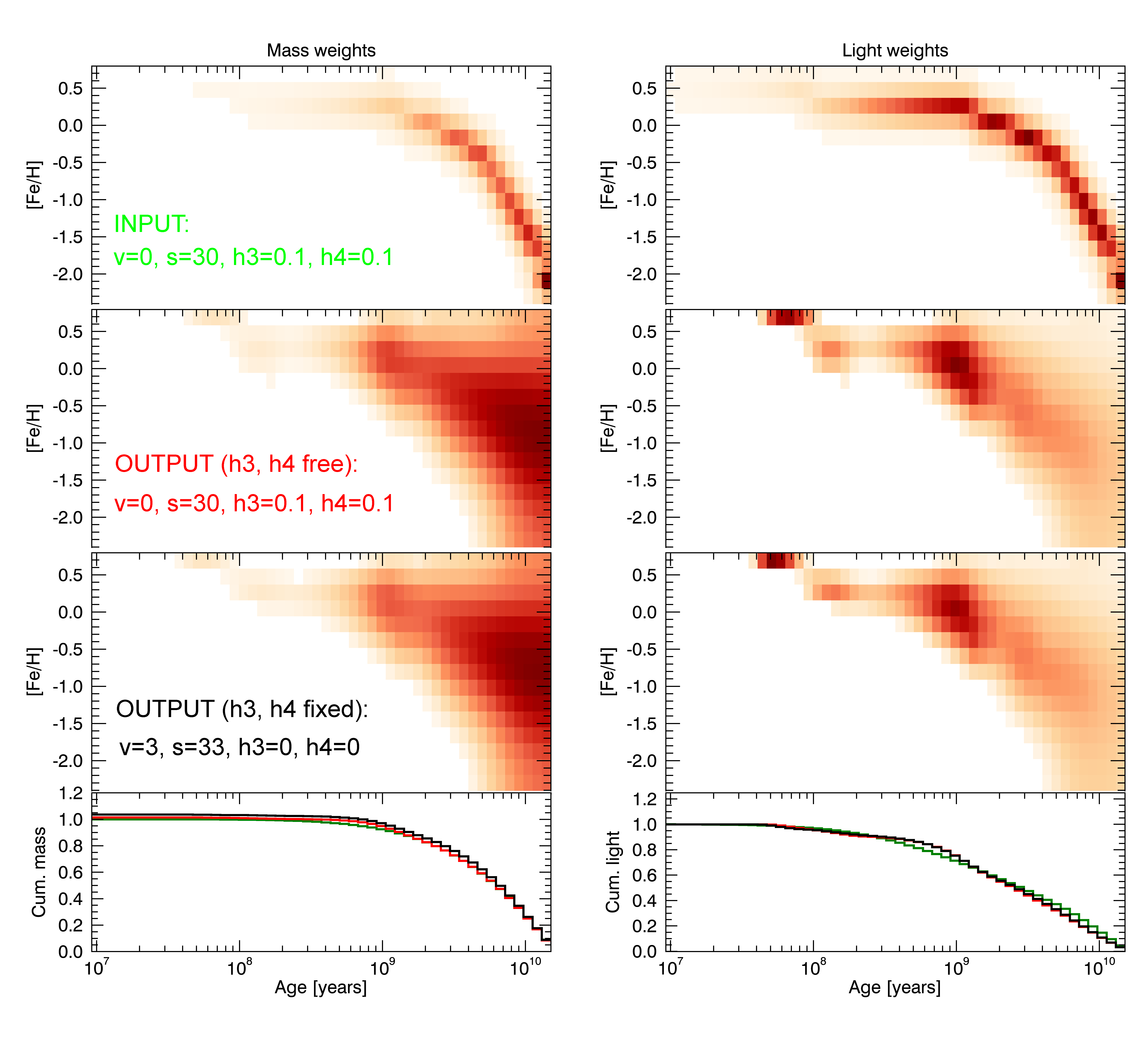}
    \caption{Mock SFH, where the building templates are convolved with non-zero $h3$ and $h4$ velocity moments. The input weights are shown in the {\it top left} panel with corresponding light fractions in the {\it top right} panel. The ppxf outputs when optimizing for $h3$ and $h4$, and setting $h3=h4=0$ are shown in the two {\it middle left} panels with corresponding light fractions in the {\it middle right} panels, respectively. The bottom panels show the cumulative weights and the corresponding light fractions (input - green line, output - red and black lines). The input and best fit velocity moments are given in each panel.}
    \label{fig:mock_h3h4}
\end{figure*}

In this section we investigate the effects that ignoring higher order velocity moments may have on the recovered SFHs of the NSCs.
We created a mock spectrum using the same input SFH as shown in Figure \ref{fig:mock1_ord1} and convolved it with non-zero 3rd and 4th velocity moments ($h_3=h_4=0.1$).
We fitted the spectrum with {\sc ppxf} allowing it to optimize for the first four velocity moments simultaneously.
The best fit results are in excellent agreement with the input parameters (Figure \ref{fig:mock_h3h4}).
Then we performed the fit again, this time ignoring higher order velocity moments and optimizing only for line of sight velocity and velocity dispersion.
The results of this fit are also shown in Figure \ref{fig:mock_h3h4}.
Besides the small bias in line of sight velocity and velocity dispersion, which is expected, the recovered SFHs are qualitatively and quantitatively identical.

We also fitted the observed spectra of the six nuclei allowing {\sc ppxf} to optimize for the first four velocity moments simultaneously.
The kinematic results are presented in Table \ref{tab:h3h4_free}.
We confirmed that the SFHs and the parameters pertaining to the stellar populations of the analysed NSCs remain unchanged.

\begin{table*}
	\centering
	\caption{Kinematic fits setting free the first four velocity moments.}
	\label{tab:h3h4_free}
	\begin{tabular}{ccccc} 
		\hline
Galaxy  & $V_{helio}$ & $\sigma_V$  & $h_3$ & $h_4$  \\
        & [\kms]      & [\kms]      &       &        \\  
		\hline              
NGC\,247  & $174.0\pm0.6$ & $18.7\pm0.6$  & $-0.12\pm0.02$ & $-0.16\pm0.02$ \\
NGC\,300  & $147.2\pm0.4$ & $15.7\pm0.5$  & $-0.07\pm0.02$ & $-0.10\pm0.02$ \\
NGC\,3621 & $720.4\pm0.4$ & $40.4\pm0.6$  & $-0.01\pm0.01$ & $0.03\pm0.01$  \\
NGC\,5102 & $466.4\pm0.3$ & $42.9\pm0.6$  & $-0.04\pm0.01$ & $0.09\pm0.01$  \\
NGC\,5206 & $568.8\pm0.2$ & $32.9\pm0.2$  & $-0.03\pm0.01$ & $-0.02\pm0.01$  \\
NGC\,7793 & $224.4\pm0.4$ & $25.2\pm0.4$  & $-0.02\pm0.01$ & $-0.08\pm0.01$  \\
		\hline
	\end{tabular}
\end{table*}

\section{Mock test with weights distributed in accordance with a first order regularization prior}

\begin{figure*}
	\includegraphics[width=15cm]{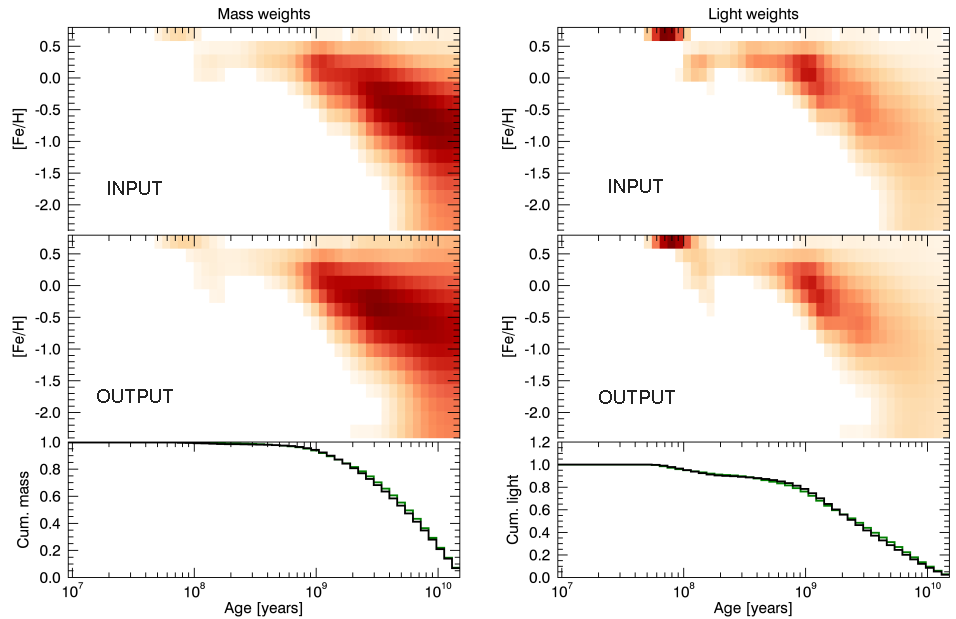}
    \caption{Mock SFH that is convolved with the regularization prior. The input weights are shown in the {\it top left} panel, the corresponding light fractions are shown in the {\it top right} panel. The ppxf output weights are shown in the {\it middle left} panel with corresponding light fractions in the {\it middle right} panel, respectively. The bottom panels show the cumulative weights and the corresponding light fractions (input - green line and output - black line). The fit is performed on the mass weights.}
    \label{fig:mock_prior_sfh}
\end{figure*}

The mock spectrum used for this test (Figure \ref{fig:mock_prior_sfh}) was created by adding noise to the best fit solution for the SFH from Figure \ref{fig:mock1_ord1} with $R=100$.
In this way, the input spectrum is already convolved with the regularization prior with neighbouring input mass weights deviating from the prior by a factor $\sim1/R$.
Then we fit the spectrum again with {\sc ppxf} applying the same regularization factor ($R=100$).
In this special case, the weights distribution of the solution is almost exactly the same as the distribution of the input weights.

\section{Mock data tests with second order regularization}
\begin{figure*}
	\includegraphics[width=15cm]{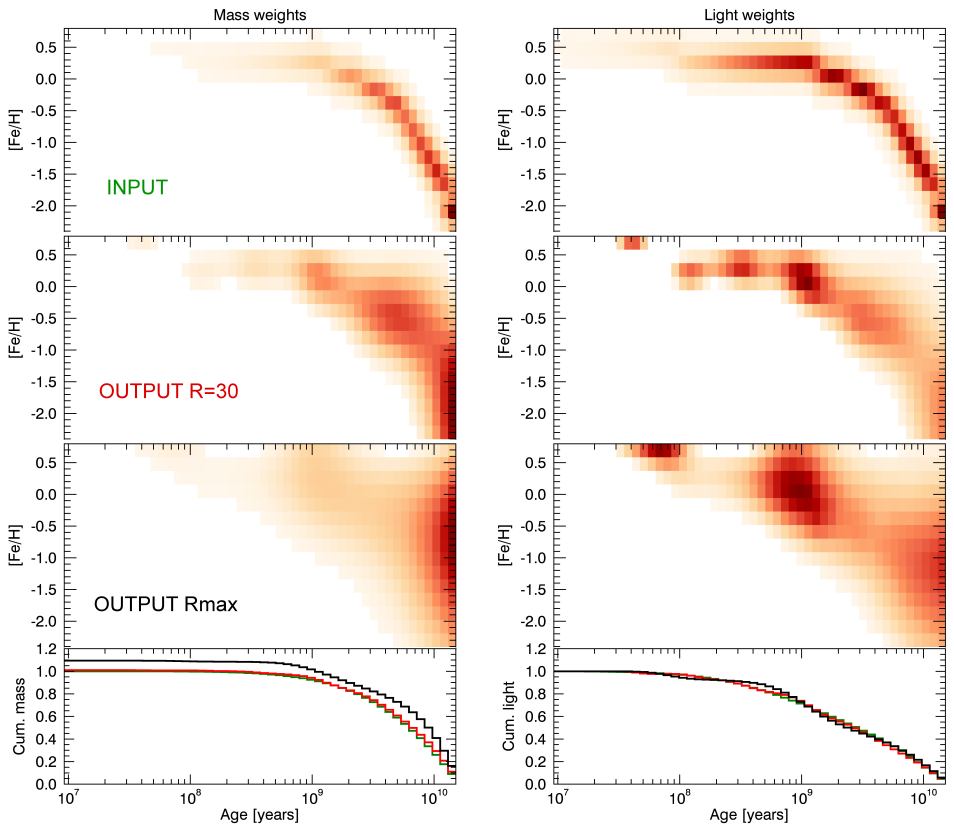}
    \caption{Continuous SFH recovery using second order regularization. The notations are the same as in Figure \ref{fig:mock1_ord1}.}
    \label{fig:mock1_ord2}
\end{figure*}

\begin{figure*}
	\includegraphics[width=15cm]{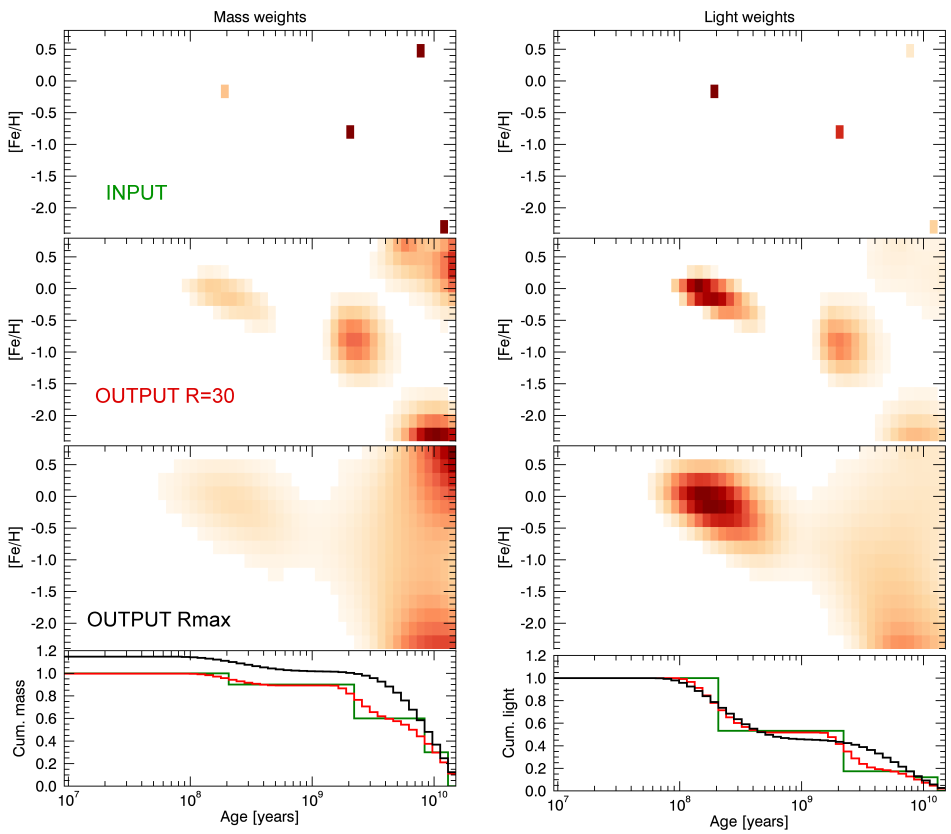}
    \caption{Merger SFH recovery using second order regularization. The notations are the same as in Figure \ref{fig:mock1_ord1}.}
    \label{fig:mock3_ord2}
\end{figure*}



\bsp	
\label{lastpage}
\end{document}